\RequirePackage{rotating}
\documentclass[iop]{emulateapj}
\usepackage{natbib,enumitem,rotating}

\newcommand{\msol}{\,\textrm{M}_\sun}                

\setlist[enumerate]{noitemsep}
\accepted{15 June 2018}
\begin{document}

\title{Resolving Quiescent Galaxies at $z \gtrsim 2$: I. Search for Gravitationally Lensed Sources and Characterization of their Structure, Stellar Populations, and Line Emission}
\shorttitle{Resolving Quiescent Galaxies at $z \gtrsim 2$: I.}
\shortauthors{Newman, Belli, Ellis, and Patel}
\author{Andrew B. Newman$^1$, Sirio Belli$^2$, Richard S. Ellis${}^{3,4}$, and Shannon G. Patel$^1$}
\affil{$^1$ The Observatories of the Carnegie Institution for Science, Pasadena, CA, USA; \href{mailto:anewman@carnegiescience.edu}{anewman@carnegiescience.edu}}
\affil{$^2$ Max-Planck-Institut f\"ur Extraterrestrische Physik (MPE), Giessenbachstr.~1, D-85748 Garching, Germany}
\affil{$^3$  Department of Physics and Astronomy, University College London, Gower Street, London WC1E 6BT, UK}
\affil{$^4$ European Southern Observatory (ESO), Karl-Schwarzschild-Strasse 2, D-85748 Garching, Germany}

\begin{abstract}
Quiescent galaxies at $z\gtrsim2$ are compact and have weak or absent emission lines, making it difficult to spatially resolve their kinematics and stellar populations using ground-based spectroscopy. Gravitationally lensed examples provide a promising route forward, but such objects are very rare. We describe a search in the fields of 232 galaxy clusters that has uncovered five  bright ($H_{\rm AB}<20$) lensed galaxies with red near-infrared colors. These include MRG-M0138, the brightest lensed galaxy known in the near-infrared. Analysis of near-infrared spectra and multiband photometry confirms that all are quiescent galaxies at $z=1.95$-2.64 with stellar ages of 0.5-1.4~Gyr (corresponding to formation epochs $z_{\rm form}\simeq3$-4) and stellar masses of $10^{11.6-12.8}\mu^{-1}\msol$, where $\mu$ is the magnification. In three cases we derive lens models and reconstruct the source structure; these galaxies are massive ($M_* \gtrsim10^{11.0}\msol$) and follow the mass--size relation defined by unlensed samples. In two of these three galaxies, the main structural component is an inclined disk. Weak emission lines are detected in four of five galaxies with high ratios [\ion{N}{2}]/H$\alpha\simeq2$-6 that are inconsistent with a star formation origin. Based on the line ratios, the H$\alpha$ equivalent widths, and the distribution and kinematics of the gas, we infer that shocks are likely present in at least two galaxies and could be present in all of the line emitters. We speculate that these could be analogs of local galaxies in which AGN jet-driven outflows are thought to heat the interstellar medium and suppress star formation. In further papers we will present spatially resolved measurements of the stellar populations and kinematics of this unique sample.
\end{abstract}


\section{Introduction}

Stellar archaeological studies \citep[e.g.,][]{Thomas10} and the evolution of the fundamental plane \citep[e.g.,][]{Kelson97,Treu05,vanderWel05} indicate that the stars in the most massive galaxies were formed at $z \gtrsim 2$. With the advent of deep, wide near-infrared imaging surveys, the emergence of quiescent galaxies is now charted directly. Quiescent galaxies constitute a $10$-20\% minority of the population at $z \gtrsim 3$, even among the highest-mass galaxies, but they become the majority of $\gtrsim 10^{11}\msol$ galaxies by $z \simeq 1.5-2$ \citep{Muzzin13}. 

Although the stellar populations of these early quiescent galaxies may have evolved more-or-less passively after quenching, their structures have evolved dramatically in the intervening 10~Gyr. Numerous studies have shown that the typical size of $z\sim2$ quiescent galaxies is 3-5$\times$ smaller than local counterparts of the same stellar mass \citep[e.g.,][]{Trujillo06,vanDokkum08,Buitrago08,Toft09,Damjanov11,Newman12}. Importantly, the number density of the most compact quiescent galaxies has been declining since $z\sim1.5$ \citep{vanderWel14}, showing that part of this remarkable evolution must arise from the continued growth of massive galaxies after the cessation of star formation \citep{Belli15}. 

Observations of quiescent galaxies at $z \gtrsim 2$ are still rather crude and have largely been confined to bulk properties: number densities, sizes, colors, S\'{e}rsic indices, and shapes. Spectroscopy of the more massive examples has been enabled by near-infrared (NIR) spectrographs on large telescopes (Keck/MOSFIRE, Magellan/FIRE, VLT/X-Shooter) and with grisms on the \emph{Hubble Space Telescope} (\emph{HST}; e.g., \citealt{Whitaker13,Bedregal13,Krogager14,Newman14,Fumagalli16,Lee-Brown17}). These data have been used to measure velocity dispersions, stellar ages, and chemical abundances. These observations are still very demanding due to the faintness of the stellar continuum compared to the NIR background. Within the literature we find only 10 distinct quiescent galaxies beyond $z=2$ that have been observed with the spectral resolution and depth needed to measure stellar kinematics \citep{Kriek09,vandeSande13,Toft12,Newman15a,Kriek16,Hill16,Belli17,Toft17}.

More detailed information is needed to address open questions regarding the formation and evolution of the quiescent population. For example, cosmological simulations predict multiple paths to form a compact quiescent galaxy at $z \sim 2$. In these simulations, some compact galaxies formed very early when the universe was more dense and remained compact; others were once more extended and then ``shrank'' in half-light radius due to centrally concentrated star formation \citep{Zolotov15,Wellons15}. The relative importance of these scenarios could be constrained by spatially resolving the star formation histories in a sample of objects. Another example concerns morphology and dynamics. Observations have indicated a rise in the proportion of flattened quiescent galaxies toward higher redshifts, leading to the inference that quiescent galaxies are more disk-like at early epochs \citep{vanderWel11,Chang13}. Measurements of rotation are needed to observe this directly, but this requires the stellar kinematics to be spatially resolved. Such observations could also help to link recently quenched galaxies to star-forming progenitor populations, whose gas distribution and kinematics are now being measured in the ionized \citep{Barro14,Nelson14,vanDokkum15} and molecular phases \citep{Barro16,Barro17,Tadaki17}.

These examples motivate the need for \emph{spatially resolved} spectroscopy of the stellar continuum for quiescent galaxies at $z \gtrsim 2$. Unfortunately  their small angular sizes (half-light radii $R_e \sim 0\farcs2$) make this impractical in ground-based seeing. Observations with the \emph{HST} grisms or ground-based adaptive optics instruments have the necessary angular resolution, but lack the necessary spectral resolution in the former case and sensitivity in the latter. The best way to resolve the stellar continuum of high-redshift quiescent galaxies using current facilities is to locate gravitationally lensed examples.

Numerous lensed \emph{star-forming} galaxies have been identified and used to spatially resolve the distribution of star formation, gas kinematics, and metals \citep[e.g.,][]{Stark08,Swinbank09,Jones10,Leethochawalit16,Leethochawalit16b}. These lensed sources are optically bright and have been identified either through targeted imaging of massive clusters or panoramic optical imaging surveys \citep{Allam07,Cabanac07,Hennawi08,Stark13}. 

The study of lensed \emph{quiescent} galaxies at high redshifts, on the other hand, is a recent development. \citet{Oldham17} located 14 lensed early-type galaxies at $z \sim 0.6$ whose spectrum was blended with that of the lens galaxy in the Sloan Digital Sky Survey (SDSS). At higher redshifts, only a few examples have been discovered. \citet{Geier13} identified two quiescent galaxies magnified by foreground clusters: A1413-1 at $z = 1.71$ and M2129-1 at $z=2.15$. They used the lensing magnification to study the source structure and spectral energy distribution. \citet{Hill16} studied the intermediate-mass quiescent galaxy COSMOS 0050+4901 which is quadruply-imaged by a foreground galaxy \citep{Muzzin12}. The flux amplification enabled a measurement of velocity dispersion at a remarkably high redshift $z=2.76$. \citet{Ebeling18} recently published the discovery of a very highly magnified quiescent galaxy at $z = 1.6$. In none of these studies was the spectrum spatially resolved in the analysis. This is partly because two of these sources (A1413-1 and COSMOS 0050+4901) are still rather faint ($H_{\rm AB} \gtrsim 21$) and compact even when magnified.

The paucity of known lensed quiescent galaxies, especially at high redshifts, arises from their relatively low density on the sky combined with their optical faintness. Locating lensed quiescent galaxies at $z \gtrsim 2$ requires NIR imaging (since the Balmer/4000~\AA~break is redshifted beyond $1.2 \mu{\rm m}$) that covers a large source plane area.

Motivated by the utility of a such a sample, we embarked on a NIR imaging survey designed to locate particularly bright examples that are extended enough to be spatially resolved from the ground during good seeing conditions. In this paper we describe a search in the fields of 232 massive galaxy clusters using both archival \emph{HST} images and a new imaging survey with the FourStar camera \citep{Persson13} at the Magellan Baade telescope. We have located five magnified galaxies that are exceptionally bright ($H_{\rm AB} < 20$) and have colors consistent with $z \gtrsim 2$ quiescent galaxies. These include what we believe to be the NIR-brightest giant arc known ($H_{\rm AB} = 16.5$). Analysis of their rest-frame optical spectra and ultraviolet-to-NIR photometry confirms that these are quiescent galaxies at $z=1.95$-2.64. Prior to the survey, only one of these five galaxies was known (M2129-1; \citealt{Geier13}). In four cases we are able to spatially resolve the stellar continuum in ground-based NIR spectra. These objects are rare and valuable resources that we will use to investigate the spatially resolved star formation histories and stellar kinematics, which is not currently possible for any other sample.

In \citet{Newman15a} we presented a pilot study of one galaxy in our sample (RGM0150, named MRG-M0150 in the scheme used in this paper). This was the first galaxy beyond $z \simeq 1.1$ for which spatially resolved stellar kinematics were measured. We showed that MRG-M0150 rotates rapidly compared to its likely descendants and therefore must ``spin down'' between $z=2.64$ and the present. \citet{Toft17} showed that the lensed galaxy M2129-1 is a massive quiescent system at $z=2.15$ that is nearly a pure disk and is also rotating surprisingly rapidly. In this and subsequent papers in the series, we will present new spectroscopic observations of this galaxy (MRG-M2129 in our naming scheme) and compare our measurements to those of \citet{Toft17}. 

In this paper we begin presenting results from the survey. In Section~2 we introduce the imaging program used to locate the five lensed quiescent galaxies. Sections~3 and 4 describe the follow-up imaging and spectroscopic observations. In Section~5 we derive lens models for three systems and use these to reconstruct the sources. In Section~6 we analyze the unresolved spectra and photometry of the lensed galaxies in order to establish their bulk properties, including redshifts, stellar ages, and quiescence. In Section~7 we demonstrate the nearly ubiquitous presence of low-level line emission in our sample and discuss its possible origins. In Section 8 we discuss the representative nature of our lensed galaxy sample, the implications of their structures and emission line properties for their evolutionary histories, and the utility of the sample for future spatially resolved studies.

In the companion Paper~II, we measure the resolved stellar kinematics of four galaxies in the sample. Further papers will discuss the resolved stellar ages and chemical abundances.

Throughout we refer to magnitudes on the AB system and assume a flat $\Lambda$CDM cosmology with $\Omega_m = 0.3$ and $H_0=70$~km~s${}^{-1}$~Mpc${}^{-1}$.

\begin{figure*}
\centering
\includegraphics[width=0.8\linewidth]{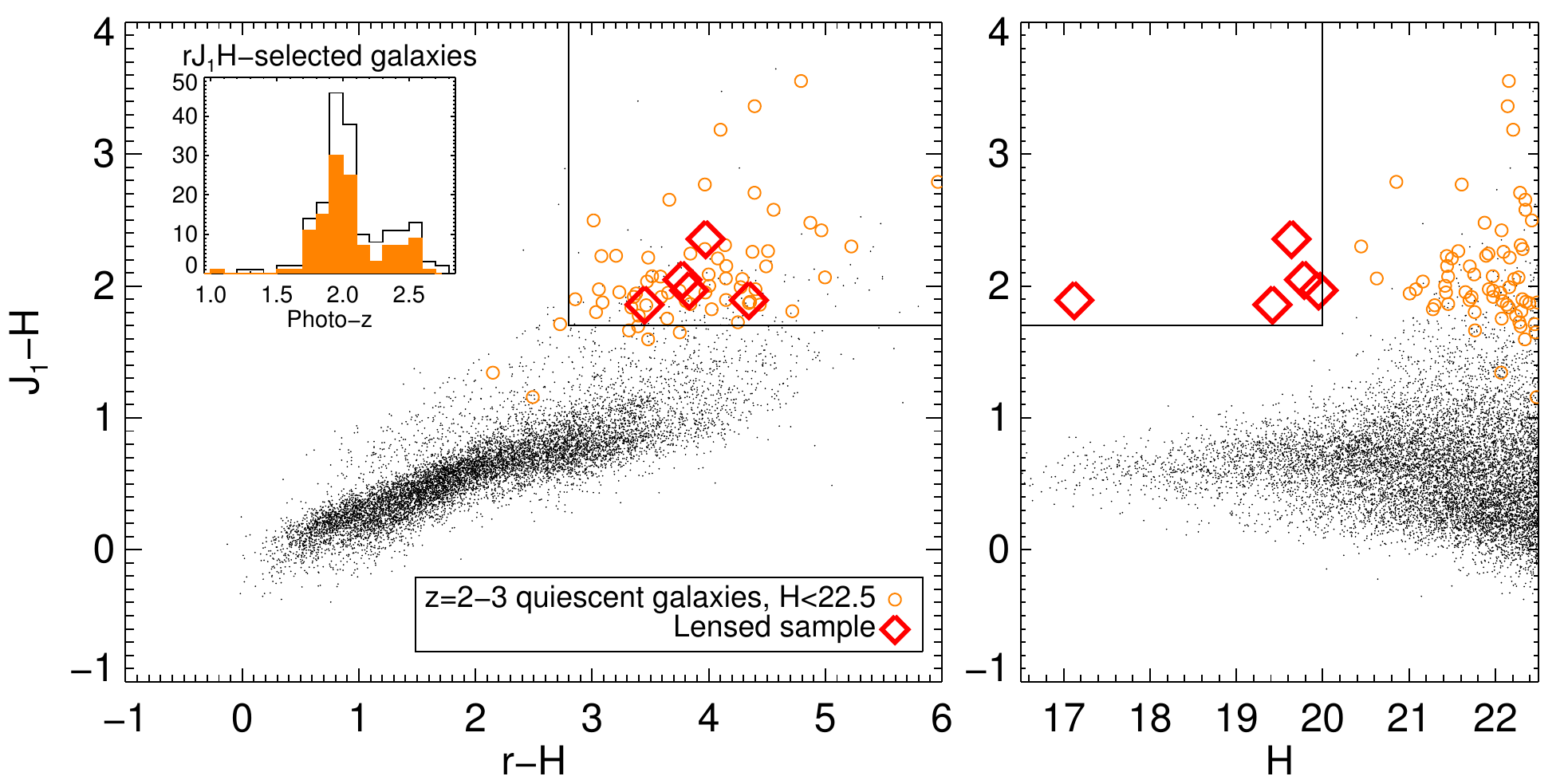}
\caption{Color-color (left panel) and color-magnitude (right) diagrams showing all $H < 22.5$ galaxies selected from the NMBS/COSMOS catalog as black points. The targeted subpopulation, consisting of $UVJ$-quiescent galaxies with $2 < z_{\rm phot} < 3$, are shown as orange circles. Black lines enclose the selection boxes used for our lensed galaxy search. The inset in the left panel shows the photometric redshift distribution of NMBS galaxies falling in the color selection box, with the $UVJ$-quiescent subset filled. In both panels the five lensed galaxies identified in our search (two from the FourStar and three from the \emph{HST} components) are plotted as red diamonds. The Subaru/SuprimeCam $r^+$ filter was used for the NMBS galaxies. For the lensed galaxies, which were observed with various filters, we integrated the fitted SED to synthesize $r^+J_1H$ magnitudes.\label{fig:YH}}
\end{figure*}

\section{Search for Gravitationally Lensed Quiescent Galaxies at $z \gtrsim2$}

We now describe our search for lensed quiescent galaxies at $z \gtrsim 2$ in the fields of 232 massive galaxy clusters. After motivating the need to image a large number of clusters, we derive an efficient color selection, outline its application to archival \emph{HST} data and new Magellan/FourStar imaging, and present our yield of five lensed quiescent galaxies.

\subsection{Basic Considerations\label{sec:considerations}}

Massive quiescent galaxies at high redshifts are relatively rare. The probability to find such a galaxy magnified by a given cluster is the product of the population's surface density and the source plane area magnified by the cluster above some minimum magnification of interest. The typical value of this area depends on the cluster sample, so only rough estimates are possible. An upper limit can be estimated from the Hubble Frontier Fields, which were chosen to be among the most powerful cluster lenses. For these clusters, \citet{Johnson14} derive a typical source plane area of $\simeq 0.3$~arcmin${}^{2}$ magnified by $\mu > \mu_{\rm min} = 3$. This area scales approximately as $\mu_{\rm min}^{-2}$. These figures enable a rough estimate of the frequency of lensed sources magnified above some limiting flux. We consider a source population of quiescent galaxies at $z=1.8-3$ whose magnitude distribution we estimate using the NEWFIRM Medium Band Survey (NMBS) catalogs \citep{Whitaker11}. Here quiescence is defined using the $UVJ$ criterion in the form presented by Whitaker et al. We find a probability of 0.02 per cluster to find such a galaxy magnified to $H < 20$.  As we will show, this flux limit is of interest because it is brighter than  unlensed examples found in the widest extragalactic deep fields. Although the estimated probability is uncertain and should be considered an upper limit, it shows that one must search hundreds of clusters to find a few very bright, lensed quiescent galaxies at these redshifts.

\subsection{Color Selection}

Although they are rare, massive quiescent galaxies at $z \simeq 2$-3 can be efficiently identified with two or three filter imaging, and their high NIR surface brightnesses imply that relatively shallow exposures are sufficient. This makes it feasible to search large numbers of clusters.

Quiescent galaxies at high redshifts can be identified on the basis of their red colors at NIR wavelengths, which are produced by the redshifted Balmer/4000~\AA~break \citep{Franx03}. The addition of an optical magnitude probing the rest-frame UV further helps to eliminate dusty star-forming galaxies \citep{Daddi04}. We have applied this basic strategy to two imaging surveys of massive clusters: (1) a dedicated campaign with the FourStar camera at the Magellan Baade telescope, and (2) a search of archival data from Wide Field Camera 3 (WFC3) onboard \emph{HST}. These surveys will be described in Sections~\ref{sec:fourstarsample} and \ref{sec:hstsample}.

\begin{deluxetable*}{lcccclcl}
\tablecolumns{8}
\tablewidth{0pt}
\tablecaption{Sample of Lensed Quiescent Galaxies}
\tablehead{\colhead{Lensed Galaxy} & \colhead{R.A.~(hr)} & \colhead{Dec.~(deg)} & \colhead{$z_{\rm spec}$} & \colhead{$H$ (mag)} & \colhead{Lensing Cluster} & \colhead{$z_{\rm lens}$} & \colhead{Source of $z_{\rm lens}$}}
\startdata
MRG-M0138 & 01:38:03.9 & -21:55:49 & 1.95 & 17.1 & MACSJ0138.0-2155 & 0.338 & This paper\\ 
MRG-M0150 & 01:50:21.0 & -10:05:14 & 2.64 & 19.6 & MACSJ0150.3-1005 & 0.365 & SDSS DR12 \citep{Alam15} \\ 
MRG-P0918 & 09:18:34.1 & -81:03:08 & 2.36 & 19.4 & PSZ1 G295.24-21.55 & 0.61 & \citet{Ade13} \\
MRG-S1522 & 15:22:53.6 & +25:35:49 & 2.45 & 19.8 & SDSSJ1522+2535 & 0.58 & SDSS DR12 (photometric) \\
MRG-M2129 & 21:29:22.3 & -07:41:31 & 2.15 & 20.0 & MACSJ2129.4-0741 & 0.589 & \citet{Ebeling07}
\enddata
\tablecomments{In cases of multiple images, listed coordinates and magnitudes are those of the brightest. \label{tab:sample}}
\end{deluxetable*}

Using the NMBS catalogs, we experimented with color cuts that efficiently identify a target population of $UVJ$-quiescent galaxies at $z=2$-3 with $H < 22.5$, which corresponds to the desired lensed magnitude limit of $H < 20$ for a maximum expected magnification of $\mu = 10$. (The precise limit is not relevant since the colors do not vary drastically with magnitude.) The goal was to balance a high completeness for selecting this target population (orange circles in Figure~\ref{fig:YH}) with minimal contamination from star-forming or lower-redshift galaxies (black points). The FourStar filter set includes the broad-band $J$, $H$ and $K_s$ filters and the medium-band $J_1$, $J_2$, $J_3$, $H_s$, and $H_l$ filters \citep{Persson13}. We found that the combination $J_1 - H > 1.7$ is nearly optimal, as shown in Figure~\ref{fig:YH}. A second cut of $r - H > 2.8$ is also plotted. For this filter combination, the optical-IR cut removes only a few galaxies and so is not necessary to identify bright quiescent candidates in our FourStar imaging; $J_1-H$ is sufficient. For our \emph{HST} archival search, we adapted these color cuts based on the available filters as described in Section~\ref{sec:hstsample}.  The optical-IR cut plays a greater role for some of these filter combinations.

The color cuts $J_1 - H > 1.7$ and $r-H > 2.8$ select 85\% of the target population, so our selection is reasonably complete. Furthermore, 31\% of the color-selected galaxies are quiescent galaxies at $z > 2$, and of the remainder, an additional 33\% are quiescent galaxies at slightly lower redshifts $z = 1.6$-2 which are still of interest. This level of purity makes it possible to pursue efficient spectroscopic follow-up to confirm the redshifts and quiescent nature of the sources. As outlined below, we have examined the fields of 232 clusters and have located five color-selected galaxies that are magnified above the $H < 20$ flux limit. Fainter color-selected galaxies were also identified, but in this paper we confine ourselves to $H < 20$ sample, for which our spectroscopic follow-up is complete.

\subsection{Magellan/FourStar Search\label{sec:fourstarsample}}

We imaged 131 clusters through the $J_1$ and $H$ filters with FourStar over the five semesters from 2014A to 2016A. Targets were drawn from several sources. Approximately 40\% were X-ray--selected clusters from the Massive Cluster Survey (MACS; \citealt{Ebeling01,Repp17}); these were included in the target lists of the \emph{HST} snapshot programs described in Section~\ref{sec:hstsample} but had not yet been observed with WFC3-IR. Another $\simeq40\%$ were selected from the first \emph{Planck} catalog of Sunyaev-Zel'dovich sources \citep{Ade13}; to maximize the lensing efficiency, we considered only sources that had been confirmed as clusters at redshifts $0.3 < z < 0.8$ and gave priority to those with higher signal-to-noise ratios in the \emph{Planck} maps. Finally, $\simeq20\%$ of our FourStar targets were optically selected clusters drawn primarily from the redMaPPer DR8 catalog \citep{Rykoff14} and prioritized by richness. A handful were drawn from the \citet{Wong13} catalog of fields suggested to be powerful lenses.

The typical observing sequence was to move to the target, correct the focus and mirror figure with the facility Shack-Hartmann system, obtain $5 \times 5.8$~sec unguided exposures in the $H$ band at each of 11 random dither positions within a $90''\times90''$ box, followed by $2 \times 32$~sec exposures at 13 random dither positions through the $J_1$ filter. The total wall clock time per cluster was approximately 30~min in ordinary conditions.

The data were reduced using the automated system described by \citet{Kelson14}, which produces stacked images for each cluster and filter with astrometry tied to 2MASS. Photometric calibration was obtained using stars in the 2MASS point source catalog \citep{Skrutskie06}. This is straightforward for the $H$ and $K_s$ filters, which are present in the 2MASS catalog. For the medium-band filters, we used the \citet{Pickles98} stellar library to derive mean relations that relate the 2MASS magnitudes of a star to its $J_1$, $J_2$, and $J_3$ magnitudes.

Applying the cuts $J_1-H>1.7$ and $H < 20$ to this sample of 131 clusters yielded five sources. Two of these are clearly magnified by the clusters PSZ1-G295.24-21.55 and MACSJ0138.0-2155. As the right panel of Figure~\ref{fig:YH} shows, these objects occupy a region of color--magnitude space that is virtually empty in extragalactic field surveys. The other three sources are compact and are located at least 4~arcmin from the cluster center. The FourStar field of view is much larger than the high-magnification region of a cluster, so these three sources are likely unmagnified systems on the very bright tail of the luminosity function. Their presence is not unexpected given that the FourStar survey encompasses 4.3 deg${}^2$, but since they are not magnified, we will not discuss these sources further. The yield of $2/131\simeq 0.02$ lensed quiescent galaxies per cluster is consistent with the upper limits roughly estimated in Section~\ref{sec:considerations}.

\begin{sidewaysfigure*}
\centering
\vspace{9cm}
\includegraphics[width=0.32\linewidth]{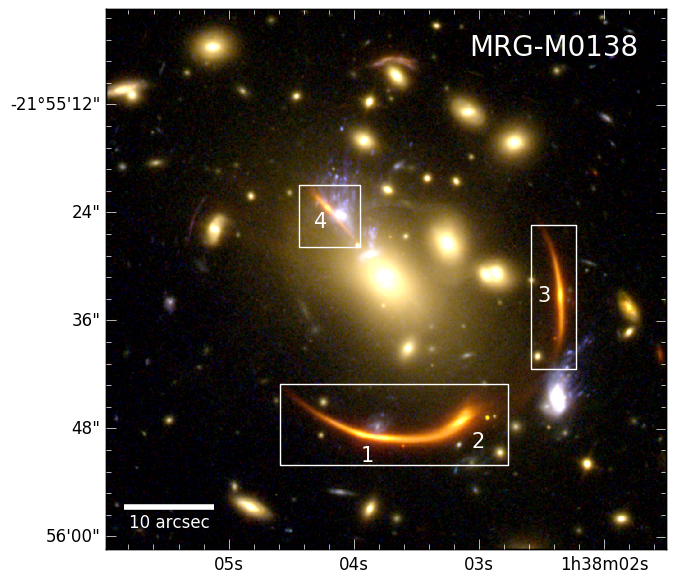}
\includegraphics[width=0.32\linewidth]{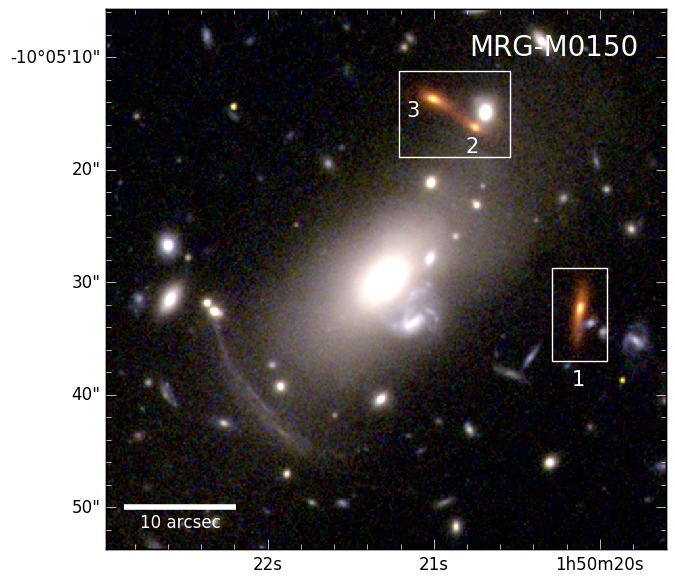}
\includegraphics[width=0.32\linewidth]{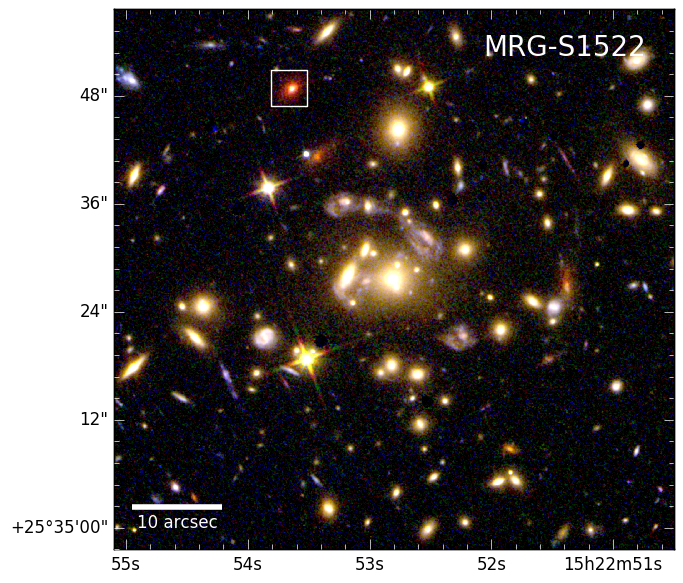} \\
\includegraphics[width=0.47\linewidth]{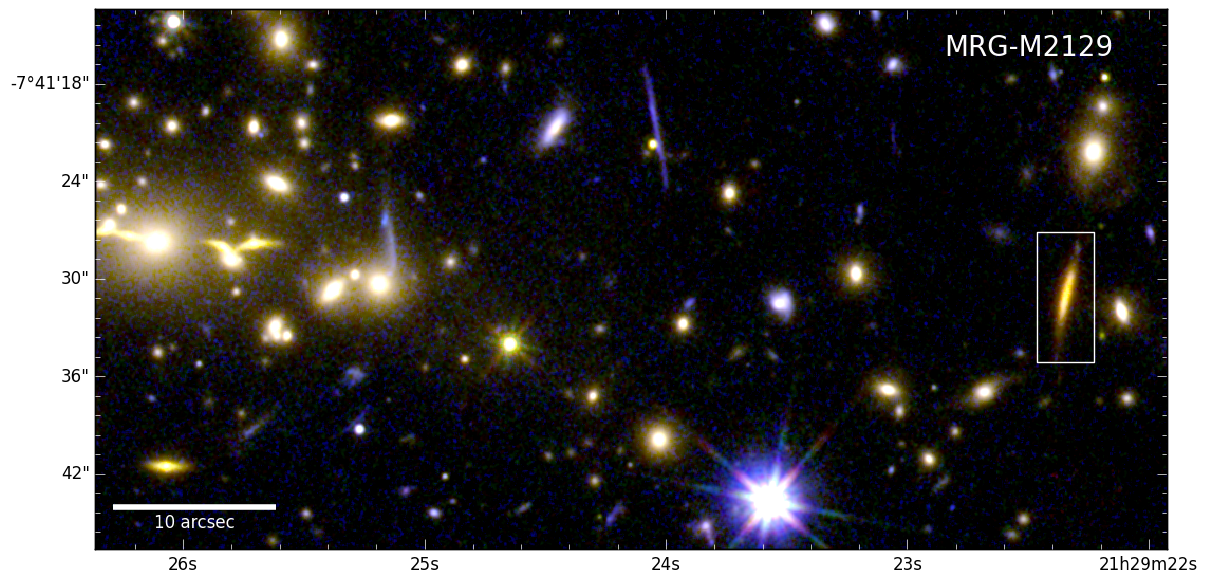}
\includegraphics[width=0.47\linewidth]{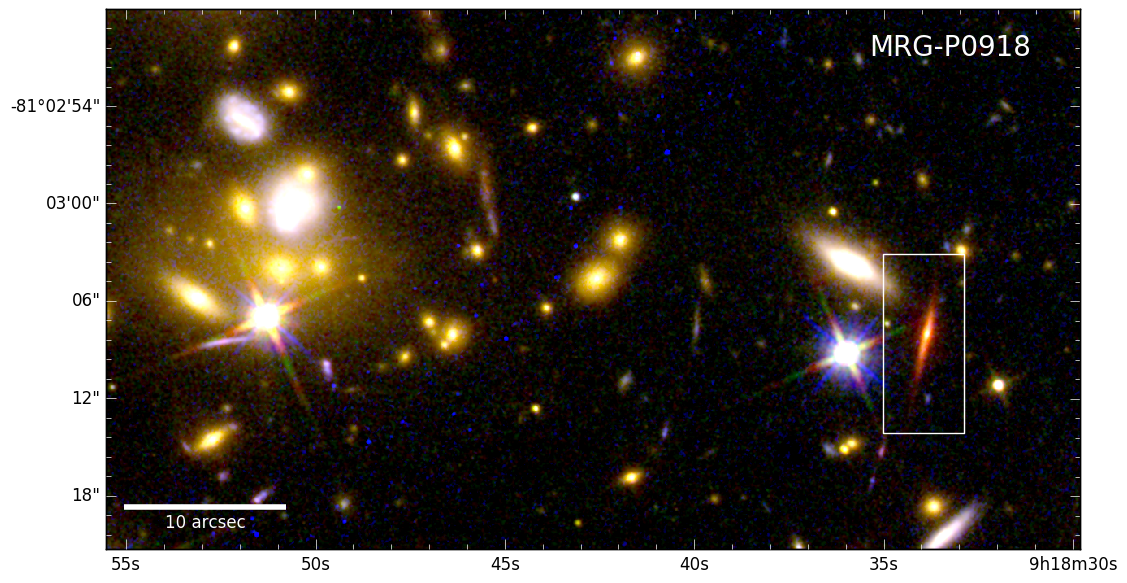} \\
\caption{Composite \emph{HST} images of the sample displayed with a logarithmic stretch. The lensed quiescent galaxies are identified by boxes. Multiple images are numbered in the cases of MRG-M0150 and MRG-M0138. (The central image of MRG-M0138 is not labeled.) Depending on the available filters, one of F555W, F606W, or F814W was used for the blue channel, F105W or F125W for the green, and F160W for the red.
\label{fig:hstimages}}
\end{sidewaysfigure*}

\subsection{HST Archival Search\label{sec:hstsample}}

Many galaxy clusters have been observed with \emph{HST}, but a much smaller subset have been observed with WFC3-IR through two filters and with ACS or WFC3-UVIS through at least one optical filter, as required to implement our color criteria. At the time of our search (2014-2016) the vast majority of such observations had been undertaken through one of three programs: (1) a snapshot imaging campaign based on the Massive Cluster Survey (MACS), led by P.I.~H.~Ebeling in Proposal IDs 10491, 10875, 12166, and 12884; (2) multi-band imaging of lensing clusters identified in the Sloan Digital Sky Survey, led by P.I.~M.~Gladders in Proposal ID 13003, and (3) the Cluster Lensing and Supernova survey with Hubble (CLASH), a multi-cycle treasury program led by P.I.~M.~Postman in Proposal ID 12065.

For all clusters included in these programs that, at the time of the analysis, had been observed through two WFC3-IR filters and one or more ACS or WFC3-UVIS filters, we produced multi-band photometric catalogs and applied color and magnitude criteria to search for lensed quiescent galaxies. Since the surveys used various filters, we adapted the color criteria shown in Figure~\ref{fig:YH} for each survey. To do so, for all galaxies in the NMBS catalogs used to construct Figure~\ref{fig:YH}, we integrated the spectral energy distribution fit to produce synthetic magnitudes in the relevant \emph{HST} filters. We then reproduced Figure~\ref{fig:YH} with the appropriate filter combinations and adjusted the color thresholds to match the balance of completeness and contamination shown in the Figure.

Within the MACS cluster sample, we processed 46 clusters with WFC3-IR and ACS imaging. The color criteria became (${\rm F110W}-{\rm F140W} > 0.7$) and (${\rm F814W-F140W} > 2$ or ${\rm F606W} - {\rm F140W} > 2.5$). We also imposed a flux limit of ${\rm F140W} < 20.7$, which reflects our fiducial $H< 20$ cut adjusted by the mean ${\rm F140W}-H$ color of the target population. We identified one lensed galaxy behind the cluster MACSJ0150.3-1005.

For the Gladders sample, we processed 30 clusters which had been imaged with WFC3-IR and WFC3-UVIS. The color criteria became (${\rm F125W}-{\rm F160W} > 0.7$ or ${\rm F110W}-{\rm F160W} > 1.0$ or ${\rm F105W}-{\rm F160W} > 1.4$) and (${\rm F606W} - {\rm F160W} > 3$) and ${\rm F160W} < 20.3$. We identified one source magnified by the cluster SDSSJ1522+2535.

For the CLASH sample, 16 bands of imaging are available for 25 clusters. The color criteria became ${\rm F110W} - {\rm F160W} > 0.9$ and ${\rm F814W} - {\rm F160W} > 2.5$ and ${\rm F160W} < 20.3$. We identified one source magnified by MACSJ2129.4-0741.

\begin{figure}
\centering
\includegraphics[width=\linewidth]{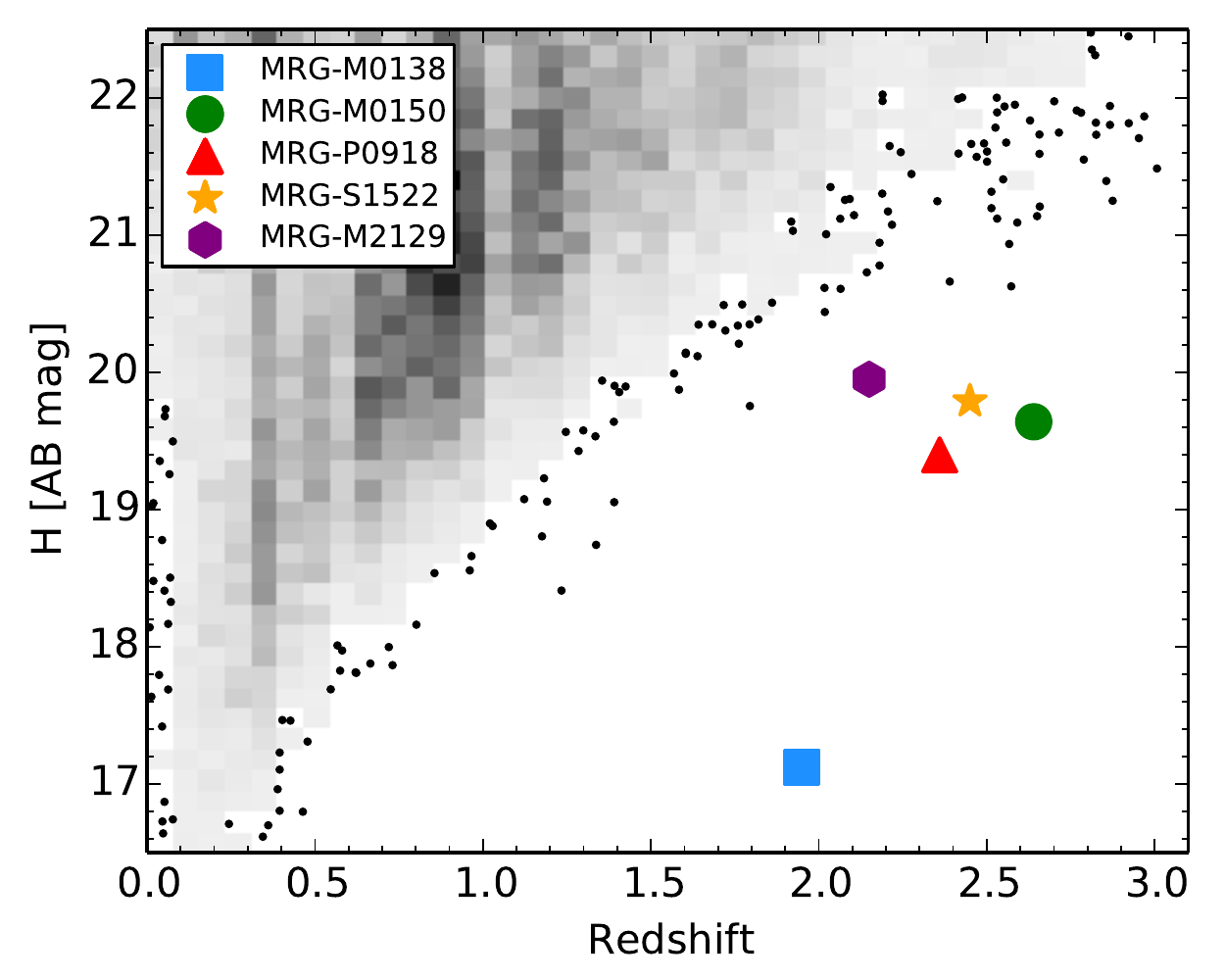}
\caption{The magnitude distribution of $UVJ$-quiescent galaxies drawn from the UltraVISTA survey \citep{Muzzin13catalog} is indicated by the grey histogram, with individual black points shown where the density is low. The magnified red galaxy (MRG) sample is significantly brighter than the brightest unlensed galaxies found over the 1.6~deg${}^2$ field.\label{fig:mag_redshift}}
\end{figure}

\subsection{The Lensed Quiescent Galaxy Sample}

The coordinates and basic properties of the five lensed color-selected galaxies located in our survey are listed in Table~\ref{tab:sample}. We name these galaxies MRG-M0138, MRG-M0150, MRG-P0918, MRG-S1522, and MRG-M2129 based on the name of the lensing cluster, with the prefix MRG denoting Magnified Red Galaxy. Of these objects, MRG-M2129 was previously identified by \citet{Geier13} and studied by \citet{Toft17}, while the other four were discovered in this survey; a pilot study of MRG-M0150 was presented by \citet{Newman15a}.

Figure~\ref{fig:hstimages} shows \emph{HST} images of the five lensed galaxies. We now briefly describe each of these galaxies, their lensing configurations, and the lensing clusters. We note that the distinctive colors and high surface brightnesses of the lensed quiescent galaxies makes it straightforward to identify multiple images, which are confirmed by lens modeling in Section~\ref{sec:lensmodels}.

MRG-M0138 is a remarkable system that presents five multiple images: two merging images forming a giant arc southward of the cluster center, a single image on the western side, a radial arc near the brightest cluster galaxy (BCG), and a central image (not visible in Figure~\ref{fig:hstimages}). The brightest image (Image 1) has $H = 17.1$, which is astoundingly bright for this redshift:  Figure~\ref{fig:mag_redshift} shows that this image is $\sim 3$~mag brighter than the brightest unlensed quiescent galaxies at similar redshifts found in the 1.6~deg${}^2$ UltraVISTA field \citep{Muzzin13catalog}! Summing the two merging images 1 and 2, the total AB magnitudes of the giant arc are $H = 16.5$ and $K_s = 16.1$. We believe this is the brightest giant arc known at NIR wavelengths, at least among $z \gtrsim 1$ sources. It is 1.1~mag brighter at $K_s$ than the giant arc discovered by \citet{Dahle16} and 2.3 mag brighter than the source described by \citet{WuytsEva10}. For the lens MACSJ0138.0-2155, we measured a redshift of 0.338 based on a Magellan/LDSS-3 spectrum of the brightest cluster galaxy.

MRG-M0150 is also a multiple image system, presenting 3 multiple images in a ``naked cusp'' configuration analyzed by \citet{Newman15a}. MRG-P0918, MRG-S1522, and MRG-M2129 are all singly imaged. The images of MRG-P0918 and MRG-M2129 are both highly elongated, whereas that of MRG-S1522 is bright but only modestly elliptical. The clusters lensing MRG-P0918 and MRG-S1522 are not well known and, as we will discuss below, we do not have the constraints needed to determine the magnification factor or reconstruct the source in these cases. MACSJ2129.4-0741, on the other hand, is one of the most X-ray--luminous clusters (\citealt{Ebeling07}; $z=0.589$) and is a well-studied lens with multiple public lens models.

Figure~\ref{fig:mag_redshift} compares the observed (lensed) magnitudes of this sample to unlensed quiescent galaxies in the UltraVISTA field \citep{Muzzin13catalog}. All sources in our lensed sample are substantially brighter than even the brightest quiescent galaxies located at comparable redshifts, even within the largest deep fields. This makes them premier targets for absorption line spectroscopy in the near-infrared and enables a number of new observations, including their spatially resolved stellar kinematics (Paper~II) and populations. In the following sections, we will describe follow-up observations verifying the quiescent nature and redshifts $z=1.95$-2.64 of these sources (Sections 3 and 4) before deriving the lensing magnifications (Section 5) and the source intrinsic properties (Sections 6-7).

%
%
%
%

\section{Imaging Data and Photometry}
\label{sec:imagingphotometry}

We observed our lensed galaxy sample using various ground- and spaced-based telescopes to characterize their spectral energy distributions from optical to near-infrared wavelengths. The data described in this section will be used to construct lens models and to study the stellar populations and structures of our sample in Sections~\ref{sec:lensmodels}-\ref{sec:stellarpops}.

\subsection{HST Observations\label{sec:hstobs}}

For the lensed galaxies located in our \emph{HST} search, a variety of archival data exists with sources listed in Section~\ref{sec:hstsample}. For MRG-M0150, we used archival WFC3-IR/F140W, WFC3-IR/F110W, ACS/F814W, and WFPC2/F606W images as described in \citet{Newman15a}. In order to obtain deeper images with improved sub-pixel sampling and a filter combination that better straddles the Balmer break, we obtained additional WFC3-IR images through the F160W and F125W filters, each at half-orbit depth (Proposal ID 14205, P.I.~A.~Newman). For MRG-S1522, we used WFC3-IR/F160W, WFC3-IR/F105W, and WFC3-UVIS/F606W archival images from the Gladders program (see Section~\ref{sec:hstsample}). For MRG-M2129, we used archival CLASH imaging through 13 ACS and WFC3 filters (see Table~\ref{tab:phot}).

No archival data existed for the two lensed galaxies located in our FourStar survey. For MRG-P0918, we therefore obtained WFC3-IR/F160W and F105W images, each with half-orbit depth, and ACS/F814W and F555W images, each with one-orbit depth, through the aforementioned program. For MRG-M0138, observations were undertaken through a mid-cycle program (Proposal ID 14496, P.I.~A.~Newman). Two orbits were split between WFC3-IR/F160W (1.6~ks) and F105W (3.6~ks) exposures, and two further orbits were devoted to an F555W exposure.

All new \emph{HST} observations employed standard sub-pixel dither patterns and were reduced using the {\tt MultiDrizzle} package \citep{Koekemoer03}. For the source reconstructions we will perform in Section~\ref{sec:lensmodels}, we require an estimate of the WFC3 F160W point spread function (PSF). For MRG-M0150 and MRG-M2129, we constructed PSFs from stars in the mosaic. Suitably bright and isolated stars were not present for MRG-M0138, so we instead generated a synthetic TinyTim \citep{Krist11} PSF and convolved it with a circular gaussian to best match a (slightly broadened) star in the mosaic. As expected, the curves of growth for all three PSFs are very similar.

\subsection{Spitzer/IRAC Observations\label{sec:spitzerobs}}

We analyzed archival images from the Infrared Array Camera (IRAC) onboard the \emph{Spitzer} Space Telescope for three galaxies: MRG-P0918 was observed in Program ID 90233 by P.I.~C.~Lawrence, MRG-S1522 was observed in Program ID 70154 by P.I.~M.~Gladders, and MRG-M2129 was observed in Program ID 90009 by P.I.~M.~Brada\v{c}. We obtained new IRAC images of MRG-M0138 via a Director's Discretionary Time program (Program ID 12127) and of MRG-M0150 via a joint \emph{HST} program in cycle 23 (Program ID 12003, P.I.~A.~Newman). 

The IRAC images cover the 3.6$\mu$m and 4.5$\mu$m channels with the exception of MRG-P0918. This galaxy falls near the edge of the 4.5$\mu$m mosaic and suffers from various artifacts; we therefore excluded the image from our analysis. We performed photometry using the standard calibrated mosaics produced by the IRAC pipeline with a $0\farcs6$ pixel scale.
 
\subsection{Ground-based Observations\label{sec:groundobs}}

We imaged the entire sample through various filters using FourStar at the Magellan Baade telescope. For MRG-M0138, MRG-P0918, and MRG-S1522, we obtained images through the $J_1$, $J_2$, $J_3$, $H$, and $K_s$ filters, using the medium bands to improve the sampling around the Balmer/4000~\AA~break. For MRG-M0150 and MRG-M2129, due to the greater number of \emph{HST}/WFC3-IR observations, we obtained FourStar images in $J$, $H$, and $K_s$ for the former and only in $K_s$ for the latter.

At optical wavelengths, we observed MRG-M0138 through the $g$, $r$, and $z$ filters using the LDSS-3 imaging spectrograph at the Magellan Clay telescope. We also observed MRG-S1522 through the $z$ filter. Photometric calibration was obtained using images of stellar fields in the Sloan Digital Sky Survey.

\subsection{Photometry\label{sec:phot}}

In each field, we produced pixel registered images with matching PSFs following the procedures described by \citet{Newman12}. Briefly, the PSF in each image was constructed by combining suitable bright stars. The ground-based and \emph{HST} images were convolved to match the image with the worst seeing, which was always $\leq 1''$.  Since the IRAC PSFs are considerably broader, we did not convolve all images to match the IRAC resolution. Instead the arc flux measured in the IRAC images was scaled by the fraction of light lost from the photometric aperture when the F160W image was convolved to match the IRAC PSF (see \citealt{Newman12}).

We measured colors in rectangular apertures aligned with the direction of magnification that had the following dimensions: $4'' \times 1\farcs5$ on MRG-M0138 Image 3, $1\farcs6 \times 1\farcs2$ on MRG-M0150 Image~1, $2'' \times 1\farcs5$ on MRG-P0918, $1\farcs5 \times 1\farcs5$ on MRG-S1522, and $3\farcs2 \times 1\farcs5$ on MRG-M2129. The lengths of these apertures (in the direction of maximum magnification) approximately match the spectroscopic apertures described in Section~\ref{sec:spec_ex} and therefore are appropriate for a joint analysis of the spectra and photometry.\footnote{Where slight aperture differences are present, these are to avoid contamination from foreground objects that is negligible in the NIR spectra but may be significant in bluer filters.} The rectangular apertures are wider ($1\farcs5$) than the spectrograph slit ($0\farcs6$-$0\farcs75$), since we preferred to avoid selecting an aperture much smaller than the IRAC PSF. However, this does not introduce a significant mismatch between the spectroscopic and photometric apertures because the images are narrow in the direction of minimum magnification. 

For each galaxy, the fluxes were then uniformly scaled to match the total F160W flux within a large aperture and corrected for Galactic extinction following \citet{SF11}. Uncertainties of 5\% (3\% for \emph{HST} measurements) were added in quadrature to account for uncertainties in the PSF matching and photometric calibration. The resulting photometric measurements are listed in Table~\ref{tab:phot}.

For MRG-M0138, the colors were measured on Image~3, which is the most isolated and affords the cleanest photometry. In Table~\ref{tab:phot} we have scaled these fluxes to match the total observed F160W flux of the brightest image and our spectroscopic target, Image~1. (Although there is some ambiguity in separating the merging Images 1 and 2, this affects only the total fluxes in Table~\ref{tab:phot} and has no consequence for any of the inferred source properties, which will ultimately be scaled based on a model of the source.)

\begin{deluxetable}{llc}
\tablecolumns{3}
\tablewidth{0pt}
\tablecaption{Photometry\label{tab:phot}}
\tablehead{\colhead{Instrument} & \colhead{Filter} & \colhead{AB mag.}}
\startdata
\multicolumn{3}{c}{MRG-M0138} \\[0.5ex]
LDSS & $g$ & $22.61 \pm 0.19$ \\
ACS & F555W & $22.26 \pm 0.20$ \\
LDSS & $r$ & $21.49 \pm 0.15$ \\
LDSS & $i$ & $20.80 \pm 0.13$ \\
LDSS & $z$ & $19.64 \pm 0.12$ \\
FourStar & $J_1$ & $19.06 \pm 0.07$ \\
WFC3-IR & F105W & $18.93 \pm 0.04$ \\
FourStar & $J_2$ & $18.40 \pm 0.07$ \\
FourStar & $J_3$ & $17.76 \pm 0.06$ \\
WFC3-IR & F160W & $17.28 \pm 0.03$ \\
FourStar & $H$ & $17.08 \pm 0.06$ \\
FourStar & $K_s$ & $16.67 \pm 0.06$ \\
IRAC & Ch. 1 & $16.21 \pm 0.05$ \\
IRAC & Ch. 2 & $16.00 \pm 0.05$ \\[0.5ex]
\multicolumn{3}{c}{MRG-M0150} \\[0.5ex]
WFPC2 & F606W & $23.89 \pm 0.16$ \\
ACS & F814W & $22.87 \pm 0.06$ \\
WFC3-IR & F110W & $21.52 \pm 0.04$ \\
FourStar & $J$ & $21.27 \pm 0.08$ \\
WFC3-IR & F125W & $21.17 \pm 0.03$ \\
WFC3-IR & F140W & $20.43 \pm 0.03$ \\
WFC3-IR & F160W & $19.90 \pm 0.03$ \\
FourStar & $H$ & $19.63 \pm 0.06$ \\
FourStar & $K_s$ & $19.16 \pm 0.06$ \\
IRAC & Ch. 1 & $18.79 \pm 0.07$ \\
IRAC & Ch. 2 & $18.33 \pm 0.06$ \\[0.5ex]
\multicolumn{3}{c}{MRG-P0918} \\[0.5ex]
ACS & F555W & $23.59 \pm 0.14$ \\
ACS & F814W & $22.36 \pm 0.05$ \\
FourStar & $J_1$ & $21.22 \pm 0.09$ \\
WFC3-IR & F105W & $21.41 \pm 0.04$ \\
FourStar & $J_2$ & $20.87 \pm 0.08$ \\
FourStar & $J_3$ & $20.38 \pm 0.08$ \\
WFC3-IR & F160W & $19.54 \pm 0.03$ \\
FourStar & $H$ & $19.36 \pm 0.06$ \\
FourStar & $K_s$ & $19.17 \pm 0.06$ \\
IRAC & Ch. 1 & $19.05 \pm 0.07$ \\[0.5ex]
\multicolumn{3}{c}{MRG-S1522} \\[0.5ex]
WFC3-UVIS & F606W & $23.91 \pm 0.05$ \\
LDSS & $z$ & $22.65 \pm 0.12$ \\
FourStar & $J_1$ & $21.80 \pm 0.07$ \\
WFC3-IR & F105W & $21.95 \pm 0.04$ \\
FourStar & $J_2$ & $21.64 \pm 0.08$ \\
FourStar & $J_3$ & $20.90 \pm 0.09$ \\
WFC3-IR & F160W & $19.96 \pm 0.03$ \\
FourStar & $H$ & $19.75 \pm 0.06$ \\
FourStar & $K_s$ & $19.41 \pm 0.06$ \\
IRAC & Ch. 1 & $19.14 \pm 0.06$ \\
IRAC & Ch. 2 & $18.99 \pm 0.06$ \\[0.5ex]
\multicolumn{3}{c}{MRG-M2129} \\[0.5ex]
ACS & F435W & $25.13 \pm 0.74$ \\
ACS & F475W & $24.53 \pm 0.29$ \\
ACS & F555W & $24.53 \pm 0.21$ \\
ACS & F606W & $24.16 \pm 0.22$ \\
ACS & F625W & $23.91 \pm 0.23$ \\
ACS & F775W & $23.22 \pm 0.15$ \\
ACS & F814W & $23.16 \pm 0.08$ \\
ACS & F850LP & $22.55 \pm 0.13$ \\
WFC3-IR & F105W & $21.90 \pm 0.05$ \\
WFC3-IR & F110W & $21.14 \pm 0.04$ \\
WFC3-IR & F125W & $20.78 \pm 0.03$ \\
WFC3-IR & F140W & $20.31 \pm 0.03$ \\
WFC3-IR & F160W & $20.06 \pm 0.03$ \\
FourStar & $K_s$ & $19.59 \pm 0.06$ \\
IRAC & Ch. 1 & $19.11 \pm 0.05$ \\
IRAC & Ch. 2 & $19.02 \pm 0.05$
\enddata
\tablecomments{Total fluxes for MRG-M0138 and MRG-M0150 are normalized to Image~1. The giant arc in MRG-M0138 (merging Images 1 and 2) is 0.5~mag brighter.}
\end{deluxetable}

%
%
%
%

\section{Spectroscopic Data}
\label{sec:specdata}

\begin{deluxetable*}{lccccc}
\tablecolumns{6}
\tablewidth{0pt}
\tablecaption{Spectroscopic Observing Log\label{tab:speclog}}
\tablehead{\colhead{Target} & \colhead{Instrument} & \colhead{Dates} & \colhead{Exposure time (hr)} & \colhead{Seeing} & \colhead{Slit PA (deg)}}
\startdata
MRG-M0138 Image 1 & MOSFIRE $J$ & 2015 Nov 3 & 1.0 & $0\farcs73$ & 82 \\
MRG-M0138 Image 1 & MOSFIRE $H$ & 2015 Nov 3, 6 & 2.0 & $0\farcs79$ & 82 \\
MRG-M0138 Image 2 & FIRE & 2016 Sep 8-9 & 5.5 & $0\farcs48$ & -60 \\
MRG-M0150 Image 1 & MOSFIRE $H$ & 2014 Nov 26-27 & 4.3 & $0\farcs67$ & -7.5 \\
MRG-M0150 Image 1 & FIRE & 2014 Nov 1, 3 & 6.5 & $0\farcs55$ & -7.5 \\
MRG-P0918 & FIRE & 2014 Apr 14-15 & 7.0 & $0\farcs42$ & 176 \\
MRG-S1522 & FIRE & 2014 Feb 28-29, & 9.0 & $0\farcs57$ & 136 \\
& & 2014 Apr 13-15 & & & \\
MRG-M2129 & FIRE & 2015 Sep 27-28, & 16.7 & $0\farcs49$ & -13 \\
 & & 2016 Sep 8-9 & & &
\enddata
\end{deluxetable*}

Each of the five lensed galaxies in our sample was observed using FIRE, a near-infrared echellette spectrograph mounted on the Magellan Baade telescope \citep{Simcoe13}. MRG-M0138 and MRG-M0150 were also observed with the near-infrared spectrograph MOSFIRE at the Keck 1 telescope \citep{McLean12}. Here we describe the observing strategy and  reduction procedures for the spectroscopic data.

\subsection{FIRE Observations}

We generally used the $0\farcs75$-wide slit, which provides a spectral resolution of $\sigma_{\rm inst}=33$~km~s${}^{-1}$. For a portion of the MRG-M2129 observations conducted in excellent seeing, we used the $0\farcs60$-wide slit. To minimize read noise, we operated the detector in the up-the-ramp sampling mode for integration times of 20-30~min.

During twilight we obtained sky exposures to measure the illumination of the slit. Although some of the target arcs could be acquired directly on the slit-viewing acquisition camera, we usually acquired a nearby star and offset the telescope to the target. After the offset, we compared the position of the offset star and other sources to their expected pixel coordinates in images taken with the acquisition camera. In some cases these differed by up to $0\farcs4$, and we offset the telescope to place the sources at the expected positions. We monitored these positions throughout the exposure sequence and corrected gradual drift when it occurred.

The $6''$-long FIRE slit was oriented in the direction of elongation of the target lensed galaxy (see P.A.~in Table~\ref{tab:speclog}). Observations were made in an AB pattern with short dithers of $0\farcs8$-$2\farcs5$ depending on the angular size of the target. Due to the resulting overlap of the extended target in the two dither positions, and also to the long exposures needed to minimize the read noise, a simple A-B subtraction is not feasible. Instead the spectra of the sky and target must be modeled in each exposure, as we describe below. Exposures of the internal quartz and ThAr lamps were interspersed throughout the FIRE observations and used for flat fielding and wavelength calibrations. To remove telluric absorption, A0V stars were observed before and after each target and usually in the middle of longer exposure sequences.

The seeing is an important ingredient in our dynamical modeling. To measure the seeing, we monitored stars on the acquisition camera and on the facility guide camera. The seeing was also estimated from the science spectra themselves through a comparison with \emph{HST} images (see Section~\ref{sec:fireredux}). By comparing these methods, we estimate the uncertainty in the seeing is $\lesssim0\farcs1$. The mean seeing during the observations of each galaxy is listed in Table~\ref{tab:speclog} and ranges from $0\farcs42$-$0\farcs57$.

\subsection{FIRE Data Reduction\label{sec:fireredux}}

The FIREHOSE pipeline\footnote{\url{http://www.firespectrograph.org}} was used to flat field the data and to provide an initial wavelength solution and initial rectification of each spectral order. Since FIREHOSE was designed primarily for the reduction of point sources, for subsequent steps we relied on custom IDL routines that were based on the FIREHOSE code.

First, traces of bright stars were used to make small corrections to the rectifications in each order. For each science exposure, we then masked all orders to isolate the inter-order background. We fit and subtracted a smooth variation with column within each of the four amplifier regions, which was necessary to remove discontinuities at the boundaries. We then modeled and subtracted a smooth scattered light background. Low-order corrections to the initial FIREHOSE wavelength solution, which was derived from ThAr lamp exposures, were then made using the OH lines. Once these low-order corrections were derived for one science exposure in a series, cross-correlations were used to correct for instrumental flexure.

Each spectral order of each exposure was then modeled as the sum of sky and galaxy emission using iterative bspline techniques as employed by FIREHOSE (see, e.g., \citealt{Kelson03}). This method requires that the spatial distribution of the emission be specified. For the sky background, we found that the residual intensity variation along the slit can be modeled as a quadratic polynomial. For the targeted galaxy, we first measured its spatial profile in the $H$ band using the initial FIREHOSE sky model. This estimate is imprecise due to the difficulty of separating an extended source from the sky background over a short slit with no prior information on the source structure. We therefore used the \emph{HST} WFC3/F160W image to measure the expected galaxy flux profile along the slit, taking into account its width and orientation. This \emph{HST}-based flux profile was then shifted and convolved by a Gaussian PSF to best match the profile in the spectrum. This procedure provides an estimate of the seeing and the position of the target in each exposure, and it produces a galaxy profile suitable for accurate sky modeling.

Observations of A0V stars were analyzed with the {\tt xtellcorr} package \citep{Vacca03}, as implemented within FIREHOSE, to provide flux calibration and removal of telluric absorption. While the A0V observations are needed to track the temporal variation of telluric features, we found that the relative flux calibration could be improved through observations of white dwarf standards. These were reduced in the same way as the galaxy spectra and used to derive low-order corrections to the flux calibration in each order. Additionally, in a few cases, we made low-order corrections to the continuum shape based on comparisons to the fitted stellar population synthesis models (Section~\ref{sec:contfit}).

For each science exposure, two-dimensional rectified spectra were then produced from each order. The exposures were then normalized to a common flux level, measured in $H$ band. All exposures in each order were then spatially registered and averaged using inverse variance weighting. Residual outlier pixels were then identified and interpolated over. One-dimensional spectra were then extracted in each order in a specified aperture. Small multiplicative offsets were applied to ensure that spectra extracted in adjacent orders have consistent fluxes within the wavelength regions of overlap. Finally, the orders were combined into a single one-dimensional spectrum with a scale of 12.5 km~s${}^{-1}$ pixel${}^{-1}$.

\subsection{MOSFIRE Observations and Reduction}

We observed two targets with MOSFIRE: Image~1 of MRG-M0150 in $H$ band, and Image~1 of MRG-M0138 in both $J$ and $H$ bands. We formed a $0\farcs7$-wide long slit on each target and on several stars in the field, which were used to align the mask and to measure the seeing. The observations were conducted with an AB dither pattern. Since subtraction of consecutive dithered exposures is fundamental to the MOSFIRE Data Reduction Pipeline (DRP), we took care to ensure that the dither distance ($5\farcs4$ for MRG-M0150 and $9\farcs2$ for MRG-M0138) was sufficiently wide to avoid self-subtraction of the extended arcs. 

The observations and reduction of MRG-M0150 were described by \citet{Newman15a}. To reduce the MRG-M0138 observations, we used the DRP to produce coadded 2D spectra for each night of observations (see Table~\ref{tab:speclog}). Relative flux calibration was then performed using twilight observations of the white dwarf GD71. Given the very high signal-to-noise ratio of the MRG-M0138 spectrum, we performed telluric absorption corrections differently from the other observations. In the $H$ band, we iteratively modeled the galaxy stellar continuum and the telluric absorption using the radiative transfer code {\tt molecfit} \citep{Smette15}. A similar procedure was described by \citet{Newman17}. In the $J$ band, where the signal-to-noise ratio is lower, we instead used {\tt molecfit} to model the telluric absorption in observations of GD71 and divided this synthetic absorption spectrum from the galaxy observations. Finally, observations made on different nights were registered, scaled to a common flux level, and averaged with inverse variance weighting.

%
%
%
%

\begin{deluxetable*}{lcccc}
\tablecolumns{5}
\tablewidth{0.9\linewidth}
\tablecaption{S\'{e}rsic Model Parameters\label{tab:sersic}}
\tablehead{\colhead{Quantity} & \colhead{Units} & Single S\'{e}rsic Model & \multicolumn{2}{c}{Double S\'{e}rsic Model}}
\startdata
\sidehead{\emph{MRG-M0138}}
$m_{\rm F160W}$ & mag & $20.0 \pm 0.1 \pm 0.4$ & $20.4 \pm 0.1 \pm 0.4$ & $21.5 \pm 0.1 \pm 0.4$ \\
$R_{e, \rm maj}$ & arcsec & $0\farcs57 \pm 0\farcs07 \pm 0\farcs12$ & $0\farcs85 \pm 0\farcs07 \pm 0\farcs18$ & $0\farcs10 \pm 0\farcs02 \pm 0\farcs02$ \\
$R_{e, \rm maj}$ & kpc & $4.8 \pm 0.6 \pm 1.0$ & $7.1 \pm 0.6 \pm 1.5$ & $0.8 \pm 0.2 \pm 0.2$ \\
$n$ & \ldots & $2.9 \pm 0.7$ & $1.3 \pm 0.3$ & 1${}^\dagger$ \\
$b/a$ & \ldots & $0.26 \pm 0.04$ & $0.19 \pm 0.03$ & $0.67 \pm 0.16$ \\
PA & deg & $36 \pm 2$ & \multicolumn{2}{c}{$36 \pm 1$} \\

\sidehead{\emph{MRG-M0150}}
$m_{\rm F160W}$ & mag & $21.3 \pm 0.1 \pm 0.4$ & & \\
$R_{e, \rm maj}$ & arcsec & $0\farcs21 \pm 0\farcs03 \pm 0\farcs04$ & & \\
$R_{e, \rm maj}$ & kpc & $1.7 \pm 0.2 \pm 0.3$ & & \\
$n$ & \ldots & $3.5 \pm 0.4$ & & \\
$b/a$ & \ldots & $0.87 \pm 0.05$ & & \\
PA & deg & $-6 \pm 10$ & & \\

\sidehead{\emph{MRG-M2129}}
$m_{\rm F160W}$ & mag & $21.8 \pm 0.2$ & $22.0 \pm 0.2$ & $23.4 \pm 0.1$ \\
$R_{e, \rm maj}$ & arcsec & $0\farcs29 \pm 0\farcs02$ & $0\farcs27 \pm 0.02$ & $0\farcs47 \pm 0.05$ \\
$R_{e, \rm maj}$ & kpc & $2.4 \pm 0.2$ & $2.2 \pm 0.2$ & $3.9 \pm 0.4$ \\
$n$ & \ldots & $1^{\dagger}$ & $1^\dagger$ & $1^\dagger$ \\
$b/a$ & \ldots & $0.29 \pm 0.03$ & $0.24 \pm 0.03$ & $0.80 \pm 0.07$ \\
PA & deg & $-39 \pm 5$ & \multicolumn{2}{c}{$-39 \pm 5$} \\
Gaussian $m_{\rm F160W}$ & mag & $25.9 \pm 0.5$ & \multicolumn{2}{c}{$26.1 \pm 0.3$} \\
Gaussian $R_e$ & arcsec & $<0\farcs007$ (95\%) & \multicolumn{2}{c}{$<0\farcs008$ (95\%)}
\enddata
\tablecomments{For the extrinsic parameters (magnitudes and $R_e$) of the multiply imaged systems, we first list the uncertainty derived from the scatter among the multiple images (Section~\ref{sec:stellarstructure}) and second the systematic uncertainty arising from the overall magnification uncertainties (Section~\ref{sec:magerrors}). ${}^\dagger$ Edge of prior.}
\end{deluxetable*}

\section{Lens Models and Source Structures}
\label{sec:lensmodels}

For three of the galaxies in our sample, we are able to construct lens models to measure the magnification and reconstruct the source. This is necessary to estimate the stellar masses of the galaxies (Section~\ref{sec:stellarpops}) and to compare their sizes, ellipticities, and S\'{e}rsic indices with those of unlensed samples and thereby evaluate the representativeness of our lensed galaxy sample (Section~\ref{sec:representativeness}). We will also compare the galaxies' structures to those of their likely $z \sim 0$ descendants to constrain their future evolution (Section~\ref{sec:evolution}) and use the lens models to interpret our stellar kinematic data (Paper~II).

For MRG-M0150 and MRG-M0138, we constrained the lens mass distribution and the source light distribution using the detailed structure of the multiple images of the quiescent galaxies themselves. MRG-M2129 is singly imaged, but since the lens is a well-studied cluster, we can rely on published lens models which are constrained by many multiple image systems. In this section, we describe the construction of these models. For the other two galaxies in our sample, the singly imaged MRG-S1522 and MRG-P0918, we cannot construct a meaningful lens model. MRG-S1522 lies at radii beyond the known multiple images, and consequently the mass distribution is not well constrained. Although the cluster magnifying MRG-P0918 produces several multiple image systems, their redshifts are not yet known. For these two systems, we do not estimate the magnification and will confine our analysis to magnification-independent quantities (e.g., $\sigma$, $V/\sigma$, age, specific star formation rate, and emission line ratios). 

\begin{figure*}
\centering
\includegraphics[width=\linewidth]{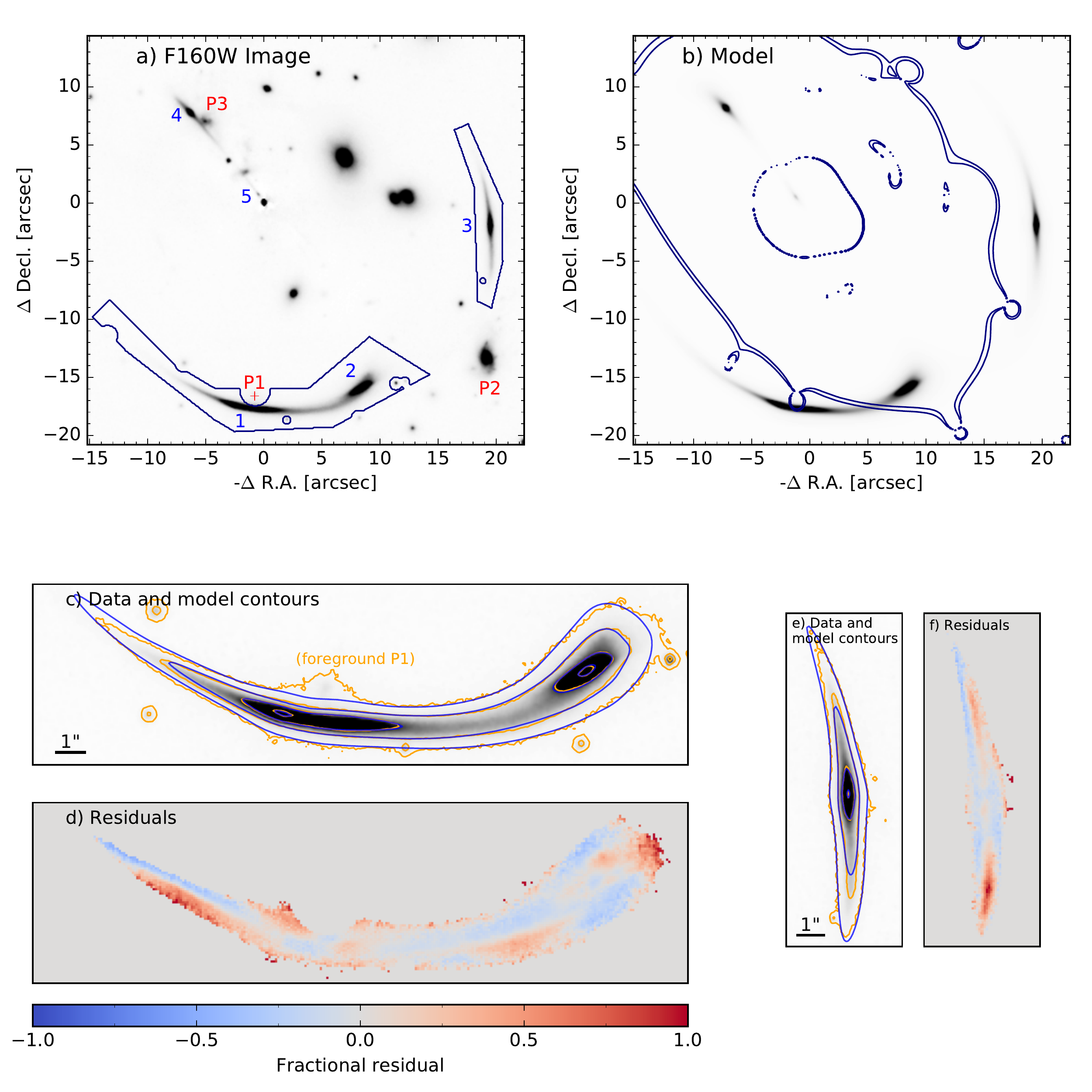}
\caption{Lens model of MRG-M0138. Panel (a) shows the \emph{HST}/WFC3 F160W image with the BCG subtracted, displayed with a linear stretch. The regions used to constrain the lens and source model are outlined. Blue labels number the images. Coordinates are relative to the BCG center. Panel (b) is the model of the image plane produced by a double S\'{e}rsic model of the source traced through the lensing potential and convolved by the PSF.  Colored curves enclose the critical line. Panels (c) and (e) show zooms of panel (a) with orange and blue contours of the data and model image, respectively. Note that the orange contours include flux from several foreground galaxies, including P1, which were masked during the fit. Panels (d) and (f) show the fractional residuals.\label{fig:m0138_lensmodel}}
\end{figure*}

\begin{figure}
\centering
\includegraphics[width=0.8\linewidth]{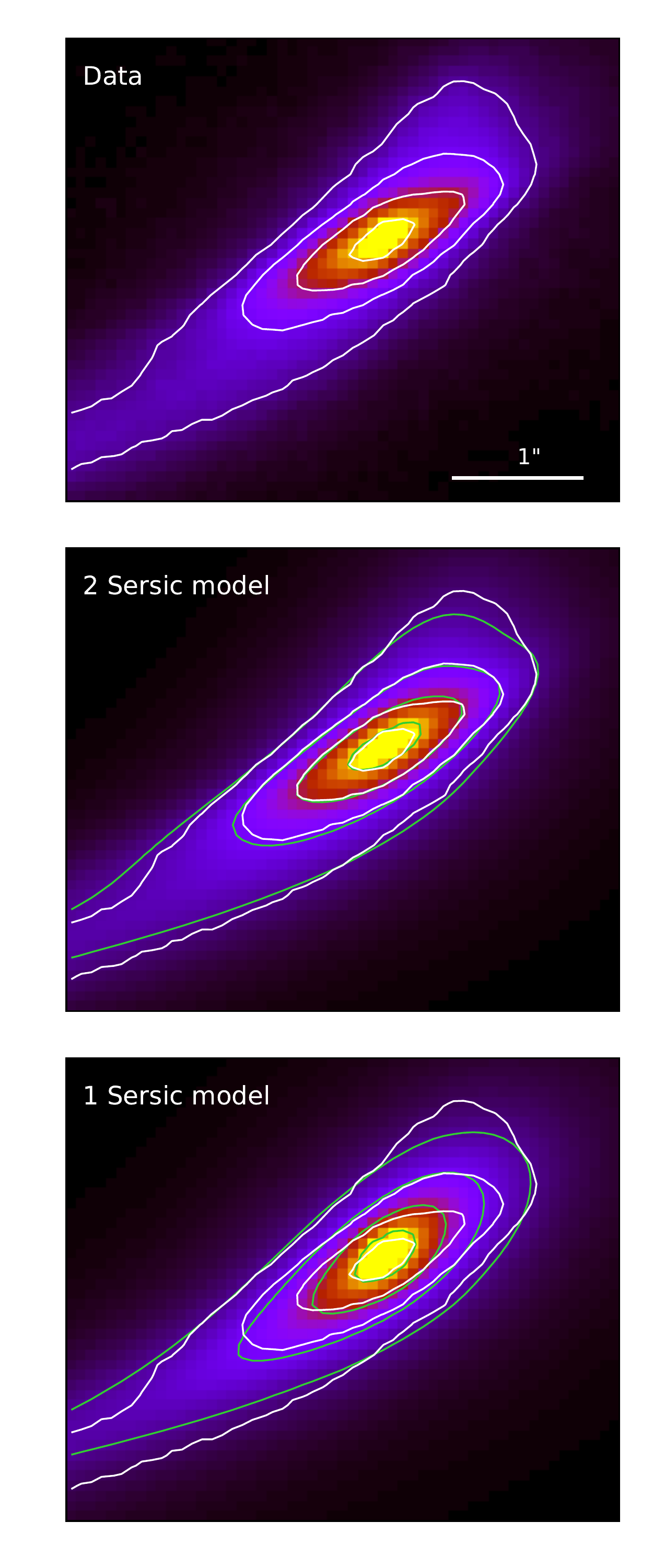}
\caption{Comparison of Image~2 of MRG-M0138 to two models. The two-component source model (middle panel) matches the data (top panel) much better than a single-component source model (bottom panel), particularly the flattened and tilted inner structure. White contours in the upper panel are repeated in the middle and lower panels, where green contours trace the models.\label{fig:m0138_prove2comp}}
\end{figure}

\subsection{MRG-M0138}

\subsubsection{Lens and Source Modeling Methods}

We used the ray tracing code introduced by \citet{Newman15b} and employed by \citet{Newman15a} to fit a simply parameterized model of the lens mass distribution and the source light distribution to the pixel-level data. The regions of the WFC3/F160W image that were used to constrain the model are outlined in blue in Figure~\ref{fig:m0138_lensmodel}a. The BCG light has been modeled and subtracted in this image.

The lensing cluster was modeled as a superposition of dual pseudo-isothermal elliptical (dPIE) mass distributions (see Appendix of \citealt{Eliasdottir07}). Each dPIE was described by a center, PA, ellipticity, two characteristic radii $r_{\rm core}$ and $r_{\rm cut}$, and a normalization $\sigma_0$. Since the lensing cluster appears to be a simple, relaxed system, we found that a single dPIE component is adequate to model the cluster dark matter halo. We left all its parameters free except for $r_{\rm cut}$, which lies well beyond the strong lensing zone and can be fixed to 1~Mpc.

We modeled the stellar mass in the BCG by fitting a dPIE profile to the surface brightness distribution to set $r_{\rm core}$ and $r_{\rm cut}$. We used Gaussian priors of ${\rm PA} = 46^{\circ} \pm 10^{\circ}$ and $b/a = 0.42 \pm 0.05$ that were informed by the photometry. The normalization $\sigma_{0,\rm BCG}$ was allowed to vary freely.

Other cluster galaxies were generally included in the mass model using scaling relations with luminosity: $\sigma_0 = \sigma_0^* (L/L_*)^{1/4}$ and $r_{\rm cut} = r_{\rm cut}^* (L/L_*)^{1/2}$, where $\sigma_0^*$ and $r_{\rm cut}^*$ are free parameters with observationally motivated priors \citep[see][]{Newman13a}. This approach ties the center, PA, and ellipticity of the mass distribution to the galaxy light and imposes a constant mass-to-light ratio. The radial distribution of the mass, encoded by $r_{\rm cut}$, can differ from the light. Similar approaches have widely been used in other parameterized lens models \citep[e.g.,][]{Jullo07,Richard10,Newman13a}.

Three galaxies either significantly perturb the critical lines or are likely to deviate from these general scaling relations and so were modeled independently of them. The galaxy labeled P1 in Figure~\ref{fig:m0138_lensmodel}a is located very close to Image~1. It significantly alters the critical line and increases the magnification. This perturbing galaxy is faint and not easily visible in Figure~\ref{fig:m0138_lensmodel}, but it can be seen in Figure~\ref{fig:hstimages}. We allowed all of the parameters describing the P1 mass distribution to vary freely. Galaxies P2 and P3 are blue cluster galaxies with spectacular examples of ram pressure stripping evident in Figure~\ref{fig:hstimages}. Given their blue colors, we do not expect them to share a common mass-to-light ratio with the other cluster members. We allowed $\sigma_0$ to vary freely for both P2 and P3 as well as $r_{\rm cut}$ for P2.  The final ingredient in the lens model is an  external shear, uniform across the image, which we ultimately found to be small, $|\gamma| = 0.05$.

We initially modeled the source light distribution using a single elliptical S\'{e}rsic profile (although, as described below, we ultimately adopted a two-component model). We denote the effective radius as $R_{e,\rm maj}$ to emphasize that it is the semi-major axis of the ellipse containing half of the light. For a given set of parameters that describe the lens model and source, we first cast the centroids of Images 1-4 back to the source plane and computed the dispersion in their positions. Models with a large dispersion were immediately discarded. For the remaining models, we traced pixels in the image plane back to the source to compute the surface brightness. Care was taken to integrate the surface brightness within the pixels near the center of the source. The model image plane was then convolved by the PSF and compared to the data. A Markov Chain Monte Carlo (MCMC) sampler, {\tt MultiNest} \citep{Feroz09}, was used to explore the 29-dimensional parameter space.

Although this model was able to reproduce the overall lens configuration, we found that it could not match the internal structure of all the images in detail. The clearest example of this is Image~2, which we show in Figure~\ref{fig:m0138_prove2comp}. This image presents a flattened inner component that is misaligned from the overall direction of the arc (upper panel). A model with a single-component S\'{e}rsic source (lower panel) is not able to reproduce this. This deficiency can be eliminated by introducing a two-component source: this model matches the morphology of the image much better in the inner parts (middle panel). Specifically, we modeled the source as the sum of two S\'{e}rsic components that share a common center and PA, but which have different axis ratios $b/a$, effective radii $R_{e, \rm maj}$, S\'{e}rsic indices $n$, and magnitudes.

To assess the fit quality, we compare the observed image to the model image plane in panels c-f of Figure~\ref{fig:m0138_lensmodel}. Although some areas of mismatch are visible, the model contours (blue) generally follow the data (orange) well. (The contours near the ``foreground P1'' label include flux from the foreground galaxy P1, which was masked during the fit.) We note that the radial images 4 and 5 were not used to constrain the model except via their approximate positions, so a close match to their detailed structure is not expected.

The magnification factors\footnote{We define the mean magnification over an image as $\langle \mu \rangle = \sum {\rm Image}(x,y)  / \sum [{\rm Image}(x,y) / \mu(x,y)]$, where the sum is over the pixels in the mask. This is equivalent to the ratio of the image and source fluxes in the limit of an infinite aperture.} for images 1, 2 and 3 are $\mu_1 = 12.5\pm5.4$, $\mu_2 = 10.3\pm3.1$, and $\mu_3 = 4.9\pm1.6$, whose uncertainties are described in the next section.

\subsubsection{Magnification Uncertainties\label{sec:magerrors}}

The information in the pixel-level \emph{HST} data over-constrains the lens model and results in minuscule formal uncertainties. However, it is known that different lens modeling assumptions and procedures can lead to different estimates of the magnification. Although these uncertainties are difficult to quantify, we attempted to estimate them by constructing a set of {\tt Lenstool} models \citep{Kneib93,Jullo07}. These models are constrained by the positions and ellipticities of the images, not the pixel-level data, but the cruder constraints make it feasible to vary several assumptions. In particular, we varied the type of constraint (image positions, positions and fluxes, source versus image plane fits), the radial density profile of the cluster (dPIE versus generalized Navarro-Frenk-White), and the inclusion of the BCG or some perturbing galaxies as separate mass components. For each of multiple images 1-3, we evaluated the maximum difference in the magnification between the fiducial model and the set of {\tt Lenstool} models. We find that the uncertainties range from 30\% to 43\%. These are comparable to the differences between lens models of Frontier Fields clusters discussed by \citet{Priewe17} and \citet{Meneghetti17}. When describing the source properties, we will conservatively use the maximum uncertainty among these multiple images as a systematic uncertainty in the luminosity and stellar mass, i.e., 43\% or 0.19~dex. Since this factor applies to the areal magnification, we approximate the fractional error in $R_e$ as half of this.

\subsubsection{Reconstructing the Source Plane\label{sec:sourcerecon}}

Our reconstruction of MRG-M0138 in the source plane is shown in Figure~\ref{fig:sourceplane}. Given the non-linear nature of the lens mapping, convolution by the PSF in the image plane can have very complex effects in the source plane. This makes it difficult to compare the consistency of reconstructions derived from different images. To address this issue, we do not directly cast the observed pixels back to the source plane. Instead, we take the unconvolved image plane model, add the residuals, and cast this image (i.e., ${\rm Data} - {\rm Model}_{\rm convolved} + {\rm Model}_{\rm unconvolved}$) back to the source plane. This technique effectively deconvolves the inner regions of the images under the assumption of a particular source model, while still allowing for deviations from the model at larger radii (see \citealt{Szomoru12} for an application in a non-lensing context). All three images consistently show a highly flattened, disk-dominated source. We note that since Images 1 and 2 merge into a giant arc, the southwestern portion of the source is not present in these images.

In Figure~\ref{fig:sourceplane}, we also compare the surface brightness profiles derived from these three source plane reconstructions. The three multiple images reproduce a single surface brightness profile (solid colored lines) with impressive consistency. This profile is well fit by the two-component S\'{e}rsic model (solid black line), leaving azimuthally averaged residuals of $\lesssim 10\%$ out to $R = 10$~kpc. The leftward arrow indicates the radius at which the effects of PSF convolution are significant; at smaller radii, the shape of the plotted profile is dominated by the assumption of the double S\'{e}rsic form.

\subsubsection{Source Structure\label{sec:stellarstructure}}

\begin{figure*}
\centering
\includegraphics[width=0.4\linewidth]{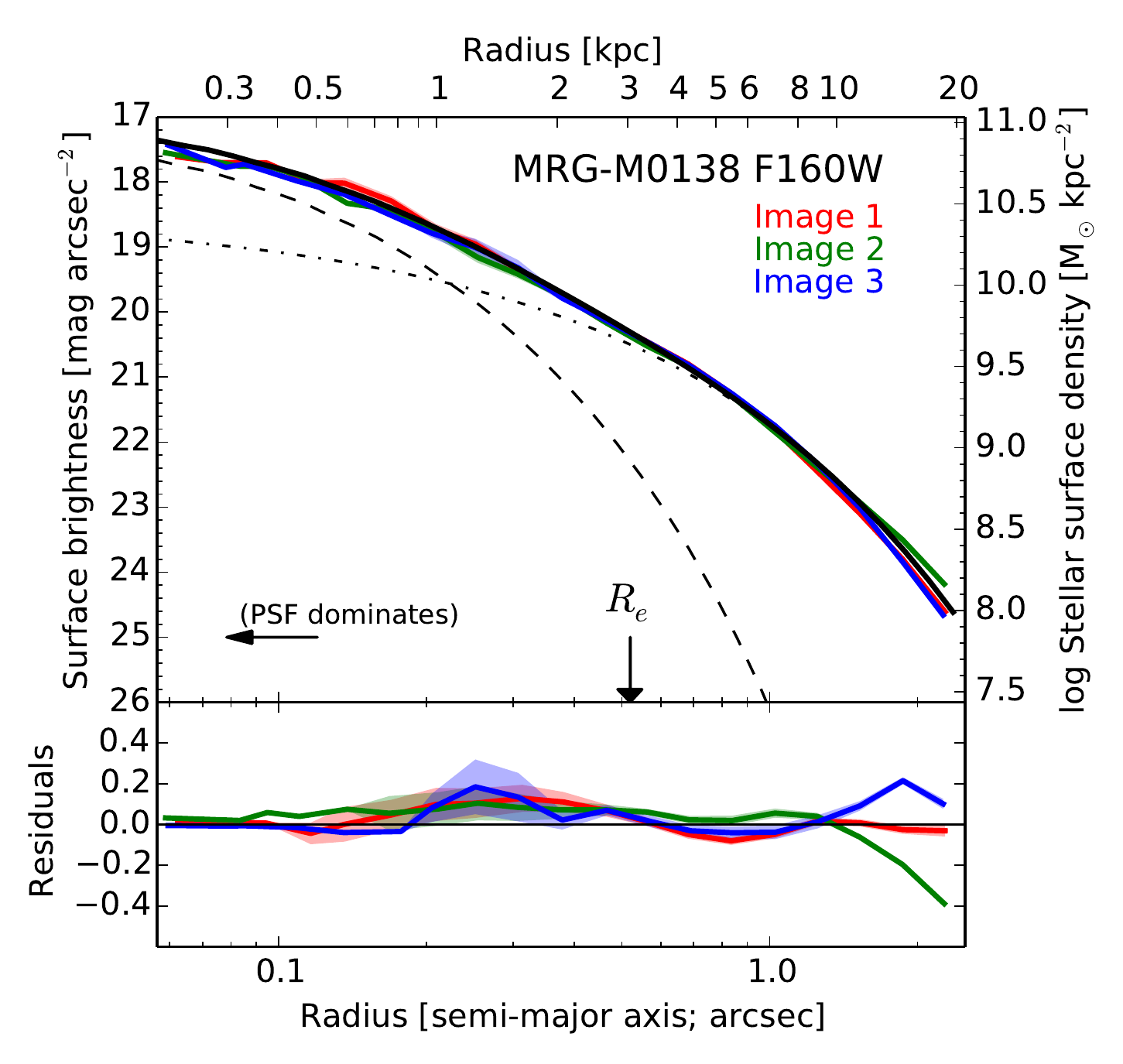} \hspace{0.75cm} \includegraphics[width=0.35\linewidth]{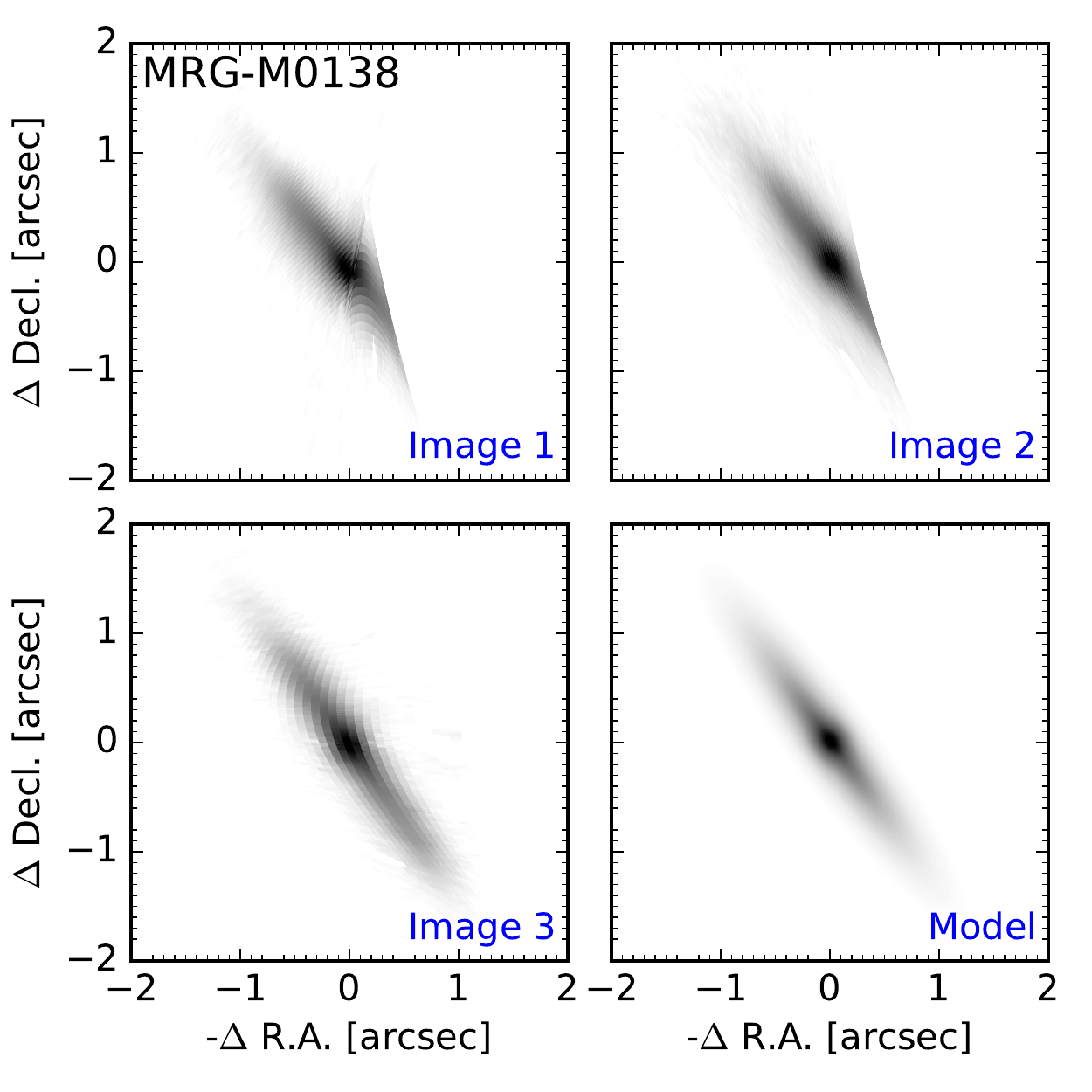} \\
\includegraphics[width=0.4\linewidth]{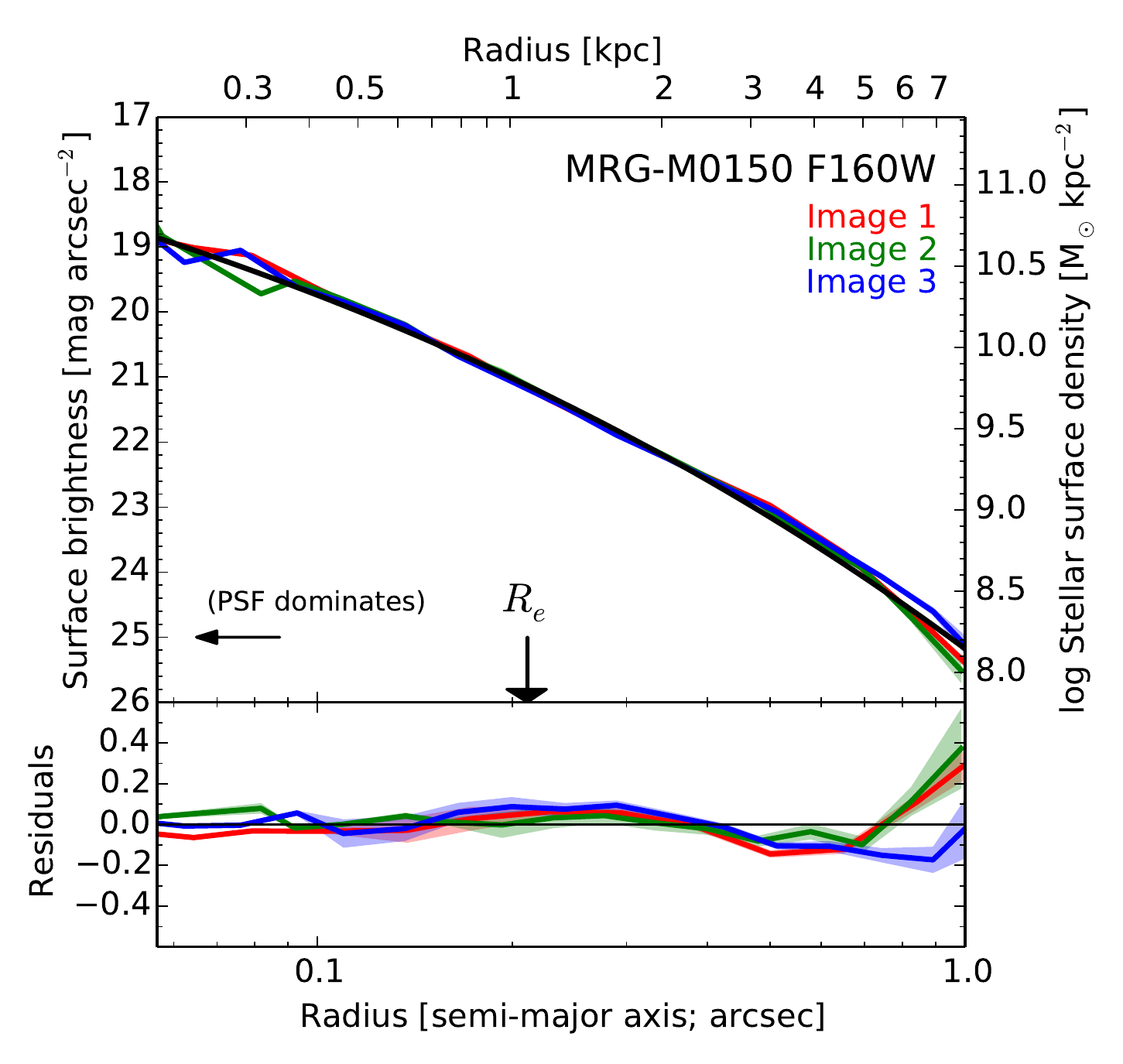} \hspace{0.75cm} \includegraphics[width=0.35\linewidth]{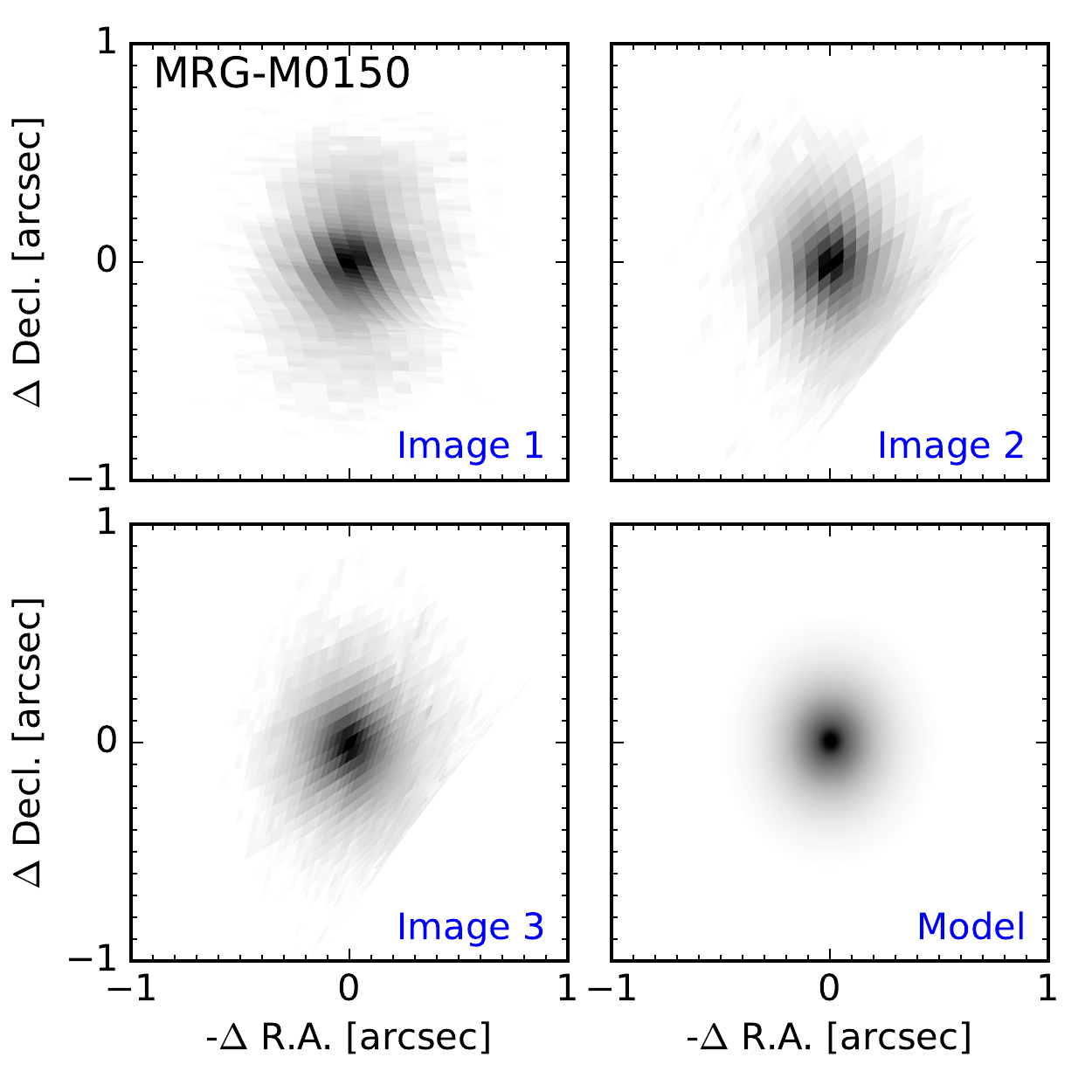} \\
\includegraphics[width=0.4\linewidth]{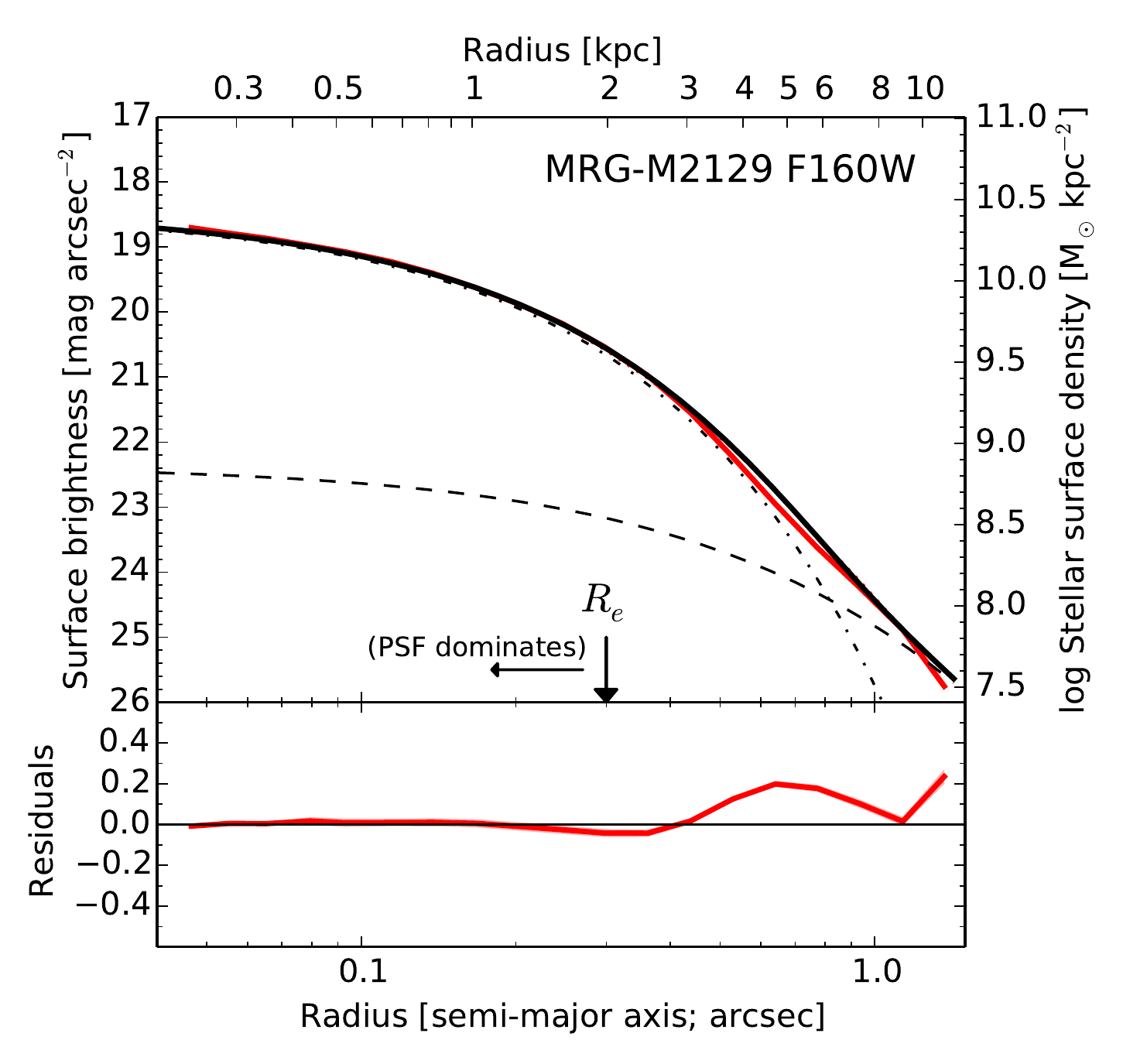} \hspace{0.75cm} \includegraphics[width=0.35\linewidth]{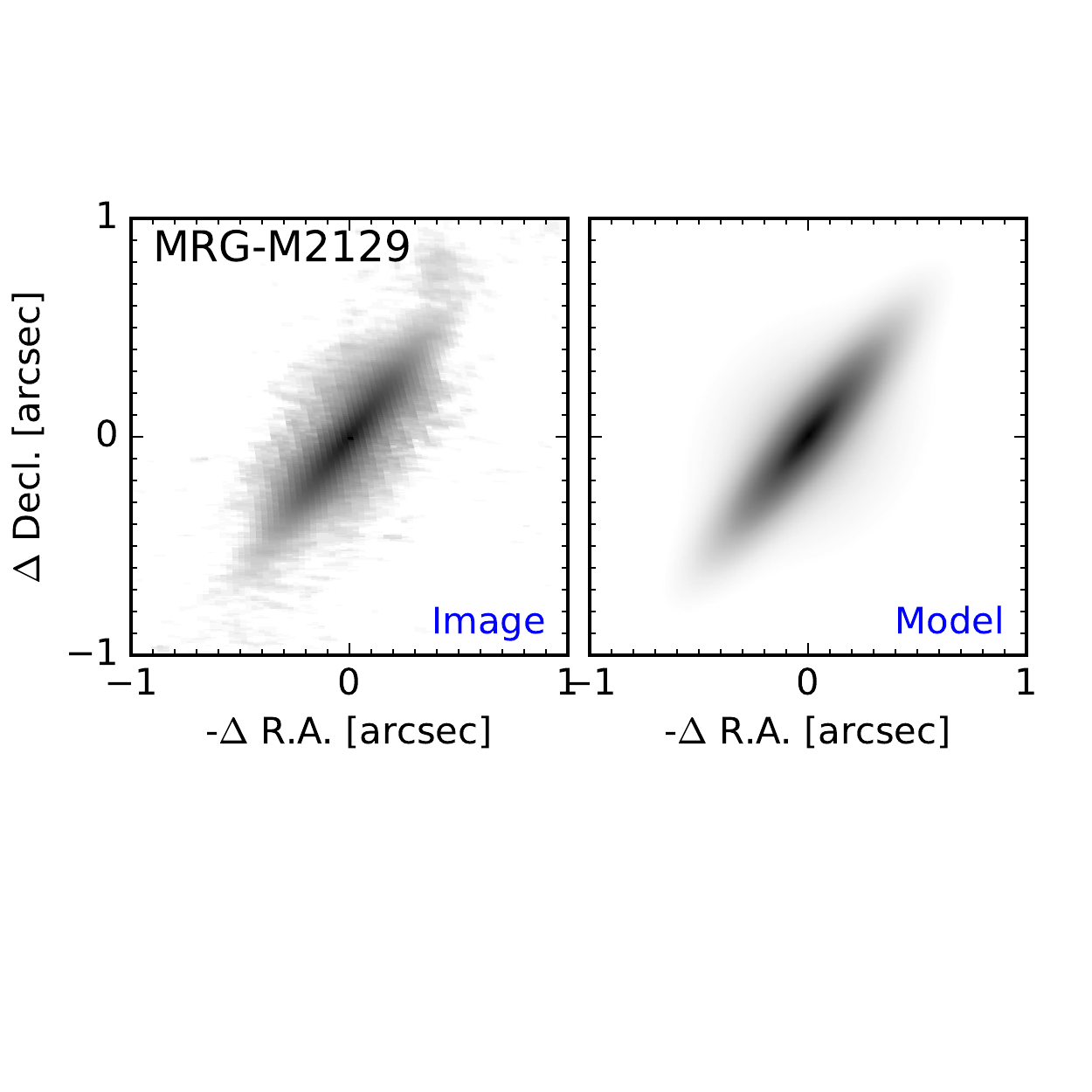}
\caption{Surface brightness profiles (left panels) and source plane reconstructions (right) of the three  galaxies with lens models. These are constructed using the deconvolved images ${\rm Data} - {\rm Model}_{\rm convolved} + {\rm Model}_{\rm unconvolved}$, as described in Section~\ref{sec:sourcerecon}. \emph{Left panels:} Solid colored lines show the profiles constructed from each multiple image. Solid back lines show the analytic profile of the best-fit single- or double-S\'{e}rsic model, with dashed and dot-dashed curves showing the individual contributions of the two components for MRG-M0138 and MRG-M2129. Residuals between the image and model are plotted in mag~arcsec${}^{-2}$. For multicomponent models, the surface brightness has been averaged within elliptical annuli whose axis ratio is that of the single-S\'{e}rsic model in Table~\ref{tab:sersic}. The right axes show the corresponding stellar surface mass density based on the global mass-to-light ratios inferred in Section~\ref{sec:stellarpops}. \emph{Right panels:} For each image, the corners of each pixel were cast back to the source plane, conserving the surface brightness, to produce these reconstructions. Images are displayed using an arcsinh stretch. Note the axis ranges are not the same for every galaxy.
\label{fig:sourceplane}}
\end{figure*}

In the best-fitting two-component model, 74\% of the stellar light is in an extended ($R_{e,\rm maj} = 7.1$~kpc), flattened ($b/a = 0.19$), disk-like component with a S\'{e}rsic index of $n = 1.3$. A fainter (26\% of total flux), rounder ($b/a = 0.67$), and much more compact ($R_{e, \rm maj} = 0.8$~kpc) component is also present. It has a S\'{e}rsic index $n = 1$, the smallest value allowed in the fits, which could indicate a structure analogous to a pseudobulge. However, we caution that the images are significantly affected by the PSF within the effective radius of the bulge-like component (Figure~\ref{fig:sourceplane}).

Table~\ref{tab:sersic} lists the S\'{e}rsic parameters and uncertainties. The uncertainties were derived from the image-to-image scatter: specifically, we fixed the lens mapping, fit the source to each of the three images individually, and measured the standard deviation of each parameter. For the total flux and $R_e$ we additionally list the systematic lens model uncertainties discussed in Section~\ref{sec:magerrors}.

\subsection{MRG-M0150}

In \citet{Newman15a} we constructed a parametrized model of the lensing cluster and source that was constrained by the archival \emph{HST} WFC3/F140W image of MACSJ0150.03-1005 that was then available. The technique was very similar to that described in the previous subsection for MRG-M0138, and its application to MRG-M0150 was detailed by \citet{Newman15a}. Subsequently we obtained a deeper WFC3 image at the slightly redder wavelength of F160W with improved sub-pixel sampling (Section~\ref{sec:hstobs}). Here we briefly describe our improved analysis using this new image.

\begin{figure*}
\includegraphics[width=\linewidth]{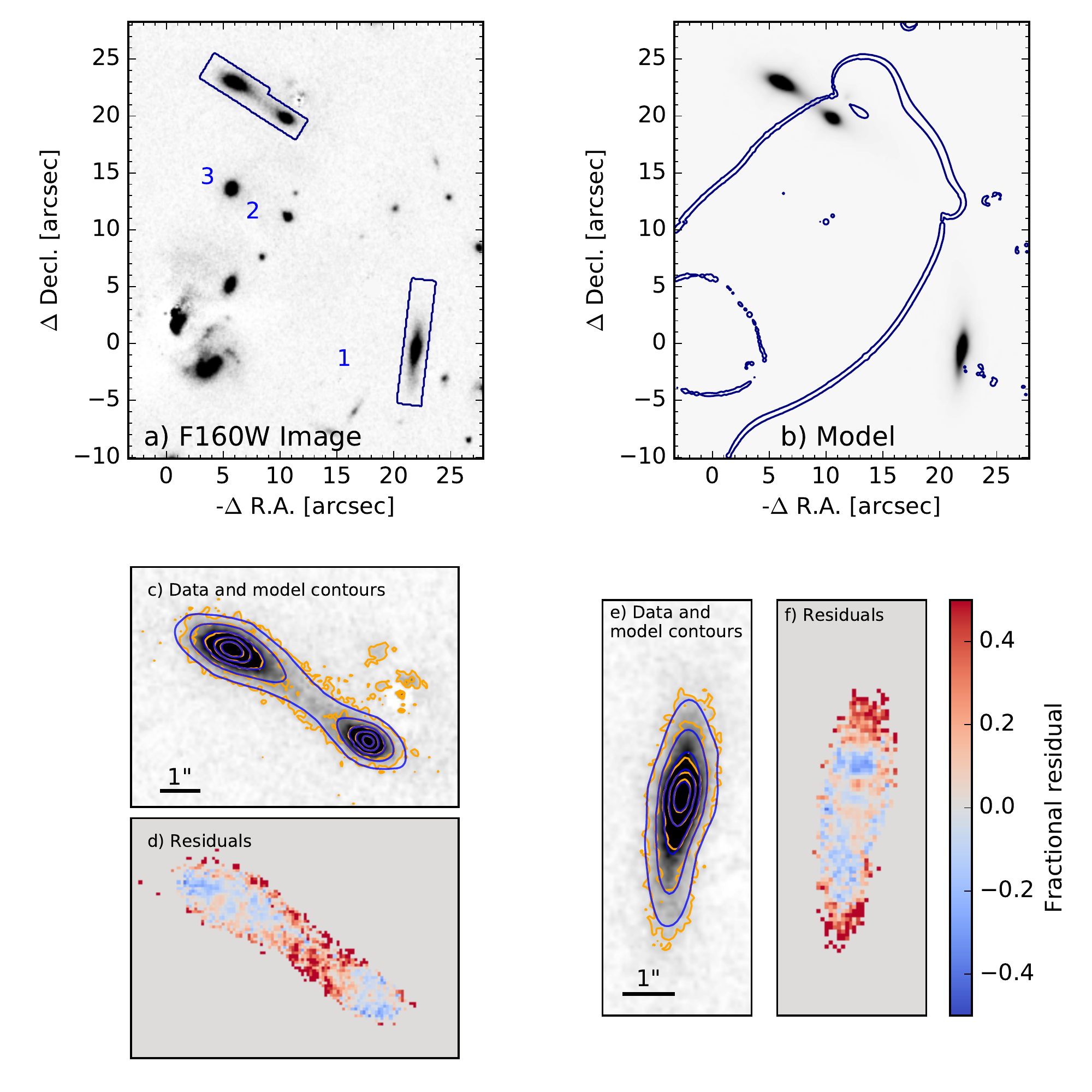}
\caption{Updated lens model of MRG-M0150. Panel (a) shows the \emph{HST}/WFC3 F160W image with the BCG subtracted, displayed with a linear stretch. The regions used to constrain the lens and source model are outlined. Blue labels number the images. Coordinates are relative to the BCG center. Panel (b) is the model of the image plane produced by the single-component S\'{e}rsic model of the source traced through the lensing potential and convolved by the PSF. Colored curves enclose the critical line. Panels (c) and (e) show zooms of panel (a) with orange and blue contours of the data and model image, respectively. Panels (d) and (f) show the fractional residuals.\label{fig:m0150_lensmodel}}
\end{figure*}

Figure~\ref{fig:m0150_lensmodel} shows the three multiple images 1-3 (panel a), the single-component S\'{e}rsic source traced through the lensing potential and convolved by the PSF (panel b), and zooms on the image regions to compare the model contours and residuals (panels c-f). The model successfully reproduces the structure of the images in detail. The magnifications of the images are $\mu_1 = 4.4\pm1.1$, $\mu_2=2.6\pm1.0$, and $\mu_3 = 4.6\pm1.3$. These uncertainties are estimated by varying the parameterization of the mass model and by comparing to results from an independent set of lens models constructed with the {\tt Lenstool} code, as described in Section~\ref{sec:magerrors} and \citet{Newman15a}.

Source reconstructions and surface brightness profiles from each multiple image are shown in Figure~\ref{fig:sourceplane}.\footnote{The surface brightness profile differs superficially from Figure~2 of \citealt{Newman15a} because we now cast the PSF-deconvolved image back to the source plane, as described in Section~\ref{sec:sourcerecon}. Jaggedness in the innermost regions is due to pixelization.} The lens model produces three consistent reconstructions of the source. Likewise the surface brightness profiles of the source from the observed multiple images are mutually consistent at the $\simeq 5\%$ level out to $4R_e$. 

The S\'{e}rsic parameters are listed in Table~\ref{tab:sersic}. The source is a compact galaxy ($R_{e,\rm maj} =1.7$~kpc) with a nearly de Vaucoulers profile ($n=3.5$) and nearly round isophotes ($b/a=0.86$). Unlike MRG-M0138, we find that a single S\'{e}rsic component is adequate to describe the source, but we note that it would be much more difficult to discern the presence of multiple components in this case since MRG-M0150 is nearly round in projection. These parameters are consistent with those measured using the shallower F140W image by \citet{Newman15a}, but the uncertainties are reduced using the new deeper data.

\subsection{MRG-M2129}
\label{sec:m2129lensmodel}

MRG-M2129 is singly imaged, so unlike MRG-M0138 and MRG-M0150, we require other multiple images to constrain the source structure. The lensing cluster MACSJ2129.4-0741 presents a large number of multiple images available to constrain the mass distribution, and several authors have published models. We used the deflection angle maps produced by several published lens models to trace a model source through the lens mapping. The model image plane can then be convolved by the PSF and fit to the \emph{HST}/WFC3 F160W image. Since this system was discovered by \citet{Geier13} and studied by \citet{Toft17}, we can also compare to their results in the Appendix.

\begin{figure}
\centering
\includegraphics[width=0.75\linewidth]{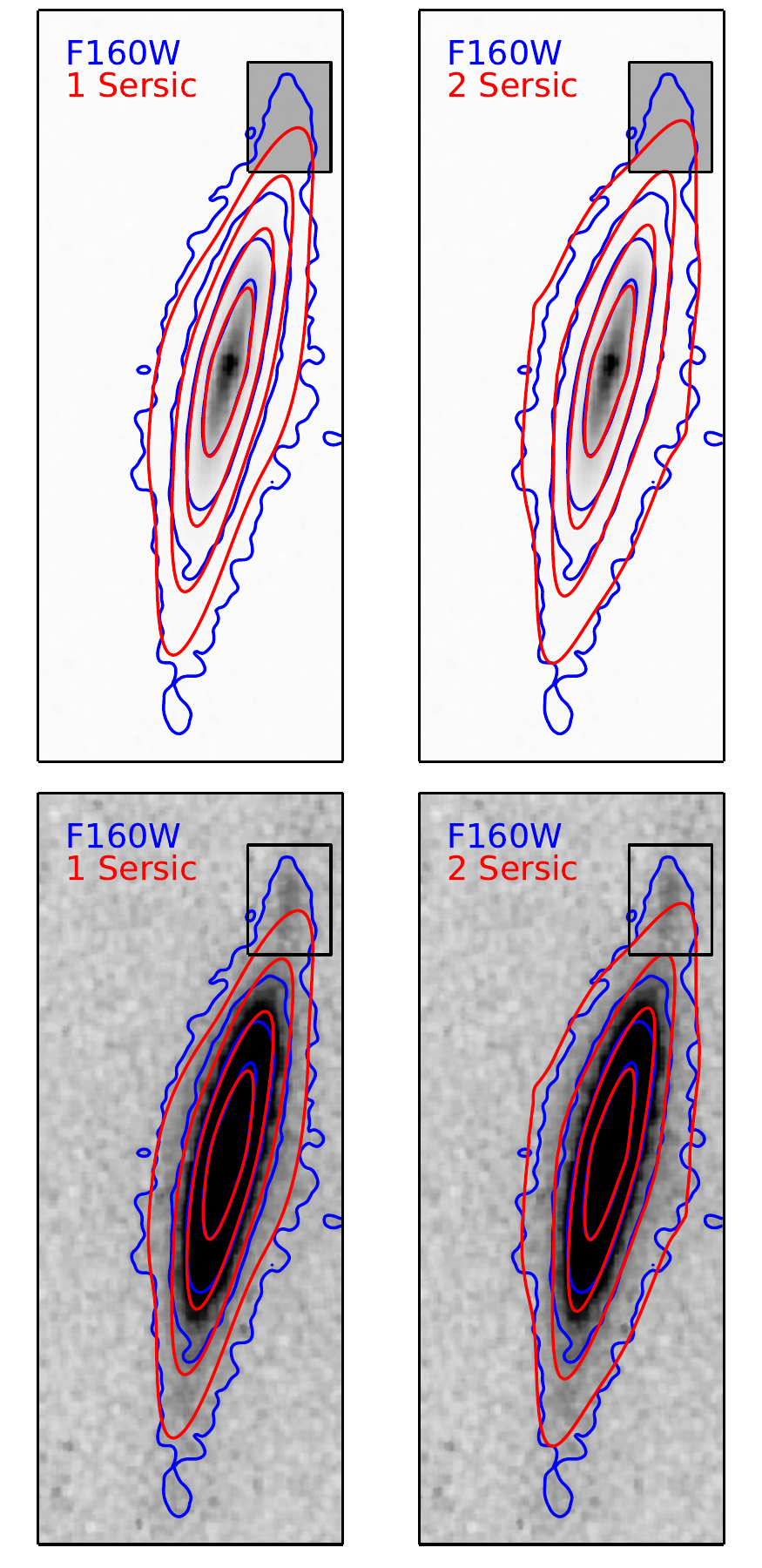}
\caption{Model of MRG-M2129. The WFC3/F160W image is shown in each panel with logarithmically spaced contours in blue. Red contours in the left and right panels show the models based on a single- and double-S\'{e}rsic source, respectively. The top and bottom rows differ only in the stretch of the image. The unresolved nuclear component is more easily seen in the top row, while a detached component (potentially a satellite galaxy) is more easily seen in the bottom row. The black box encloses the region masked during the fit due to the detached component. The double S\'{e}rsic model fits the outer contour better than the single-S\'{e}rsic model.
\label{fig:m2129_lensmodel}}
\end{figure}

We considered three lens models. \citet{Monna17} kindly supplied maps of the deflection angle for their parametric model. We also obtained two mass models constructed by A.~Zitrin \citep{Zitrin09,Zitrin13} as high-level science products of the CLASH program \citep{Postman12}. We found that the \citet{Monna17} lens model produced the lowest $\chi^2$, so we used the parameters derived from it and estimated the uncertainties in the galaxy structural parameters by comparing to those obtained using the two Zitrin et al.~lens models.

Figure~\ref{fig:m2129_lensmodel} shows the region of the \emph{HST}/WFC3 F160W image that we fit. We first modeled and subtracted a foreground cluster member located to the west of the arc; to avoid any residual contamination from this galaxy, the fit region is not centered on the arc. We then subtracted the sky background measured in an empty region to the east of the arc. In this case, the lensed galaxy is sufficiently distant from the BCG that it was unnecessary to model. A second source is visible in the upper right corner, which is disconnected from the main galaxy (see also \citealt{Geier13}). We masked this object, which is potentially a satellite galaxy.

We found that a single S\'{e}rsic model is inadequate to model MRG-M2129, both in the central and outer regions. First, there is a centrally located, point-like source visible in the top row of Figure~\ref{fig:m2129_lensmodel}. We modeled this by adding a second component to source model: a circular Gaussian at the galaxy center with a free effective radius $R_e$ and flux. The inclusion of the central source improved the fit by $\Delta \chi^2 = 759$ and so is clearly justified. Its magnitude is $m_{\rm F160W} = 25.9 \pm 0.5$ (demagnified), or $\sim1\%$ of the galaxy flux, and it is very compact: $R_e < 0\farcs007 = 60~{\rm pc}$ (95\% confidence). The compact central source could be continuum emission from a Seyfert nucleus, which is supported by the emission line ratios (Section~\ref{sec:lineorigins}). The properties of the single S\'{e}rsic+Gaussian source model are listed in Table~\ref{tab:sersic}. 

Second, we found that a single-component S\'{e}rsic model does not reproduce the shape of the outer isophotes well, which is evident in the upper-left and lower-right regions of the images in Figure~\ref{fig:m2129_lensmodel}. The addition of a second S\'{e}rsic component with a center and PA tied to those of the first component improved the fit. This can be seen qualitatively in the right panels of Figure~\ref{fig:m2129_lensmodel} and is supported quantitatively by an improvement of $\Delta \chi^2 = 1287$. Like MRG-M0138, then, MRG-M2129 shows evidence for two components with different ellipticities.

The source model consists of (1) the central source described above, (2) a flattened ($q = 0.24$) component containing 79\% of the flux in a compact ($R_{e,\rm maj} = 0\farcs27 = 2.2$~kpc) exponential disk ($n = 1$), and (3) a rounder ($q = 0.80$) component containing 21\% of the flux that is more extended ($R_{e, \rm maj} = 0\farcs47 = 3.9$~kpc) and also nearly exponential ($n = 1$). The parameters are summarized in Table~\ref{tab:sersic}. We restricted the S\'{e}rsic index to values $n > 1$ in the fits. Since both components hit this limit, their S\'{e}rsic indices should be regarded with caution. However, the presence of two components---one compact and flattened, one fainter, rounder, and more extended---seems robust and holds for all three lens models. The uncertainties in Table~\ref{tab:sersic} are derived from the standard deviation of values obtained using the three lens models.

\subsection{Summary of Galaxy Structures}

The structures of MRG-M0138 and MRG-M2129 are both dominated by a highly flattened exponential component, i.e., a nearly edge-on disk. Both galaxies also contain additional components. In MRG-M0138, we find a compact rounder component emitting 26\% of the luminosity, which is potentially a nascent bulge embedded in a very massive and extended ($R_e = 7$~kpc) disk. In MRG-M2129 we find an extended rounder component and, remarkably, an unresolved central point source that could be Seyfert nucleus. This is particularly interesting given the evidence for an active galactic nucleus (AGN) from the emission line ratios, which we will discuss in Section~\ref{sec:lineorigins}. MRG-M0150 is instead well described by a single compact component with a nearly de Vaucouleurs profile ($n = 3.5$) and round isophotes. With imaging alone we cannot tell whether MRG-M0150 is intrinsically rounder than the other two systems or is merely less inclined, but we will address this question using stellar kinematics in Paper~II.

The disk-dominated structures of MRG-M0138 and MRG-M2129 are quite distinct from massive early-type galaxies in the local universe. In Section~\ref{sec:evolution} we will compare our sample to local systems and discuss the implications for their future evolution.

%
%
%
%

\section{Unresolved Stellar Populations}
\label{sec:stellarpops}

We now use the photometric and spectroscopic data collected in Sections~\ref{sec:imagingphotometry} and \ref{sec:specdata}, together with the magnification factors derived in Section~\ref{sec:lensmodels}, to measure the redshifts, stellar masses, and ages of the lensed galaxies in our sample and to establish their quiescence. For the purposes of this paper, we will focus on representative values derived from the integrated spectra, deferring an analysis of their spatially resolved stellar populations.

\subsection{Spectral Extraction\label{sec:spec_ex}}

For each target, we examined the flux distribution along the slit in order to define the extraction region. The extraction box was chosen to cover the region where the flux profile was $\gtrsim 0.15$ times the peak intensity.\footnote{Relative to the peak position, the boundaries of these regions were $-5\farcs6$ to $2\farcs0$ for Image~1 of MRG-M0138, $\pm1\farcs25$ for Image~2 of MRG-M0138, $-1\farcs5$ to $+1\farcs2$ for Image~1 of MRG-M0150, $\pm 1\farcs0$ for MRG-P0918, $\pm0\farcs7$ for MRG-S1522, and $\pm 1\farcs6$ for MRG-M2129.} As discussed in Section~\ref{sec:phot}, these approximately match the apertures in which colors were measured. For our analysis, we consider the wavelength range from $\lambda_{\rm rest} = 3600$~\AA~to $\lambda_{\rm obs} \simeq 2.3~\mu$m. The spectrum outside the range is generally too noisy to be useful. Only for the highest-redshift source, MRG-M0150, did we reduce the spectrum out to $\lambda_{\rm obs} = 2.45~\mu$m in order to include H$\alpha$ and [\ion{N}{2}].

The spectra are shown in Figure~\ref{fig:specfits}, where they have been rebinned for display purposes as indicated in the caption. The quality is remarkable considering the redshifts of these sources. The median signal-to-noise ratio in the $H$ band generally ranges from 21-32 per 300~km~s${}^{-1}$ bin, approximately one velocity dispersion element. For the ultra-bright MRG-M0138, this figure reaches 137 and 77 for the MOSFIRE and FIRE spectra of Images~1 and 2, respectively.

\subsection{Spectral Modeling\label{sec:contfit}}

We jointly modeled the spectra and photometry using the {\tt pyspecfit} code described by \citet{Newman14}. We used the \citet[][BC03]{BC03} population synthesis models and adopted an exponentially declining star formation history ${\rm SFR} \propto \exp(-t/\tau)$, the \citet{Calzetti00} dust attenuation curve, and the \citet{Chabrier03} initial mass function. For each galaxy we fit the redshift, velocity dispersion, age,\footnote{The age refers to the time between the observation epoch and the beginning of the exponential star formation history.} $\tau$, metallicity $Z$, $A_V$, and the emission line parameters described below. As described in \citet{Newman14}, when comparing the data to a given model, we warped the spectrum continuum shape by the polynomial that minimizes the $\chi^2$. This accounts for flux calibration errors in the spectrum and anchors its continuum shape to the photometry. We allowed a separate polynomial of order $\simeq \Delta \lambda_{\rm rest} / 200$~\AA~in each of the $J$, $H$, and $K$ bands, where $\Delta \lambda_{\rm rest}$ is the unmasked rest-frame wavelength interval.

Most of the galaxies in the sample exhibit weak emission lines whose measurement requires accurate modeling of the stellar continuum. We therefore fit both simultaneously. Emission lines from [\ion{O}{2}], [\ion{O}{3}], [\ion{N}{2}], [\ion{S}{2}] and the Balmer series were modeled as Gaussians with a velocity and velocity dispersion $\sigma_{\rm em}$ that are common to all emission lines but are distinct from those of the stellar component. For each galaxy, we modeled only the lines falling in the atmospheric transparency windows. The intensity ratios [\ion{O}{3}] $\lambda 5008/\lambda 4960$ and [\ion{N}{2}] $\lambda6585$/$\lambda6550$ were fixed to 2.98 and 3.05 \citep{Storey00}, while for the Balmer series we fixed the relative intensities assuming Case B recombination. Since H$\beta$ is always in net absorption, we cannot separately constrain the emission line attenuation via the Balmer decrement. Therefore, once the emission line spectrum was constructed based on the aforementioned ratios, we attenuated it at each wavelength by the same factor as the starlight. We will consider the effects of possible differential extinction in our interpretation; these are expected to be small because the continuum attenuation is mild.

The posterior distributions were sampled using {\tt MultiNest}. In addition to the model parameters, we also computed derived quantities such as the specific star formation rate (sSFR) and emission line ratios. The priors were broad and uninformative, with two exceptions: we restricted the stellar metallicity to $Z=0.01$-0.05 (the solar value is 0.02 in the BC03 models) and placed a Gaussian prior on $v_{\rm gas} - v_{\rm stars}$ with a mean of 0 and a dispersion of 200~km~s${}^{-1}$.\footnote{For MRG-M0138, the only emission line covered in the FIRE spectrum is [\ion{O}{2}], which is not detected. In this case, we tied the gas velocity and dispersion to those of the stars to derive flux limits.}

The posterior constraints on the stellar populations and emission lines are listed in Table~\ref{tab:specmeas}. As Figure~\ref{fig:specfits} shows, these models generally fit the photometry and spectra well. For the photometry, the reduced $\chi^2_{\rm phot}$ is in the range 0.4-1.5. Since deep NIR spectra often exhibit noise somewhat above the formal estimate, we rescaled the spectral uncertainties by a constant factor in the range 1.0-2.5 so that the reduced $\chi^2_{\rm spec} \simeq 1$. We masked Mg~b since the galaxies may have non-solar abundance ratios. For MRG-M0138, we also masked Na~D which clearly has a non-solar abundance or is affected by interstellar absorption. For the MOSFIRE spectrum of MRG-M0138, we also masked the region around the G band and H$\gamma$ due to imperfect correction of telluric absorption in this region.

\begin{figure*}
\centering
\includegraphics[width=0.98\linewidth]{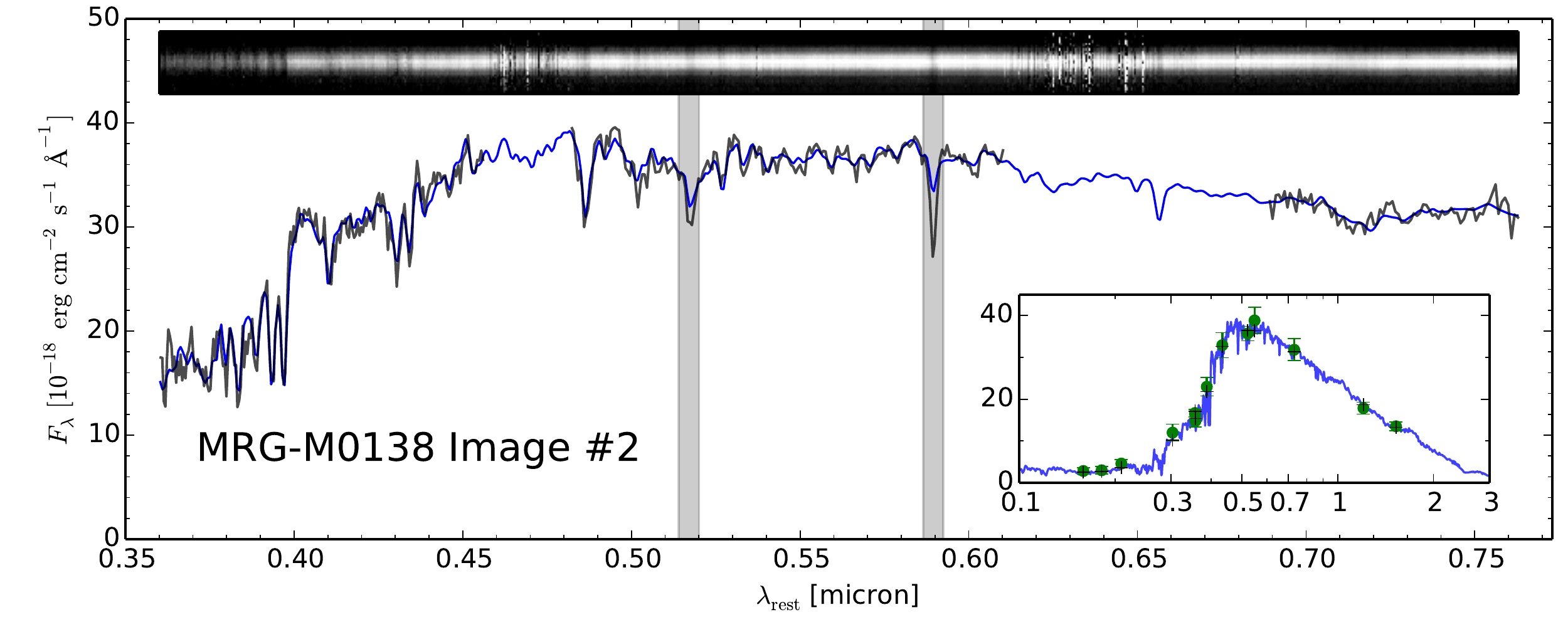}
\includegraphics[width=0.98\linewidth]{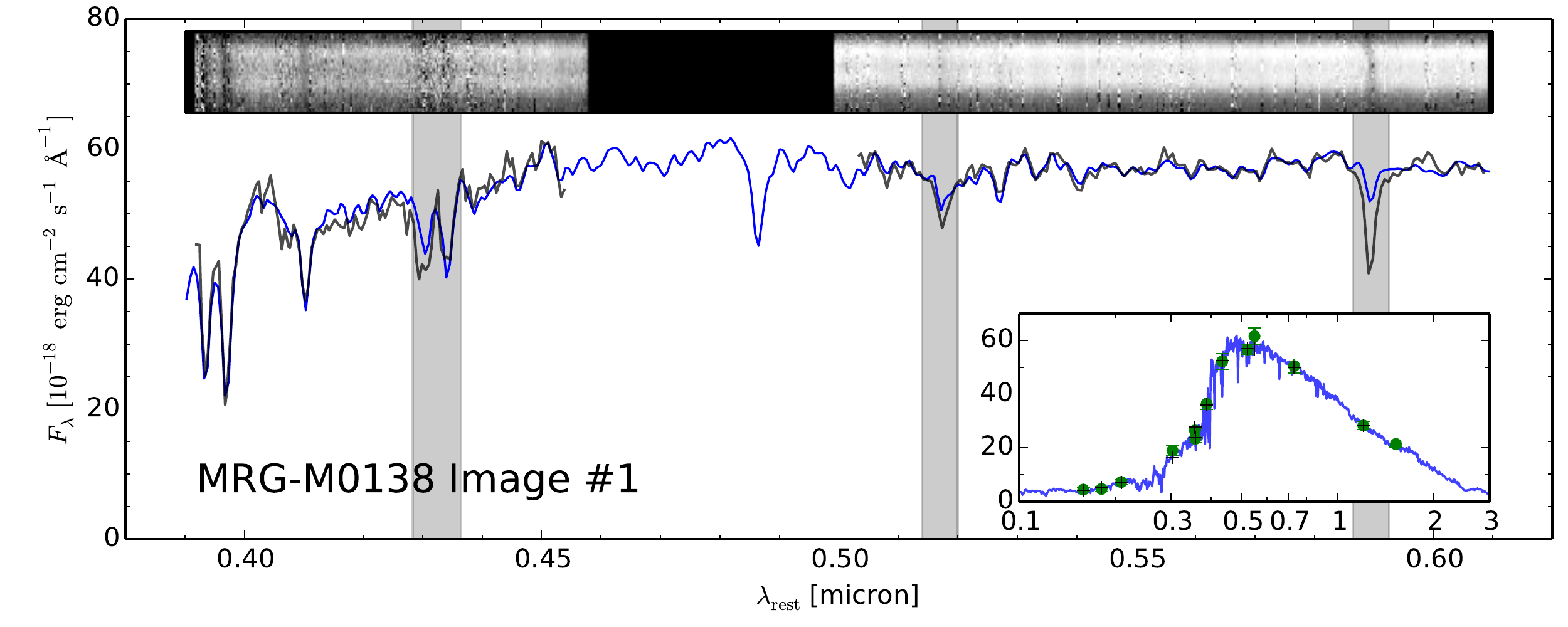}
\includegraphics[width=0.98\linewidth]{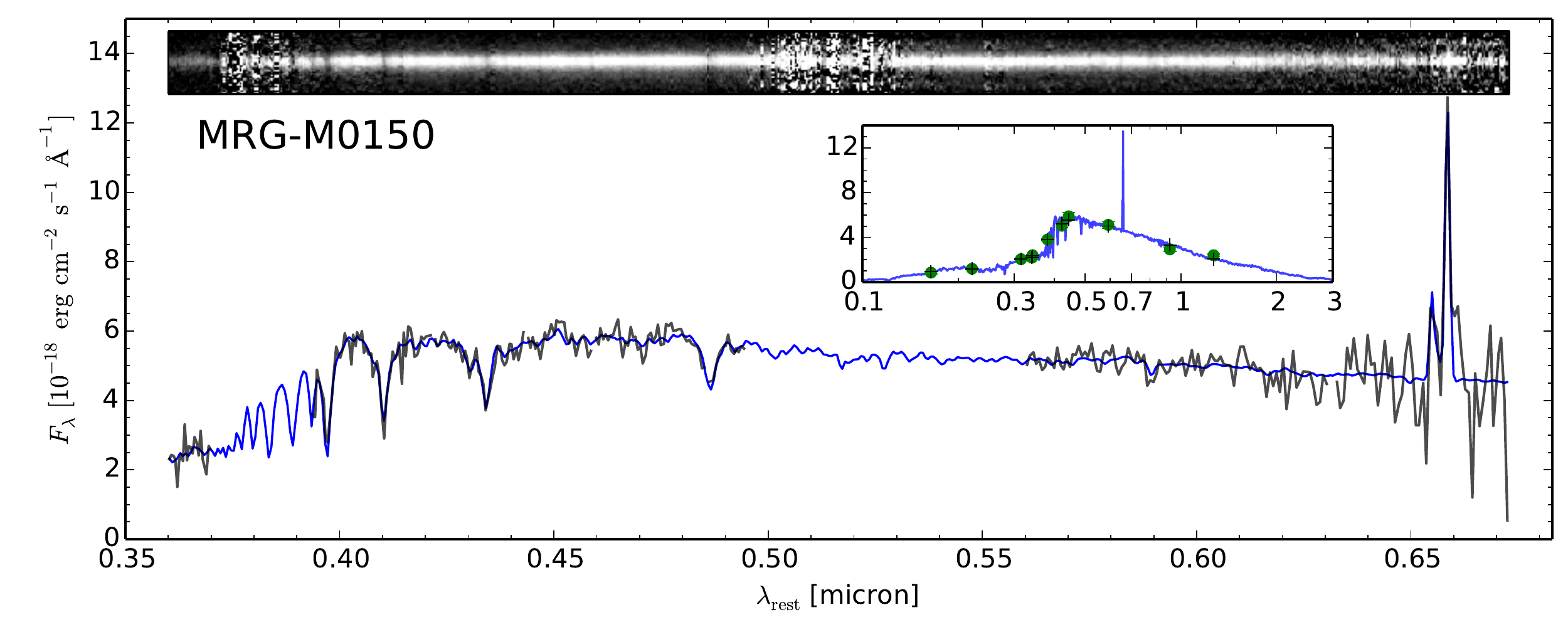}
\caption{Integrated spectra of the lensed quiescent galaxy sample. Grey curves show the high-resolution spectra after taking the inverse variance weighted mean in 30 pixel (375~km~s${}^{-1}$) wide spectral bins. The blue curve shows the best-fit stellar and emission line model described in the text with the the same rebinning applied. Grey bands indicate regions that were masked for the fit. The inset shows the broadband photometry (green circles) and the model magnitudes (crosses). The axes of the insets have the same units as the main panels. The top of each panel shows the two-dimensional spectrum.
\label{fig:specfits}}
\end{figure*}

\begin{figure*}
\centering
\includegraphics[width=0.98\linewidth]{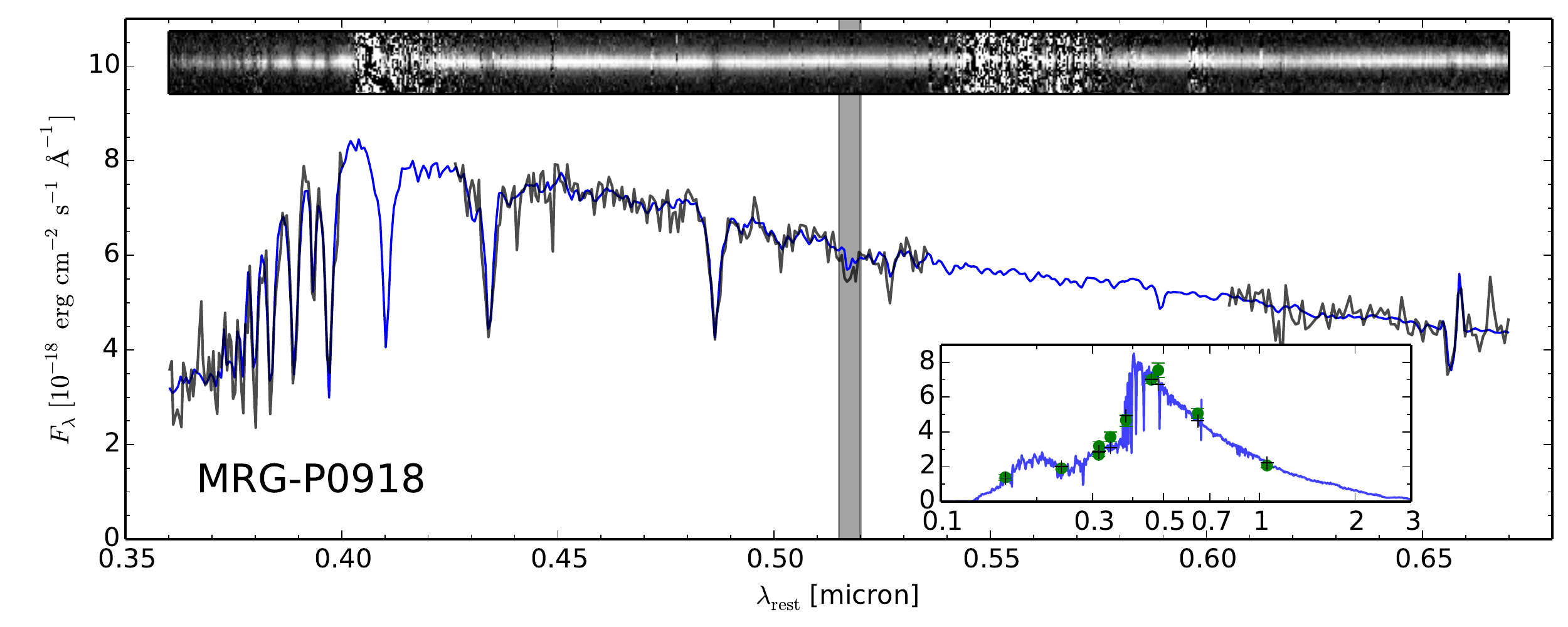}
\includegraphics[width=0.98\linewidth]{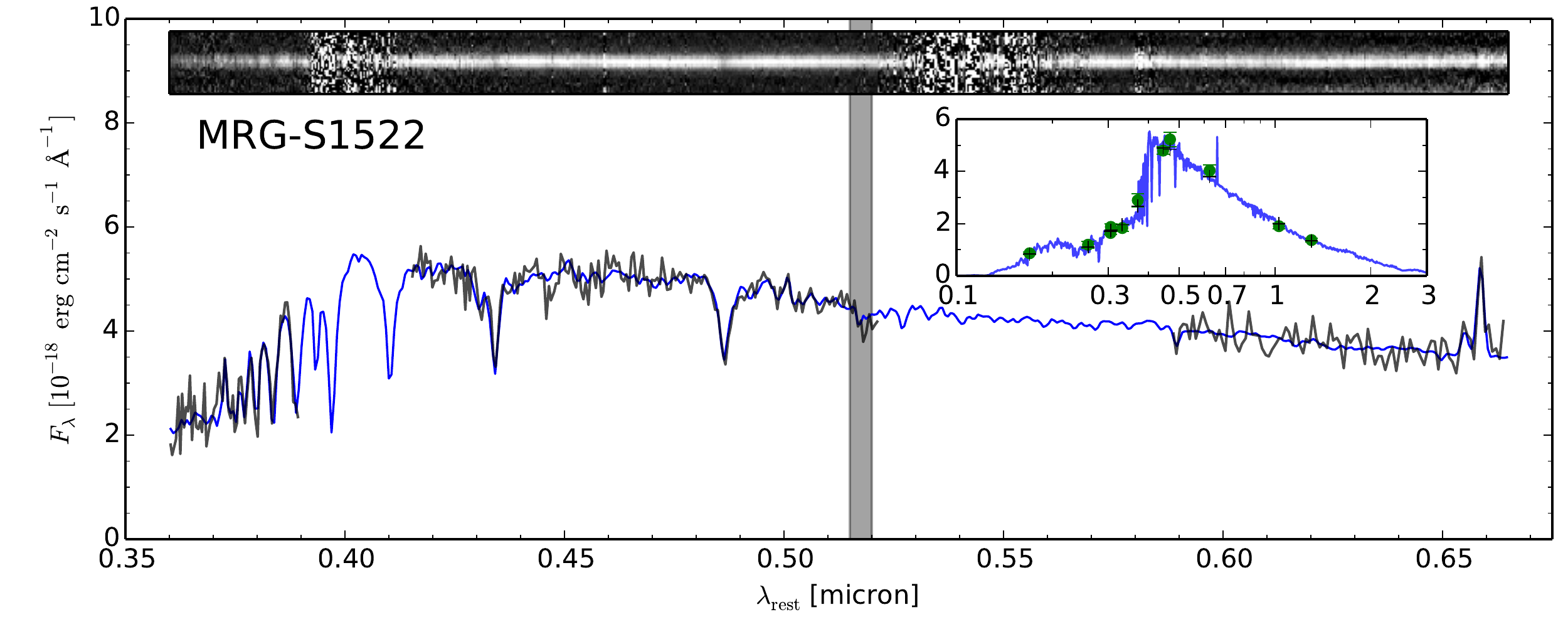}
\includegraphics[width=0.98\linewidth]{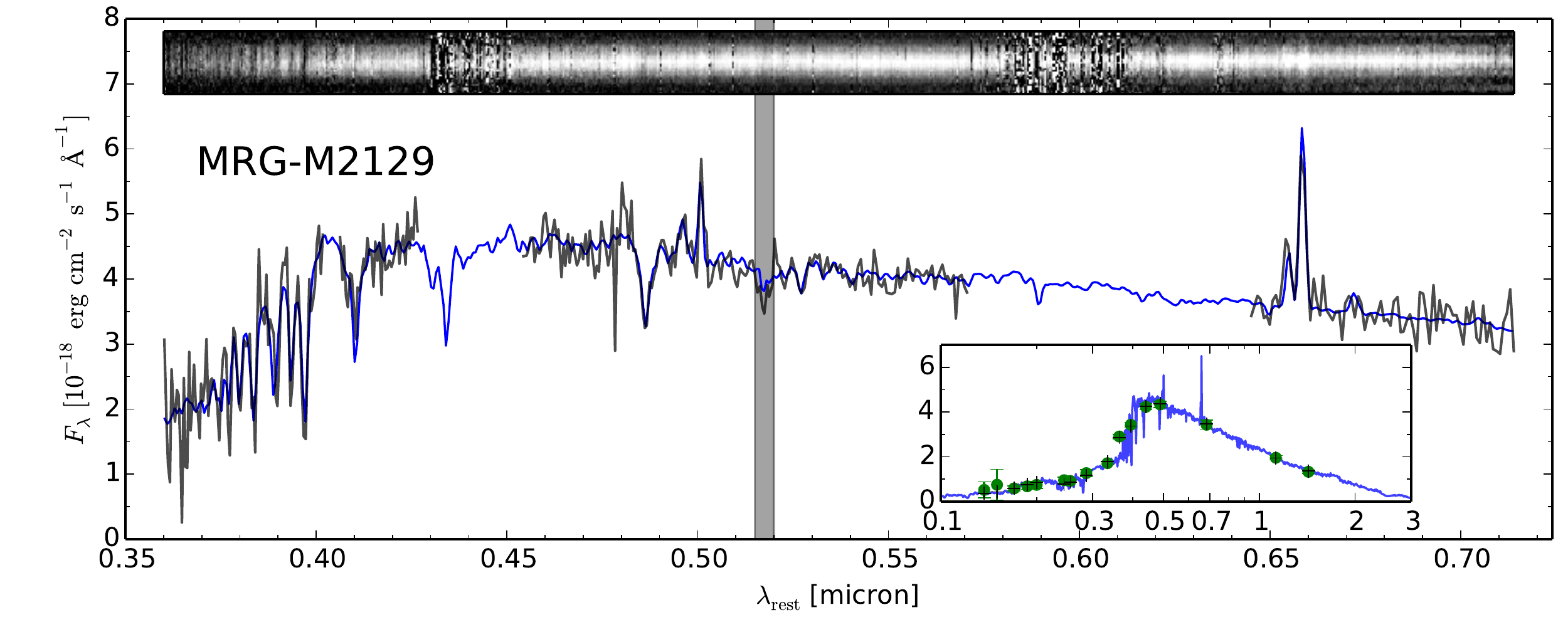}
\caption{Continuation of Figure~\ref{fig:specfits}\label{fig:specfits2}}
\end{figure*}

\subsection{Massive, Quiescent Stellar Populations}

Consistent with the goal of our color--magnitude selection, our modeling of the stellar continuum indicates high stellar masses and low sSFRs for all five lensed galaxies in the sample. Their stellar masses, uncorrected for magnification, span the range $\mu M_* = 10^{11.6-12.8} \msol$. For the three systems with estimated magnification factors $\mu$, we find $M_* = 10^{11.0-11.7} \msol$. The specific star formation rates (sSFRs), which are independent of magnification, are $10^{-10.7}$~yr${}^{-1}$ or smaller. As we will discuss in Section~\ref{sec:emlines}, emission lines (when present) do not have ratios indicative of star formation.

Inspection of the spectra immediately shows a diversity of ages from 0.5-1.4~Gyr. Coupled with the low current sSFRs, this implies a substantial decline in star formation rate over the past Gyr with $e$-folding times of $\tau \lesssim 100$-200~Myr. A mild amount of reddening, corresponding to $A_V = 0.1$-0.6, is inferred for the full sample. We find stellar abundances near solar. For the most massive and oldest galaxy, MRG-M0138, the FIRE spectrum indicates $Z \simeq 2 Z_{\odot}$. We will analyze the detailed chemical abundance pattern of this galaxy in a forthcoming paper.

Although the stellar continuum suggests only mild reddening, mid- and far-infrared observations are needed to address the possibility of highly extinguished star-forming regions. Stacking analyses have shown that quiescent galaxies at $z < 2.5$ identified by the $UVJ$ diagram or similar criteria usually do \emph{not} harbor much obscured star formation \citep{Fumagalli14,Man16}. For the galaxies in our sample, the only existing observation with adequate depth to address this question is a \emph{Spitzer} MIPS 24~$\mu$m image that covers MRG-M2129 (P.I.~M.~Yun, Program ID 50610). MRG-M2129 is not detected with a $2\sigma$ upper limit of 86~$\mu$Jy, corresponding to $\mu L_{\rm IR} < 8 \times 10^{11} {\rm L}_{\odot}$ assuming the \citet{Wuyts08} template. In the limit where the dust is heated only by star formation, this limit implies ${\rm SFR} < 18 \msol$~yr${}^{-1}$ (corrected for magnification).\footnote{We used a circular aperture with a radius of $3\farcs5$, a sky annulus extending from 7-$10''$, and an aperture correction from the MIPS Instrument Handbook. \citet{Toft17} find a more stringent $3\sigma$ limit of ${\rm SFR} < 5 \msol$~yr${}^{-1}$ apparently due to differences in the estimated noise or aperture correction.} This would still place the galaxy below the main sequence of star-forming galaxies, given its high mass; furthermore, this limit is conservative since the emission lines in MRG-M2129 show evidence for an active galactic nucleus (AGN; see Section~\ref{sec:lineorigins}) that could contribute to $L_{\rm IR}$.

\subsection{Systematic Uncertainties and Robustness Tests}

The formal uncertainties in Table~\ref{tab:specmeas} are usually very small, which is expected given the high quality of the data and the simplicity of the star formation history and other aspects of the model. We emphasize that the listed uncertainties are purely statistical. Systematic errors in the models certainly dominate \citep[e.g.,][]{Muzzin09}. As a rough estimate of these, we also analyzed the data using the FSPS v3.0 \citep{Conroy09,Conroy10} stellar population synthesis models using the MIST isochrones \citep{Choi16}. We find systematic differences of $t_{\rm FSPS} - t_{\rm BC03} = +0.2$~Gyr in age, $+50$~Myr in $\tau$, $+0.1$~mag in $A_V$, $+0.2$~dex in sSFR, and $+0.14$~dex in $M_*$, and $+0.3$~\AA~in emission line EWs. In $[$Z/H$]$ the differences are not systematic, but there is a scatter of 0.2~dex indicating that the metallicity constraints are the least robust.

For MRG-M0138 we have analyzed spectra of both Image~1 and 2, which provides a consistently test. The same colors were used in both fits, but we rescaled the flux level to match Image~2 when fitting the FIRE spectrum. Because MRG-M0138 is rotating (Paper~II) and the slit cuts through the source at very different angles with respect to the major axis for Images~1 and 2, we do not expect to measure the same velocity dispersion $\sigma$ in the two spectra. However, the stellar population parameters are reasonably consistent.

\begin{deluxetable*}{lccccccc}
\tablecolumns{8}
\tablewidth{0pt}
\tablecaption{Spectroscopic Measurements\label{tab:specmeas}}
\tablehead{\colhead{Quantity} & \colhead{Units} & \colhead{MRG-M0150} & \colhead{MRG-P0918} & \colhead{MRG-S1522} & \colhead{MRG-M2129} & \colhead{MRG-M0138} & \colhead{MRG-M0138} \\
\colhead{} & \colhead{} & \colhead{} & \colhead{} & \colhead{} & \colhead{} & \colhead{Image 1} & \colhead{Image 2}}
\startdata
\cutinhead{Stellar population properties}
$z$ & \ldots & 2.6355 & 2.3559 & 2.4503 & 2.1487 & 1.9486 & 1.9469 \\
$\sigma$ & km~s${}^{-1}$ & $261 \pm 30$ & $223 \pm 16$ & $241 \pm 18$ & $266 \pm 21$ & $298 \pm 7$ & $409 \pm 11$ \\
Age & Gyr & $0.76 \pm 0.08$ & $0.51 \pm 0.02$ & $0.61 \pm 0.06$ & $0.80 \pm 0.10$ & $1.35 \pm 0.08$ & $1.39 \pm 0.16$ \\
$\tau$ & Myr & $95 \pm 35$ & $<43$ & $<71$ & $103 \pm 24$ & $181 \pm 17$ & $178 \pm 27$ \\
$A_V$ & mag & $0.61 \pm 0.09$ & $0.18 \pm 0.06$ & $0.34 \pm 0.07$ & $0.33 \pm 0.09$ & $0.35 \pm 0.05$ & $0.11 \pm 0.05$ \\
$[$Z/H$]$ & \ldots & $<0.33$ & $0.02 \pm 0.03$ & $-0.03 \pm 0.13$ & $0.16 \pm 0.13$ & $0.01 \pm 0.04$ & $0.25 \pm 0.09$ \\
log sSFR & yr${}^{-1}$ & $<-10.72$ & $<-12.80$ & $<-11.77$ & $-11.18 \pm 0.54$ & $-11.28 \pm 0.11$ & $-11.44 \pm 0.22$ \\
$\log \mu M_*$ & ${\rm M}_{\odot}$ & $12.06 \pm 0.04$ & $11.72 \pm 0.02$ & $11.74 \pm 0.03$ & $11.62 \pm 0.05$ & $12.77 \pm 0.03$ & $12.56 \pm 0.04$ \\
$\mu$ SFR & ${\rm M}_{\odot}~{\rm yr}^{-1}$ & $<22$ & $<1$ & $<1$ & $3 \pm 2$ & $31 \pm 9$ & $13 \pm 3$ \\
$\log M_*$ & ${\rm M}_{\odot}$ & $11.50 \pm 0.17$ & \ldots & \ldots & $10.96 \pm 0.10$ & $11.69 \pm 0.19$ & $11.68 \pm 0.19$\\
SFR & ${\rm M}_{\odot}$~${\rm yr}^{-1}$ & $< 6.1$ & \ldots & \ldots & $0.6 \pm 0.4$ & $2.6 \pm 1.4$ & $1.8 \pm 0.9$\\
\cutinhead{Emission line properties}
$\sigma_{\rm em}$ & km~s${}^{-1}$ & $213 \pm 20$ & $190 \pm 49$ & $345 \pm 55$ & $364 \pm 22$ & $\ldots$ & $\ldots$ \\
$v_{\rm em}-v_{\rm stars}$ & km~s${}^{-1}$ & $-20 \pm 24$ & $27 \pm 37$ & $40 \pm 41$ & $-12 \pm 30$ & $\ldots$ & $\ldots$ \\
H$\alpha$ EW & \AA & $4.3 \pm 1.2$ & $0.3 \pm 0.5$ & $4.2 \pm 0.9$ & $2.6 \pm 0.9$ & $\ldots$ & $\ldots$ \\
$[$\ion{N}{2}$]$ $\lambda$6585 EW & \AA & $24.5 \pm 2.3$ & $3.7 \pm 0.6$ & $11.1 \pm 1.8$ & $17.1 \pm 1.1$ & $\ldots$ & $\ldots$ \\
$[$\ion{O}{3}$]$ $\lambda$5008 EW & \AA & $\ldots$ & $0.1 \pm 0.2$ & $2.0 \pm 0.3$ & $5.3 \pm 0.5$ & $\ldots$ & $\ldots$ \\
$[$\ion{O}{2}$]$ EW & \AA & \ldots & $<3.5^{\dagger}$ & $5.1 \pm 0.9$ & $2.5 \pm 1.3$ & \ldots & $-0.1 \pm 0.6$ \\
$[$\ion{S}{2}$]$ EW & \AA & \ldots & \ldots & \ldots & $<3.6^{\dagger}$ & \ldots & \ldots \\
log $[$\ion{N}{2}$]$/H$\alpha$ & $\ldots$ & $0.74 \pm 0.12$ & $>0.33$ & $0.40 \pm 0.11$ & $0.79 \pm 0.16$ & $\ldots$ & $\ldots$ \\
log $[$\ion{O}{3}$]$/H$\beta$ & $\ldots$ & $\ldots$ & $\ldots$ & $0.28^{+0.32}_{-0.11}$ & $0.86^{+0.34}_{-0.15}$ & $\ldots$ & $\ldots$ \\
\sidehead{\emph{Line fluxes uncorrected for magnification or extinction:}}
H$\alpha$ flux & $10^{-18}$~cgs & $73 \pm 20$ & $5 \pm 7$ & $51 \pm 11$ & $30 \pm 10$ & $\ldots$ & $\ldots$ \\
$[$\ion{N}{2}$]$ $\lambda$6585 flux & $10^{-18}$~cgs & $391 \pm 38$ & $54 \pm 9$ & $129 \pm 21$ & $183 \pm 13$ & $\ldots$ & $\ldots$ \\
$[$\ion{O}{3}$]$ $\lambda$5008 flux & $10^{-18}$~cgs & $\ldots$ & $2 \pm 4$ & $31 \pm 5$ & $68 \pm 6$ & $\ldots$ & $\ldots$ \\
$[$\ion{O}{2}$]$ flux & $10^{-18}$~cgs & \ldots & $<44^{\dagger}$ & $42 \pm 7$ & $17 \pm 9$ & \ldots & $-3 \pm 27$ \\
$[$\ion{S}{2}$]$ flux & $10^{-18}$~cgs & \ldots & \ldots & \ldots & $<39^{\dagger}$ & \ldots & \ldots
\enddata
\tablecomments{BC03 models and a \citet{Chabrier03} IMF are assumed. Uncertainties are statistical and do not include model systematics (see text). All upper limits are $2\sigma$. The flux magnification factor is $\mu$; hence, $\log M_*$ is demagnified and $\log \mu M_*$ is not. EWs are reported in the rest frame. Line flux units are $10^{-18}$~erg~cm${}^{-2}$~s${}^{-1}$~\AA${}^{-1}$. The [\ion{O}{2}]$\lambda\lambda$3727,3730 and [\ion{S}{2}]$\lambda\lambda$6718,6733 fluxes and EWs refer to the sum of the two doublet components. The spectra and derived extrinsic quantities, including line fluxes, are scaled to match the total photometric magnitudes. ${}^{\dagger}$These lines were formally detected at marginal significance, but since inspection of the spectra showed these may be spurious, we have chosen to quote only upper limits.}
\end{deluxetable*}

\begin{figure}
\centering
\includegraphics[width=0.9\linewidth]{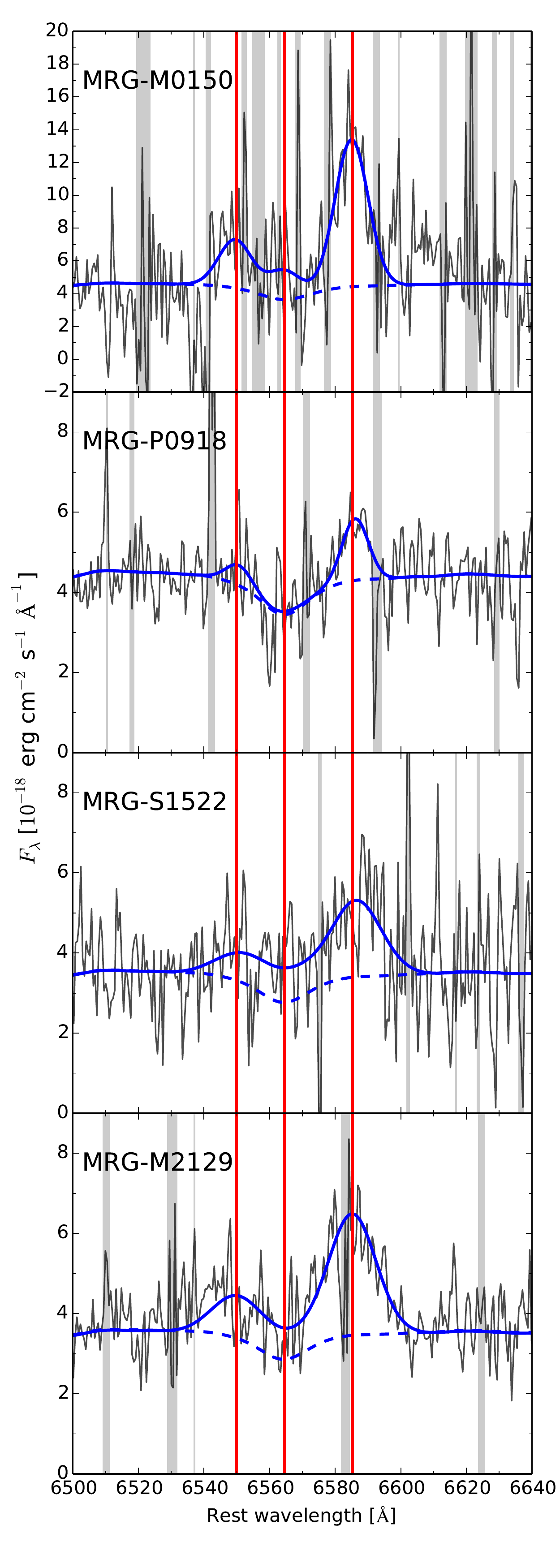}
\caption{Integrated spectra in the [\ion{N}{2}]+H$\alpha$ region binned to 25~km~s${}^{-1}$~pixel${}^{-1}$. Solid blue lines show the best-fit models, including the stellar continuum and line emission. Dashed blue lines include only the stellar continuum. Vertical red lines show the positions of the [\ion{N}{2}]$\lambda\lambda$6550,6585 and H$\alpha$ lines at the redshift of the stars. Grey bars indicate regions of bright sky emission.
\label{fig:niiha}}
\end{figure}

\begin{figure*}
\centering
\includegraphics[width=0.48\linewidth]{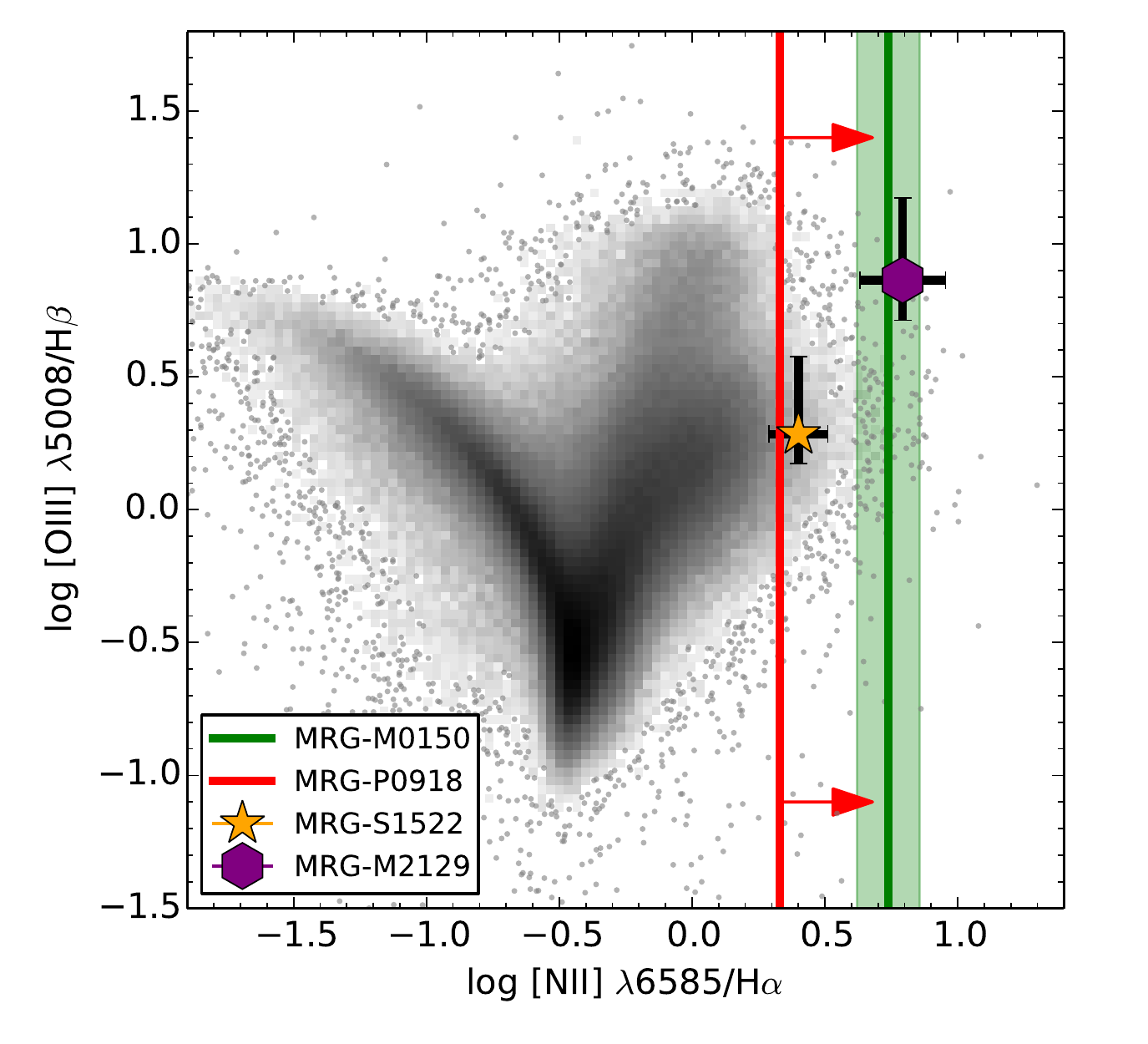}
\includegraphics[width=0.48\linewidth]{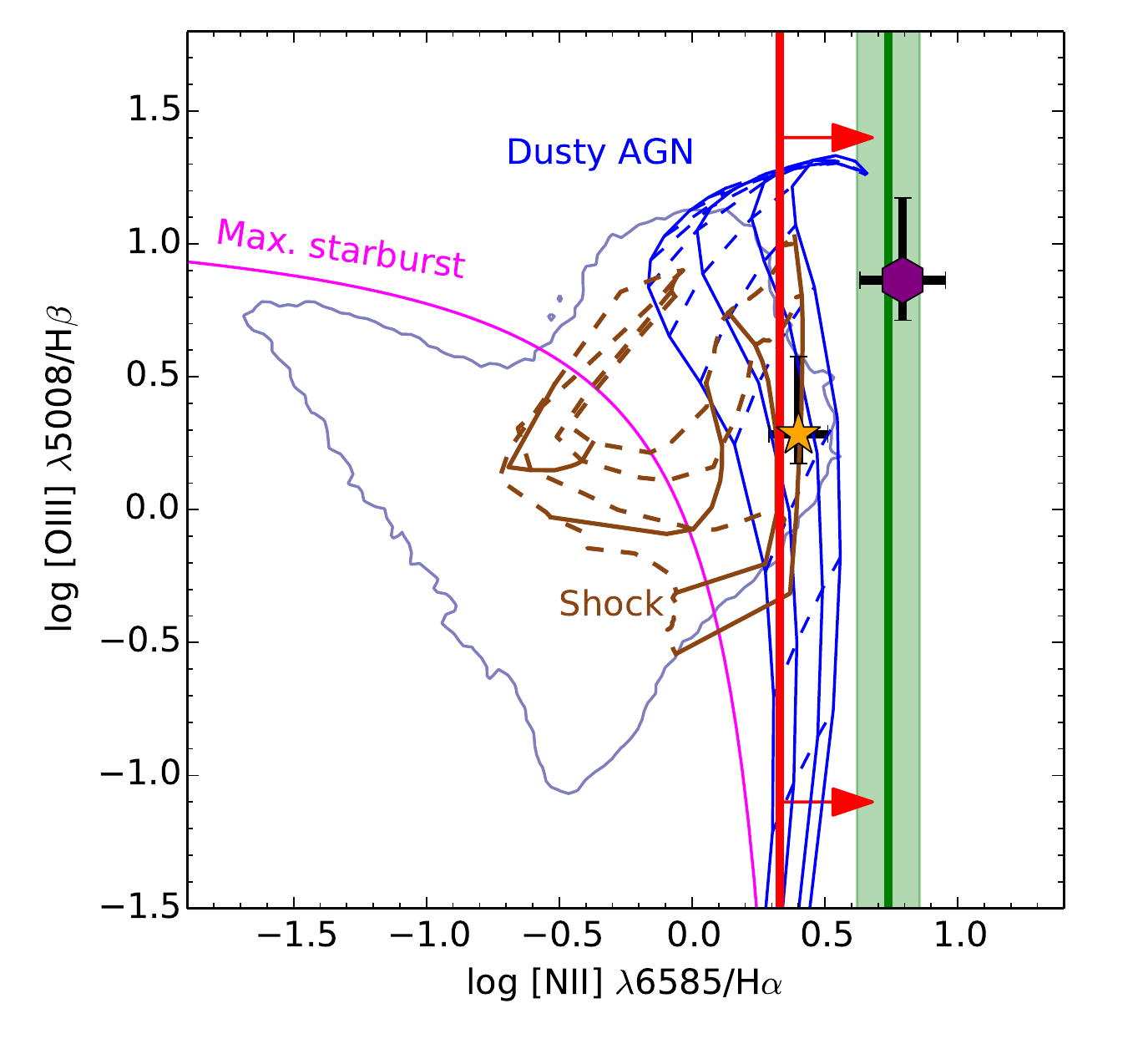}
\caption{\emph{Left:} The BPT diagram showing the lensed quiescent galaxies with detected emission lines. Note that there is no constraint on [\ion{O}{3}]/H$\beta$ for MRG-P0918 and MRG-M0150, and only a lower limit on [\ion{N}{2}]/H$\alpha$ for the former. The grey shading shows the location of local galaxies in the SDSS selected from the MPA-JHU catalog \citep{Brinchmann04} with $>2\sigma$ detections in all lines. Individual grey points are shown where the density is low. \emph{Right:} The lensed quiescent galaxies are compared to several theoretical models. The magenta line is the \citet{Kewley01} theoretical maximum starburst curve. The blue lines show a grid of dusty AGN models \citep{Groves04a,Groves04b} with $4Z_{\odot}$ metallicity, $n = 10^4$~cm${}^{-3}$, $\alpha=-2$ to -1.2, and $\log U=-4$ to 0. The brown lines show a grid of shock models \citep{Allen08} without precursor emission, $2Z_{\odot}$ metallicity, $n = 1$~cm${}^{-3}$, $B = 10^{-4}$ to 10~$\mu$G, and velocities 100-1000~km~s${}^{-1}$. The grids were generated using {\tt itera} \citep{Groves10}. The gray contour outlines the SDSS locus from the left panel for reference.\label{fig:bpt}}
\end{figure*}

\begin{figure*}
\centering
\includegraphics[width=\linewidth]{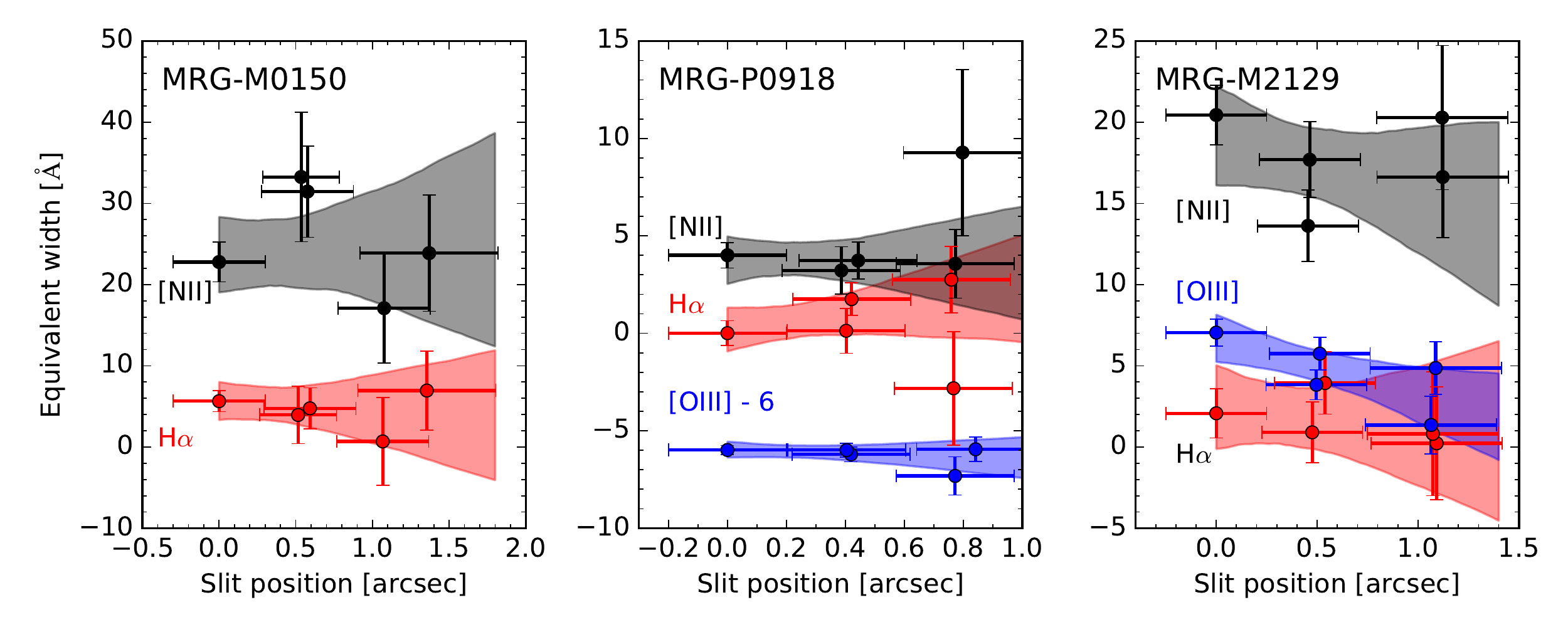}
\caption{Spatially resolved emission line rest-frame EWs as a function of distance along the slit from the center of the image. Measurements on both sides of center are plotted. The horizontal error bars indicate the bin sizes. Points have been shifted horizontally by $\lesssim 0\farcs1$ for clarity, and the [\ion{O}{3}] EW in the middle panel is shifted downward 6~\AA. Colored bands show the 95\% credible range for a linear fit. The only gradient with $>2\sigma$ significance is the decline in [\ion{O}{3}] with radius in MRG-M2129.\label{fig:resolved_ew}}
\end{figure*}

\section{Measurement and Interpretation of Emission Lines}
\label{sec:emlines}

We detected emission lines in all of the lensed galaxies except MRG-M0138 (in that case, H$\alpha$, [\ion{N}{2}], and [\ion{O}{3}] are all contaminated by telluric absorption). In all cases the strongest line is [\ion{N}{2}]. H$\alpha$ is much weaker, with rest-frame equivalent widths (EWs) of $\lesssim 4$~\AA. As seen in Figure~\ref{fig:niiha}, we detect this weak H$\alpha$ emission from the infilling of the stellar absorption in all cases except MRG-P0918, for which we are able to place only an upper limit. The ratio $[$\ion{N}{2}]~$\lambda 6585$/H$\alpha$ ranges from 2-6. Figure~\ref{fig:bpt} shows our sample in the BPT diagram \citep{Baldwin81}. Comparing to the \citealt{Kewley01} extreme starburst line (magenta), we find that the [\ion{N}{2}]/H$\alpha$ ratios are all well in excess of those producible from photoionization by massive stars. Therefore the nearly ubiquitous line emission in this sample is not indicative of low-level star formation.\footnote{Even if the weak H$\alpha$ emission were interpreted as arising from star formation, assuming that its attenuation is similar to the starlight, the inferred SFRs would generally be consistent with limits from the stellar continuum modeling: $\mu {\rm SFR} = 28 \pm 8$, $3 \pm 2$, $14 \pm 3$, and $5 \pm 2 \msol$~yr${}^{-1}$ for MRG-M0150, MRG-P0918, MRG-S1522, and MRG-M2129, respectively.} In this section we consider the robustness of the emission line measurement and explore its possible origins.

\subsection{Robustness of Emission Line Measurements}

It is clear that the stellar absorption correction significantly affects the inferred H$\alpha$ emission. Are the high inferred  [\ion{N}{2}]/H$\alpha$ ratios reliable? To test this, we fit only the [\ion{N}{2}]+H$\alpha$ region of the spectrum with the continuum fixed to a simple stellar population with an age of 400~Myr. Since the H$\alpha$ absorption is maximum around this age, the test supplies an upper limit to the H$\alpha$ flux and thus a lower limit to [\ion{N}{2}]/H$\alpha$. Due to the weak sensitivity of the H$\alpha$ absorption to age in the relevant range of ages, [\ion{N}{2}]/H$\alpha$ never shifted by more than half of its random uncertainty in this test. Systematic shifts in the line ratios that arise when using the FSPS versus BC03 stellar models are even smaller ($\simeq 0.02$~dex), with the exception that the lower limit on log~[\ion{N}{2}]/H$\alpha$ is weakened from $> 0.54$ to $>0.33$ for MRG-P0918 when using the FSPS models. We conservatively adopted this weaker constraint in Table~\ref{tab:specmeas}. In all other cases, the [\ion{N}{2}]/H$\alpha$ ratio is robust to the stellar continuum modeling details.

The [\ion{O}{3}]/H$\beta$ ratio is inferred less directly. Since the H$\beta$ and higher-order emission is much weaker than the stellar absorption, the intensity of the Balmer emission lines in the fit is driven by H$\alpha$ infilling, and we cannot constrain the Balmer decrement H$\beta$/H$\alpha$. Our spectrum model assumes that the nebular lines and the stars are equally attenuated. If the line emission is more extinguished, this would push the inferred [\ion{O}{3}]/H$\beta$ upward. \citet{Reddy15} found that the attenuation of the H$\alpha$ photons can exceed that of the stellar continuum by up to 1.5~mag in high-$z$ star-forming galaxies. We think that such a large difference is unlikely to hold in quiescent galaxies where attenuation is mild, but assuming 1.5~mag of differential attenuation as a limiting case, we would infer [\ion{O}{3}]/H$\beta$ typically 0.3~dex higher. The asymmetric error bars in Figure~\ref{fig:bpt} and Table~\ref{tab:specmeas} include this uncertainty added in quadrature.

\subsection{Origins of Line Emission}
\label{sec:lineorigins}

Determining the excitation mechanisms in the Seyfert/LIER\footnote{LIERs are low-excitation emission-line regions. Following \citet{Belfiore16} and others, we drop the ``nuclear'' designation of the more traditional LINER term since the emission is not confined to the central regions.} region of the BPT diagram is difficult, even in local galaxies with much more detailed information. Although we cannot expect to definitively identify the excitation mechanisms in every case, we can consider which scenarios are consistent with the available evidence. The ratio [\ion{N}{2}]/H$\alpha \simeq 2$-6 is high throughout our sample, but the strength of the oxygen lines and the spatial distribution and kinematics of the gas relative to the stars provide additional constraints and show more diversity.

The spatial variation of emission line EWs is shown in Figure~\ref{fig:resolved_ew}. These measurements were made from spectra extracted in bins along each image. (These bins are the same used to extract stellar kinematics in Paper~II, where further details can be found.) We then modeled each spectrum as described in Section~\ref{sec:contfit}, except that we did not include photometric constraints which are not necessary to measure the emission line EWs.

Photoionization by an AGN is one way to produce a high [\ion{N}{2}]/H$\alpha$ ratio. In this scenario we expect high [\ion{O}{3}]/H$\beta$ and [\ion{O}{3}]/[\ion{O}{2}] ratios and a negative gradient in the emission line EWs (i.e., centrally concentrated emission). MRG-M2129 is the only galaxy that fits these criteria. Its high values of log~[\ion{O}{3}]/H$\beta = 0.86 \pm 0.15$ and log~[\ion{O}{3}]/[\ion{O}{2}]${} = 0.5 \pm 0.2$ are characteristic of Seyfert rather than LIER-type emission \citep{Veilleux87,Kewley06}.\footnote{Our measurements of [\ion{O}{3}]/[\ion{O}{2}] assume that the emission line and stellar light are equally attenuated. We caution that the [\ion{O}{2}] emission in MRG-M2129 is very weak and lies in a spectral region with significant residuals. Nonetheless, at a minimum we can exclude that [\ion{O}{2}] is stronger than [\ion{O}{3}] as in LIERs.} The emitting gas extends out to radii of at least 2~kpc, with a declining [\ion{O}{3}] EW that is qualitatively consistent with photoionization by a central source. No other emission line in any of the galaxies in our sample shows $>2\sigma$ evidence for radial variation in Figure~\ref{fig:resolved_ew}. A compact central source is detected in the \emph{HST} images of MRG-M2129 (Section~\ref{sec:m2129lensmodel}), which could be a Seyfert nucleus. All of these lines of evidence point toward photoionization by an AGN as one component of the excitation. \citet{Toft17} came to similar conclusions on MRG-M2129, with some quantitative differences that we discuss in the Appendix.

Although present in MRG-M2129, AGN photoionization is likely not the dominant source of excitation in MRG-S1522 or MRG-P0918. The lower [\ion{O}{3}]/H$\beta$ ratio of MRG-S1522 is more consistent with a LIER than a Seyfert classification in the BPT diagram, although the measurement uncertainties make this ambiguous. Confirmation comes from its stronger [\ion{O}{2}] emission than [\ion{O}{3}], as seen in LIER-type spectra, with log~[\ion{O}{3}]/[\ion{O}{2}]${} = -0.2 \pm 0.1$. In MRG-P0918, we do not detect [\ion{O}{3}] to sensitive limits. Although we cannot constrain the ratio [\ion{O}{3}]/H$\beta$ and place this galaxy on the BPT diagram, the low EW of [\ion{O}{3}] and its weakness compared to [\ion{N}{2}] (with a ratio $\lesssim 0.1$) are not typical of Seyfert-type spectra. For MRG-M0150, the only strong emission lines observable from the ground are [\ion{N}{2}] and H$\alpha$, so more information on this galaxy will have to await improved spectral coverage with \emph{JWST}.

Photoionization of diffuse interstellar gas by hot evolved stars, such as post-asymptotic giant branch (post-AGB) stars, is often involved to explain LIER emission in local early-type galaxies \citep[e.g.,][]{Yan12,Singh13,Belfiore16}. At the ages relevant to quiescent galaxies at $z = 2$, i.e., $\simeq0.5$-3~Gyr, stellar population synthesis models predict H$\alpha$ EWs of $\simeq0.1$-0.8~\AA~\citep[][Figure~2]{CidFernandes11}. For MRG-M0150, MRG-S1522, and MRG-2129, we observe H$\alpha$ EWs in the range 2.6-4.3~\AA, which means there is insufficient ionizing flux from post-AGB stars to explain most of the emission. In MRG-P0918, however, the lower H$\alpha$ EW of $0.3 \pm 0.5$~\AA~is consistent with expectations for post-AGB stars. Furthermore, in this scenario the gas should share the distribution and kinematics of the stars, which we observe: the gaseous and stellar velocity dispersions are consistent (Table~\ref{tab:specmeas}) and the emission line EWs are essentially constant (Figure~\ref{fig:resolved_ew}). The main difficulty with this interpretation is that photoionization models of 3-13~Gyr old populations produce log~[\ion{N}{2}]/H$\alpha \lesssim 0.1$ \citep{Binette94,Byler17}, which is smaller than observed in MRG-P0918. Since models at the relevant age of 0.5~Gyr have not been explored and probably have uncertainties in the shape of the ionizing spectrum, we consider that post-AGB stars may still be a viable explanation for most of the line emission in MRG-P0918, but not for the other galaxies in our sample.

Shocked gas is another possible source of high-[\ion{N}{2}]/H$\alpha$ emission with low ionization. Shocks are consistent with the lower values of [\ion{O}{3}]/H$\beta$ and [\ion{O}{3}]/[\ion{O}{2}] seen in MRG-S1522. Supporting evidence comes from the high line width of the ionized gas ($\sigma =345$ km~s${}^{-1}$) relative to the stars (241 km~s${}^{-1}$), which indicates outflowing or turbulent ionized gas. We think that shocks are likely contributing to the line emission in MRG-M2129 as well, since the line width of the gas is also elevated well above that of the stars (364 versus 266~km~s${}^{-1}$). Such a situation could in general be explained by a differing spatial distribution of the gas and stars. However, in the case of MRG-M2129 the EW of the [\ion{N}{2}] emission does not change much across the image (Figure~\ref{fig:resolved_ew}), and the elevated $\sigma_{\rm gas}$ is not confined to the nucleus but extends to $R \simeq 2$~kpc. This suggests that both AGN photoionization and shocks likely contribute to the line emission in MRG-M2129.

We have explored whether the \citet{Allen08} shock models can reproduce the line ratios in Figure~\ref{fig:bpt}. For MRG-S1522, we find that a model with twice solar metallicity, $n = 1$~cm${}^{-3}$, and $B = 2~\mu$G can reproduce the [\ion{N}{2}]/H$\alpha$ and [\ion{O}{3}]/H$\beta$ ratios for shock velocities $v \gtrsim 400$~km, comparable to $\sigma_{\rm gas}$.\footnote{We note that $\sigma_{\rm gas}$ includes unresolved rotation and so does not measure purely turbulent motion.} The agreement holds only if we consider the shock emission and not the radiative precursor, however, which may be an unphysical scenario. The emitting area required to produce the observed H$\alpha$ luminosity is $\lesssim 100~(\mu / 5)^{-1}$~kpc${}^2$, which corresponds to a spherical radius $\lesssim 3~(\mu / 5)^{-1/2}$~kpc. Since this is comparable to the galaxy size it requires a large-scale shock, but the energetics are feasible. On the other hand, the high [\ion{N}{2}]/H$\alpha$ ratios of MRG-M2129 and MRG-0150 cannot be reproduced by any of the \citet{Allen08} shock models or the \citet{Groves04a,Groves04b} AGN models. This is demonstrated by the grids shown in Figure~\ref{fig:bpt}, where we have chosen parameters that maximize the [\ion{N}{2}]/H$\alpha$ ratio. Extracting more detailed physical information about these intruiging systems may require further developments in the models.

In summary, four of the five lensed galaxies in our sample show line emission with very high [\ion{N}{2}]/H$\alpha$ ratios inconsistent with a star formation origin. (For the fifth, MRG-M0138, all of the strong nebular lines except [\ion{O}{2}] are inaccessible from the ground.) MRG-M2129 shows clear evidence of AGN photoionization. Based on emission line ratios and the distribution and kinematics of the ionized gas, we argue that shocked gas is present in MRG-S1522 and MRG-M2129. Shocks could well power the line emission in MRG-M0150, but we lack the information needed to distinguish AGN photoionization, and they could also contribute in MRG-P0918, but post-AGB stars might instead be the main ionizing source in that system. The high incidence of line emission in galaxies seen soon ($<800$~Myr) after quenching, the indications of shocked gas in many (and potentially all) cases, and the lack of ongoing star formation are all consistent with the idea that AGN-driven outflows or turbulence could play an important role in maintaining the quiescence of these galaxies. We will discuss these implications further in Section~\ref{sec:disclines}.

\begin{figure*}
\centering
\includegraphics[width=0.48\linewidth]{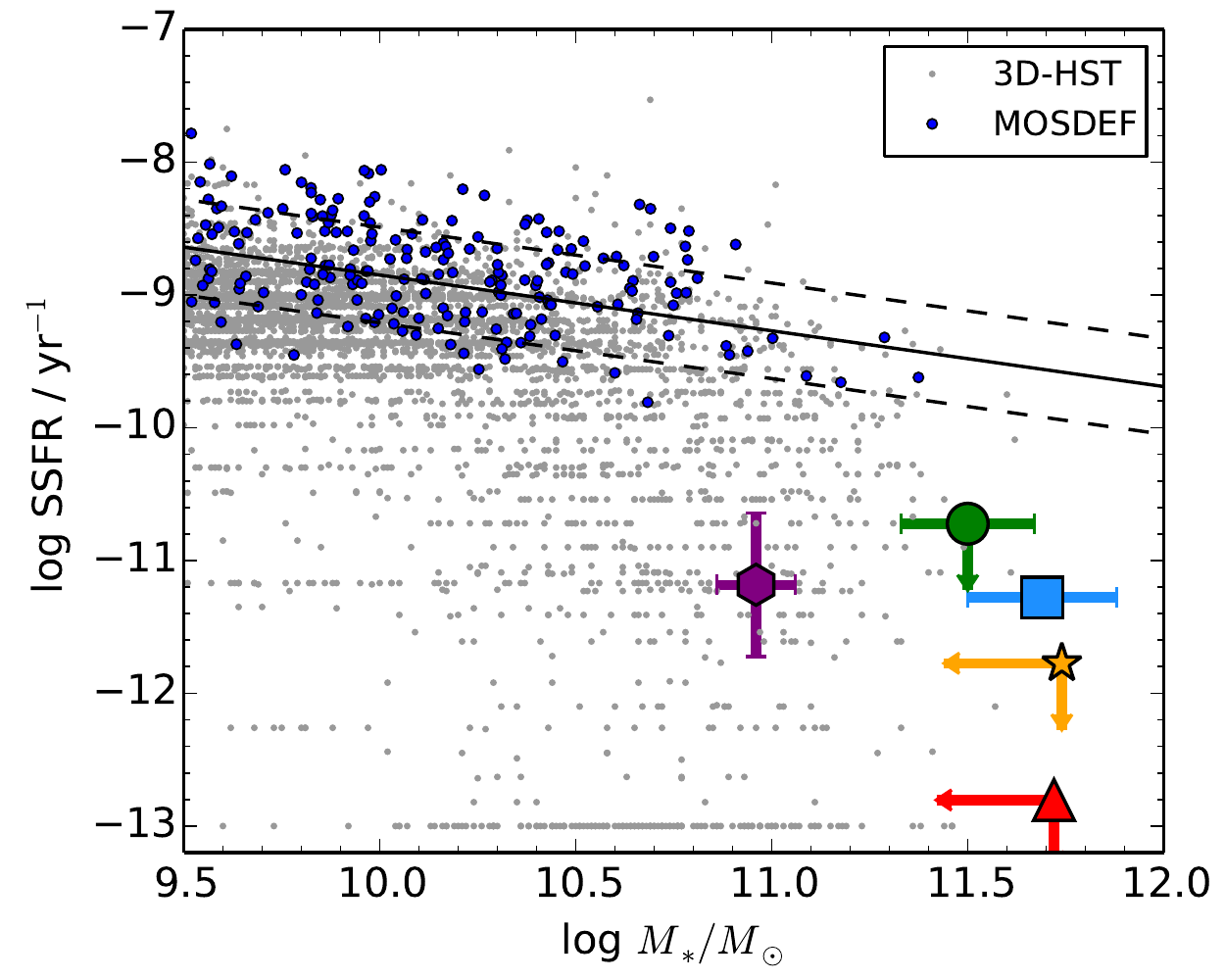}
\includegraphics[width=0.48\linewidth]{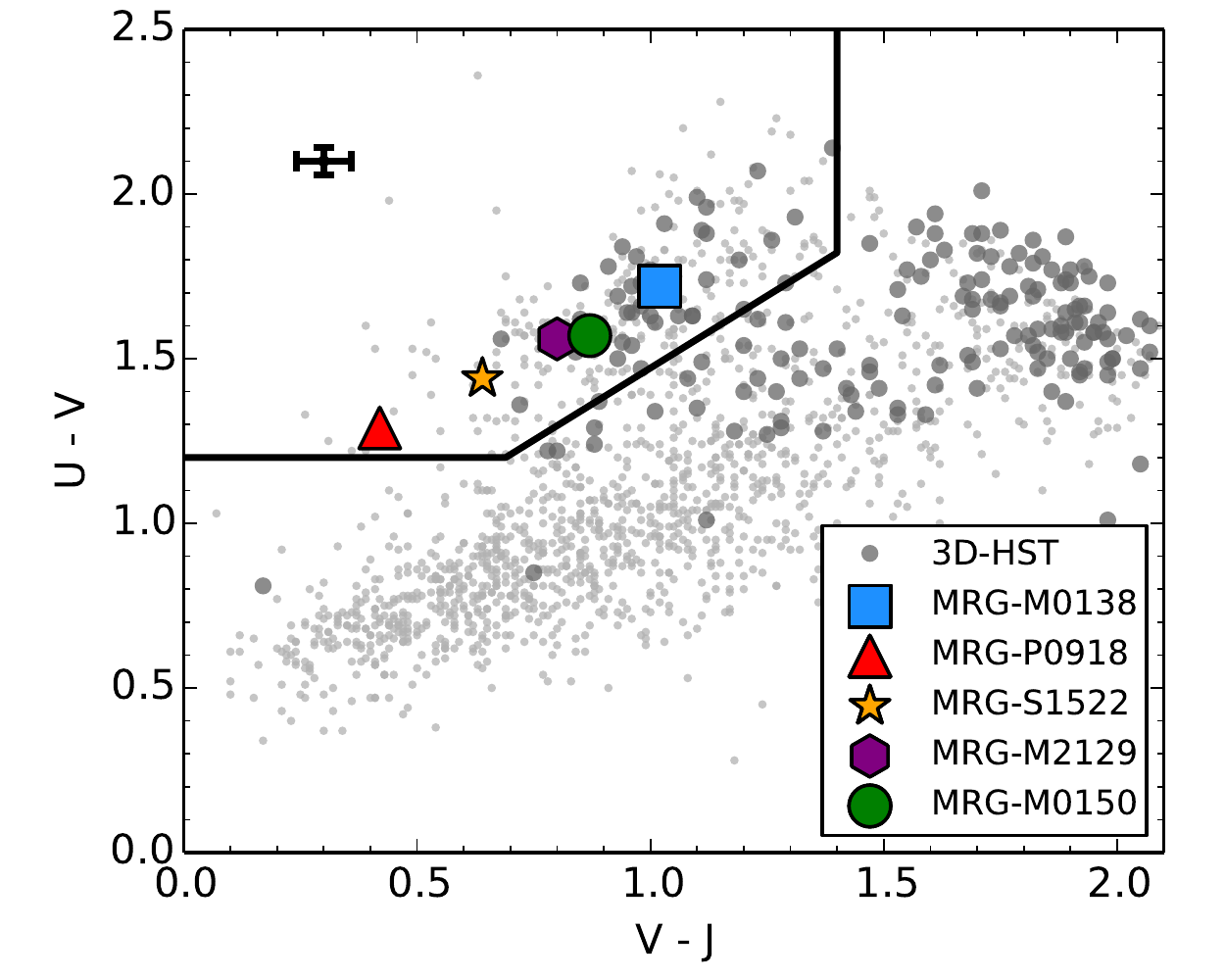}
\caption{\emph{Left:} The stellar mass--sSFR relation (the ``main sequence'') defined by (a) a stellar mass-selected sample at $z=2$-2.6 from the 3D-HST catalogs \citep{Skelton14}, where the SFRs are derived from SED fitting, and (b) H$\alpha$-based SFRs, corrected for extinction, for galaxies in the MOSDEF survey \citep{Shivaei15} at $z=2.09$-2.61. The linear fit and $1\sigma$ scatter for the MOSDEF sample is shown. The lensed galaxies (see legend in right panel) fall $\gtrsim 1.5$~dex below the ``main sequence'' of star forming galaxies. The stellar masses of lensed galaxies with unknown magnifications are plotted as upper limits. \emph{Right:} The rest-frame $UVJ$ colors of galaxies in the 3D-HST fields at $z=2$-2.6 are compared to those of the lensed sample. Smaller and larger grey symbols denote 3D-HST galaxies having $10^{10} \msol < M_* < 10^{11} \msol$ and $M_* > 10^{11} \msol$, respectively. Colors of the lensed galaxies are determined by integrating the fits in Figures~\ref{fig:specfits} and \ref{fig:specfits2}; representative uncertainties are indicated by the black error bars. The black line separates the star-forming and quiescent regions defined by \citet{Whitaker11}. The lensed sample spans the quiescent sequence, except for the reddest cases.\label{fig:mainseq}}
\end{figure*}

\begin{figure*}
\centering
\includegraphics[width=0.48\linewidth]{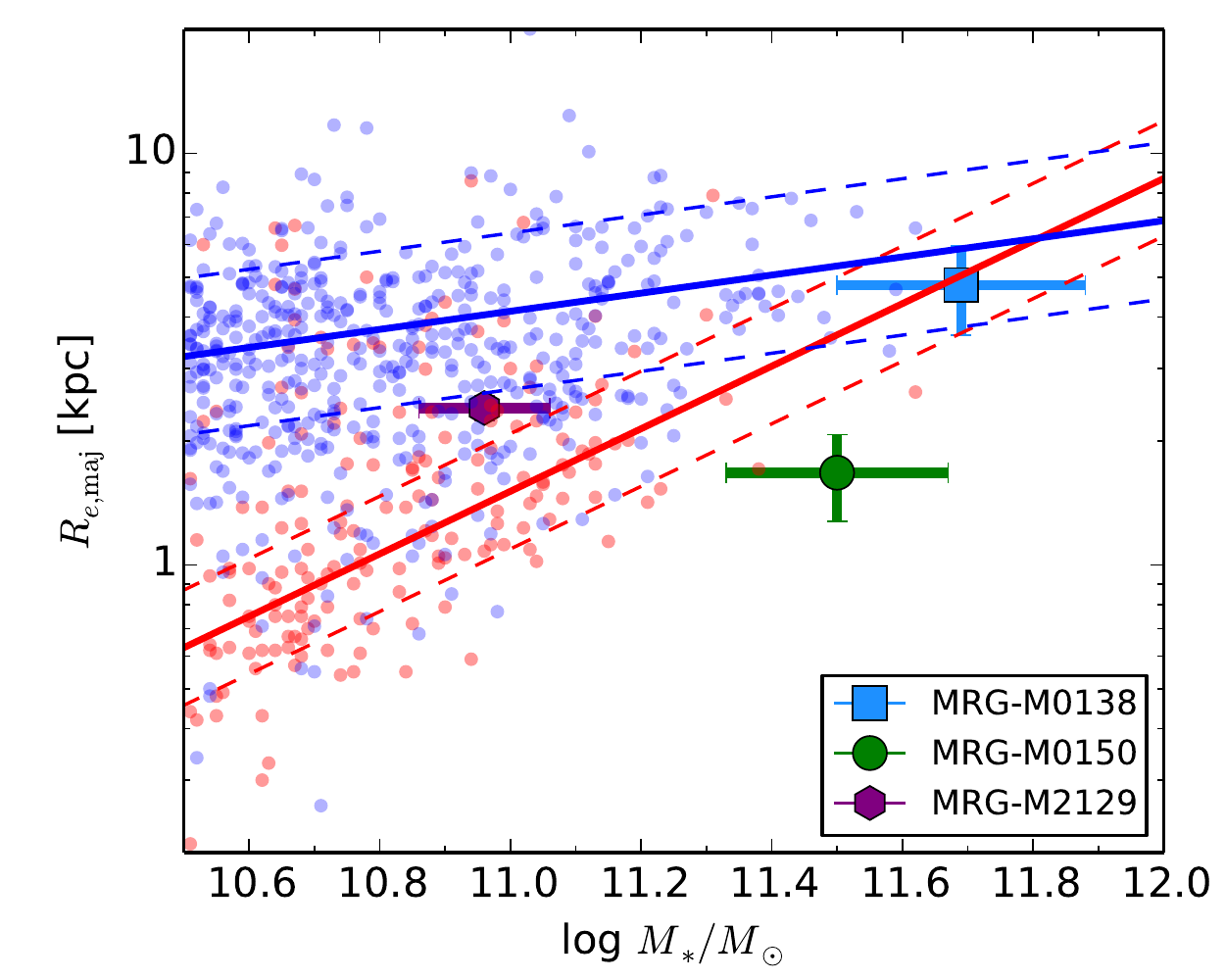}
\includegraphics[width=0.48\linewidth]{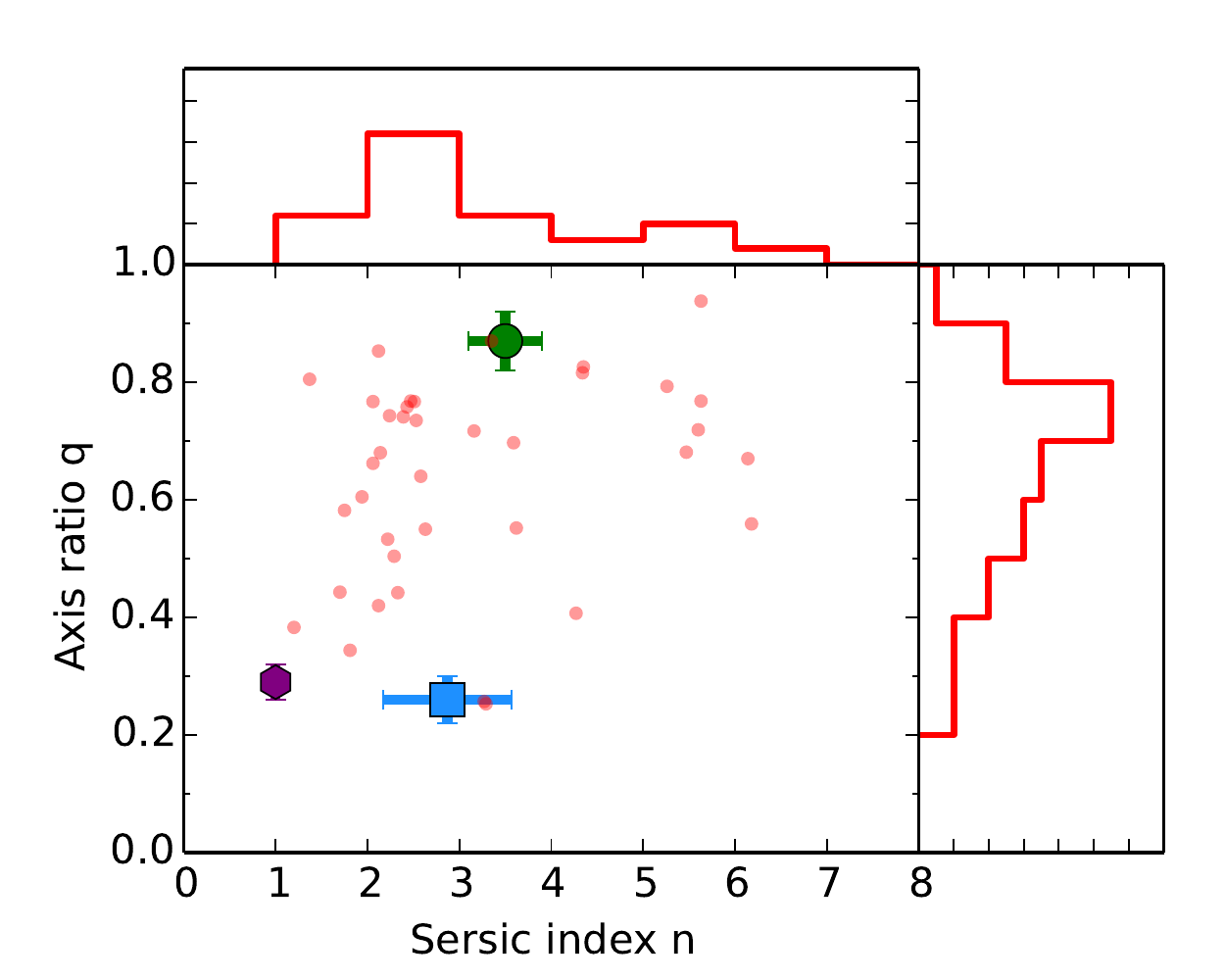}
\caption{\emph{Left:} The stellar masses and sizes of the lensed quiescent galaxies with lens models are compared to the relations defined by star-forming (blue) and quiescent (red) galaxies at $z=2$-2.6 in the 3D-HST fields, as classified by the $UVJ$ criterion \citep{Whitaker11} and measured by \citet{vanderWel14}. Linear fits to the relation at $z=2.25$ are overlaid, with dashed lines indicating the $1\sigma$ scatter. \emph{Right:} S\'{e}rsic indices $n$ and projected axis ratios $q=b/a$ for the lensed quiescent sample (see legend in left panel) are compared to those of similarly massive ($M_* > 10^{11} \msol$) quiescent galaxies at $z=2$-2.6 in the 3D-HST fields. Histograms show the marginal distributions of the 3D-HST sample. Measurements from our single-S\'{e}rsic models are shown for the lensed galaxies. The lensed sample is reasonably representative of coeval quiescent galaxies in their sizes and S\'{e}rsic indices, but high ellipticities (low $q$) may be overrepresented. \label{fig:mass_size}}
\end{figure*}

\section{Discussion}
\label{sec:discussion}

We have discovered a sample of five quiescent galaxies at $z = 2$-2.6 that are significantly magnified by galaxy clusters. These galaxies are extraordinarily bright in the near-infrared---$H_{\rm AB} < 20$ in all cases---due to their high stellar masses ($M_* \gtrsim 10^{11} \msol$ for the three galaxies with lens models) combined with lensing magnification by factors of $\mu \approx 4$-13. Observations with the Magellan/FIRE and Keck/MOSFIRE spectrographs confirmed the redshifts and evolved stellar populations (ages 0.5-1.4~Gyr) for the full sample. The integrated spectra are among the most detailed yet obtained for quiescent galaxies at these redshifts, particularly for the remarkable case of MRG-M0138, which is the NIR-brightest lensed distant galaxy yet discovered. In addition to the flux amplification, lensing affords the possibility of spatially resolving the stellar continuum with ground-based NIR spectrographs and measuring their internal kinematics and stellar population gradients. The lensed galaxies in the present sample are the only quiescent systems at $z > 2$ for which this is currently practical, making them unique and valuable resources. 

\subsection{Assessing the Representative Nature of the Lensed Galaxy Sample}
\label{sec:representativeness} 

Further papers will present spatially resolved measurements of the stellar kinematics and populations. In order to interpret these measurements, we must be able to place this sample in the context of the full galaxy population. Here we compare our lensed galaxies to coeval samples selected from deep field surveys.

Figure~\ref{fig:mainseq} (left panel) compares the sample to the star formation ``main sequence'' derived in the 3D-HST and MOSDEF surveys. The lensed sample falls $\gtrsim1.5$~dex below the main sequence. The right panel of Figure~\ref{fig:mainseq} shows that all galaxies in our sample fall in the quiescent region of the $UVJ$ diagram. Consistent with the wide range of ages spanning 0.5-1.4~Gyr that we measured from the spectra and photometry (Section~\ref{sec:stellarpops}), our sample spans nearly the full range of the $UVJ$ quiescent sequence. Only the reddest and presumably oldest galaxies are not represented. 

In Figure~\ref{fig:mass_size}, the structural properties of the three galaxies with a lens model are compared to coeval quiescent galaxies in the 3D-HST survey. The left panel shows that the lensed galaxies scatter around the mass--size relation and so can be considered typically ``compact.'' The right panel compares the axis ratios and S\'{e}rsic indices of our sample to those of similarly massive, coeval quiescent galaxies. The lensed sample displays S\'{e}rsic indices of $n \simeq 1$-4 and is consistent with being drawn from the 3D-HST distribution. On the other hand, the axis ratios of the lensed galaxies are at the round and flat extremes of the 3D-HST distribution. 

Two aspects of the sample that are perhaps surprising are the high ellipticities of two systems (MRG-M0138 and MRG-M2129) and the high stellar masses ($\log M_* = 11.69 \pm 0.19$ for MRG-M0138 and $\log M_* = 11.50 \pm 0.17$ for MRG-M0150) of two others. Considering first the ellipticities, we could suppose that errors in the lens model could lead to spuriously high values. However, we found that the ellipticity is robustly recovered across three lens models for MRG-M2129. In Paper~II we will show that their stellar kinematics imply that both MRG-M0138 and MRG-M2129 are intrinsically flat. These comparisons suggest that the high ellipticities are genuine. One might suppose that a selection effect could enhance their probability of inclusion in our sample. This might be the case if images were selected for follow-up based on their length-to-width ratio, for example, but our follow-up observations were based on a complete color-magnitude--selected sample in the central regions of the galaxy clusters in our survey. Given the small number of objects, it may be that high ellipticities are simply randomly over-represented in the three lensed galaxies with lens models. Regardless of its origin, this over-representation is important to bear in mind for our kinematic analysis in Paper~II.

Turning to the remarkably high stellar masses inferred for MRG-M0150 and MRG-M0138, we note that since we are studying galaxies on the bright tail of the luminosity function, we expect the lensed galaxies to pile up near the flux limit of our selection box. This is generally the case for the $H$ band flux (Figure~\ref{fig:YH}), with the exception of the ultra-bright MRG-M0138. In Paper~II we will show that its dynamical mass agrees with the stellar mass inferred in this paper. Since the stellar and dynamical masses have different dependences on the lens mapping, this consistency provides some reassurance that the magnification is not very far in error, i.e., well beyond the $\sim0.2$~dex uncertainty we estimated in Section~\ref{sec:magerrors}. For MRG-M0150 we will show in Paper~II that the dynamical mass is $\simeq 0.2$~dex smaller than the stellar mass. This could indicate that the magnification is underestimated and the stellar mass overestimated, but the difference is within the estimated uncertainties and, in any case, still implies a very massive galaxy.

We conclude that the sample is broadly representative of the colors and sizes of massive quiescent galaxies at $z=2$-2.6. The stellar masses and ellipticities are likely to be accurate within our estimated uncertainties, but the subset of three galaxies with lens models has a high proportion of high-ellipticity galaxies compared to unlensed samples and so is not fully representative in that property. Future lens models with additional constraints from any newly identified multiple image systems would be useful to help validate our magnification estimates.

\subsection{Disky Quiescent Galaxies and Evolution to $z=0$}
\label{sec:evolution}

The structures of the galaxies in our sample are very different from typical early-type galaxies in the local universe with a similar mass. Their sizes are comparable to other $z=2$ quiescent galaxies (Figure~\ref{fig:mass_size}) and are therefore smaller than $z=0$ galaxies of equal mass. Even more striking, however, is the presence of a dominant disk component in two of the three galaxies for which we have reconstructed the source, MRG-M0138 and MRG-M2129. Both galaxies have ellipticities $e > 0.7$. To place these shapes in the context of low-redshift galaxies, we selected galaxies with $\log M_* > 11$ and $z = 0.05$-0.25 from the GAMA survey \citep{Driver09,Taylor11,Kelvin12}. Only 4\% of the local galaxies have ellipticities $e > 0.7$. Even more rare is the apparent lack of a bulge in MRG-M2129, which is well-described by a pure exponential disk ($n=1$), as also remarked by \citet{Toft17}. In our local comparison sample, only 0.9\% of galaxies have $n \leq 1$.

These galaxies therefore need to evolve in size, shape, and concentration to resemble their $z \sim 0$ descendants. The extreme differences in structure imply that relatively little of the total transformation from star-forming galaxies into local early-type systems occurred when these galaxies were quenched. Instead these changes must have occurred later, likely through a series of major and minor mergers. The structures of the galaxies in our sample also have implications for their earlier evolution. In particular, whatever processes quenched star formation in these galaxies did not destroy the stellar disk or, in the case of MRG-M2129, even produce a significant bulge. In Paper~II we will discuss the past and future evolution of these galaxies in greater detail by bringing additional kinematic evidence to bear.

\subsection{The Nature of Emission Lines in $z\sim2$ Quiescent Galaxies} 
\label{sec:disclines}

The nearly ubiquitous presence of line emission with high [\ion{N}{2}]/H$\alpha$ ratios in our sample of massive, quiescent galaxies is striking. While massive ($M_* \gtrsim 10^{10.9} \msol$) star-forming galaxies at $z=1$-3 commonly host nuclear outflows that are thought to be driven by AGN \citep{Genzel14}, less is known about quiescent galaxies in this redshift range. \citet{Belli17b} studied the incidence of line emission in $z=0.7$-2.7 galaxies in the KMOS${}^{\rm 3D}$ survey. They detected unambiguous line emission in 17\% of the quiescent galaxies (as classified by $UVJ$ colors) and in about half of these attributed the emission to shocks. It is interesting that in our sample the incidence is much higher---emission line are detected in all four galaxies for which the strong lines are observable from the ground---and that our sample reaches [\ion{N}{2}]/H$\alpha \simeq 6$, whereas values above $\sim 1$ are absent from the Belli et al.~sample.

These differences might be explained in part by differences in the sensitivities of the observations and in the stellar masses characterizing each sample. Belli et al.~found a trend for galaxies with lower H$\alpha$ EWs and higher stellar masses to have higher [\ion{N}{2}]/H$\alpha$ ratios. Since the lensed galaxies in our sample have, on average, lower H$\alpha$ EWs and higher masses than the KMOS${}^{\rm 3D}$ quiescent galaxies, it is possible that future observations will show that the KMOS${}^{\rm 3D}$ and lensed galaxies sample different parts of a common sequence. We emphasize that we can characterize the weak emission lines in our sample of quiescent galaxies only because of the galaxies' brightness (due to their high masses combined with lensing magnification) and our accurate modeling of the stellar continuum. With shallower data we could not measure the infilling of the stellar H$\alpha$ absorption, and if a stellar redshift could not be measured then one could easily mistake the stronger [\ion{N}{2}] $\lambda6585$ line for H$\alpha$, leading to a completely different interpretation.

Shocks are most clearly present in two cases (MRG-S1522 and MRG-M2129), as judged by the line ratios, gas kinematics, and in the case of MRG-M2129, the extended distribution of the [\ion{N}{2}] emission. Although shocks could be present in MRG-P0918 and MRG-M0150 as well, other explanations remain viable in these cases because of the fewer constraints on MRG-M0150 and the low H$\alpha$ EW seen in MRG-P0918 (see Section~\ref{sec:lineorigins}). Given that the current rates of star formation are very low in MRG-S1522 and MRG-M2129 ($\lesssim 1 \msol~{\rm yr}^{-1}$ ; Table~\ref{tab:specmeas}), sources of energy are needed other than massive stars and core-collapse supernovae. Type Ia supernovae occurring at the expected rate \citep{Maoz12}, which is much higher than in local massive ellipticals, can supply kinetic power comparable only to the H$\alpha$ luminosities seen in our sample, which is expected to be $\lesssim2\%$ of the shock luminosity \citep{Allen08}. Shocks powered by type Ia supernovae ejecta therefore cannot generate the observed line emission. Mergers could produce shocks, but we do not find evidence for ongoing mergers in our sample except for a possible faint companion of MRG-M2129, and this explanation seem inconsistent with the spatial uniformity of the [\ion{N}{2}] emission. Therefore AGNs provide the most obvious energy source.

Observations of some low-redshift quiescent galaxies have revealed significant reservoirs of turbulent molecular gas with $\sim$kpc (or larger) sizes and suppressed star formation relative to the Kennicutt--Schmidt relation, often combined with warm shocked gas \citep[e.g.,][]{Alatalo15,Guillard15,Lanz16}. A significant fraction of intermediate-mass quiescent galaxies in the local universe appear to have ionized gas outflows \citep{Cheung16}. The injection of turbulence into the interstellar medium via an AGN jet has been suggested to suppress star formation in these galaxies. Our observations are at least consistent with a similar mechanism acting in MRG-S1522 and MRG-M2129, and potentially in \emph{all} of the galaxies in our sample that have ages $\lesssim 1$~Gyr.

The high fraction of massive systems in which an AGN seems to be affecting the gas---both in high-mass star-forming galaxies at $z \sim 2$ and in recently quenched galaxies---suggests than AGN activity might play an important role in quenching star formation and maintaining low star formation rates in these galaxies, although proving causality is very difficult. Our sample affords detailed observations but is small; more deep spectra of $z \gtrsim 2$ quiescent galaxies are needed to firmly establish the prevalence of emission lines and the variation of their properties with the time since quenching. Observations of the distribution and kinematics of any molecular gas in such galaxies may also be revealing.

\subsection{Future Work}

In four of lensed quiescent galaxies discussed in this paper (all but MRG-S1522), we are able to spatially resolve the stellar continuum in our NIR spectra. In the companion Paper~II, we present the resolved stellar kinematics of these galaxies. In a future paper, we plan to study their resolved stellar populations in order to dissect the star formation histories of early quiescent galaxies. We will also use observations of the extraordinarily bright MRG-M0138 to measure its multi-element stellar abundance pattern. Many of these measurements are currently possible only for quiescent galaxies that are gravitationally lensed. Observations with \emph{JWST} will only marginally resolve a typical compact quiescent galaxy at $z \sim 2$. While such data will be very valuable, lensed quiescent galaxies will continue to offer the highest-resolution views of these galaxies and to provide unique insights into their formation, even in the \emph{JWST} era.

\acknowledgments
We thank the anonymous referee for a close reading and helpful comments. We thank A.~Monna for providing her lens model of MACSJ2129.4-0741 in electronic form, I.~Shivaei for providing the MOSDEF data plotted in Figure~\ref{fig:mainseq}, and J.~Rich for helpful conversations. RSE acknowledges financial support from European Research Council Advanced Grant FP7/669253. Support for programs GO-14496 and GO-14205 was provided by NASA through grants from the Space Telescope Science Institute, which is operated by the Associations of Universities for Research in Astronomy, Incorporated, under NASA contract NAS5-26555. This paper includes data gathered with the 6.5 meter Magellan Telescopes located at Las Campanas Observatory, Chile. Some of the data presented herein were obtained at the W. M. Keck Observatory, which is operated as a scientific partnership among the California Institute of Technology, the University of California and the National Aeronautics and Space Administration. The Observatory was made possible by the generous financial support of the W. M. Keck Foundation. The authors wish to recognize and acknowledge the very significant cultural role and reverence that the summit of Maunakea has always had within the indigenous Hawaiian community.  We are most fortunate to have the opportunity to conduct observations from this mountain. This paper is also based on observations made with the NASA/ESA Hubble Space Telescope, and obtained from the Hubble Legacy Archive, which is a collaboration between the Space Telescope Science Institute (STScI/NASA), the Space Telescope European Coordinating Facility (ST-ECF/ESA) and the Canadian Astronomy Data Centre (CADC/NRC/CSA). 

\bibliographystyle{apj}
\bibliography{lensed_nuggets_I}

\appendix

\begin{figure*}
\centering
\includegraphics[width=\linewidth]{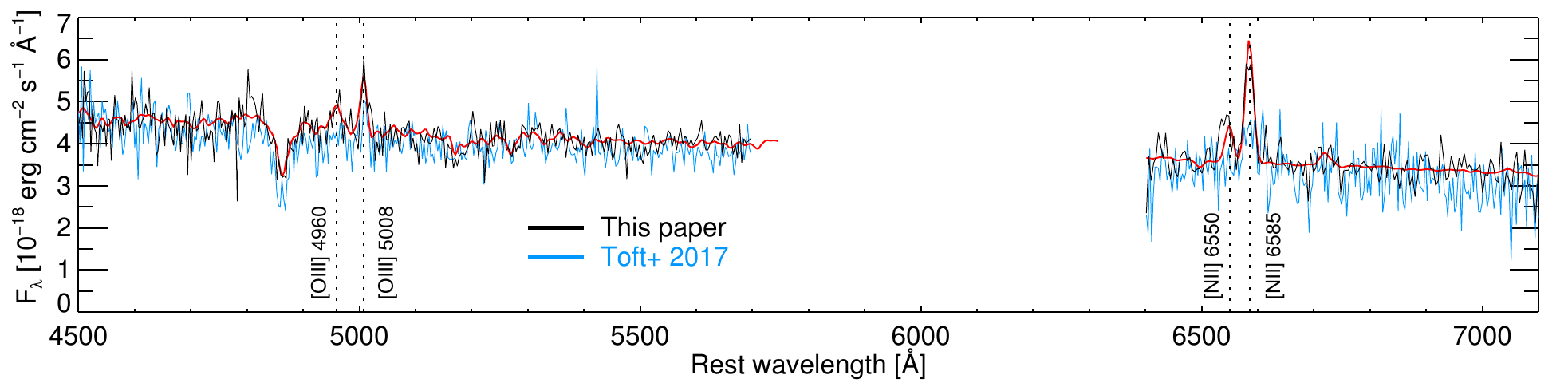}
\caption{Integrated $H$- and $K$-band spectra of MRG-M2129 from this paper (black, with the model from Figure~\ref{fig:specfits2} shown as the red curve) and from \citet[][blue]{Toft17} binned to $\simeq 200$~km~s${}^{-1}$~pixel${}^{-1}$. Significant differences in the EWs of [\ion{O}{3}] and [\ion{N}{2}] are clearly visible.
\label{fig:compare_to_toft_spec}}
\end{figure*}

Here we compare the stellar population, structure, and emission line properties we derived for MRG-M2129 with those recently published by \citet{Toft17}. Despite using quite different formulations of the star formation history, we find stellar population parameters that are consistent with Toft et al. Comparing their Extended Data Table 1 to our Table~\ref{tab:specmeas}, we find that the stellar mass and age are consistent, and although the centers of our posteriors favor a higher metallicity and lower dust attenuation than Toft et al., these too are consistent within the quoted uncertainties.

Our fiducial magnification from \citet{Monna17} is $\mu = 4.5$, which is within 3\% of the Toft et al.~model. The magnification in the more extreme of the two Zitrin models (Section~\ref{sec:m2129lensmodel}) is only 18\% higher. As described by Toft et al., the magnification is relatively well constrained in this system because it is located far from the cluster center where the mass distribution is dominated by the smooth dark matter halo.

Toft et al.~fit a single S\'{e}rsic model to the source, so we compare their parameters to our single S\'{e}rsic+Gaussian model. Like Toft et al., we infer a nearly exponential ($n=1$) surface brightness profile with a compact size. The effective radius $R_{e,{\rm maj}}$ and PA are consistent within $\simeq 6\%$ and $6^{\circ}$, respectively. The main difference is that our reconstruction gives a much flatter source with $b/a = 0.29 \pm 0.03$ compared to their $0.59^{+0.03}_{-0.09}$. In correspondence with Toft and co-workers, it was found that a likely explanation is the manner in which the PSF was treated in their analysis. Once a discrepancy in their treatment was corrected, they found a value $b/a \simeq 0.4$ that is much closer to our measurement (S.~Toft., A.~Man, et al., private communication).

Finally, Toft et al.~find a relatively high [\ion{N}{2}]/H$\alpha$ ratio coupled with an emission line velocity dispersion in excess of the stars. Like us, they interpret this as possible evidence of AGN-driven turbulence or outflows. Beyond this qualitative agreement, there are substantial quantitative differences. Toft et al.~find log~[\ion{N}{2}]/H$\alpha = -0.06 \pm 0.10$ versus our $0.79 \pm 0.16$. Figure~\ref{fig:compare_to_toft_spec} shows that this difference arises from manifestly different emission line EWs at the level of the raw spectra. Although our spectrum is deeper, the differences do not seem consistent with noise fluctuations, especially for [\ion{N}{2}]. This suggests some systematic effect in the observations or data reduction.

The presence of the [\ion{N}{2}] and [\ion{O}{3}] doublets with the correct ratios rules out the stronger lines in our spectra as arising from a simple data reduction artifact. The extraction apertures used to produce the spectra in Figure~\ref{fig:compare_to_toft_spec} are nearly matched ($\pm1\farcs6$ for this paper and $\pm 1\farcs4$ for Toft et al.). This fact, combined with the weak spatial variation in [\ion{N}{2}] EW that we see along the arc (Figure~\ref{fig:resolved_ew}), makes it hard to understand these differences in the spectra as arising either from target acquisition errors affecting one of the observations, or from unintended self-subtraction of the galaxy wings during the data reduction. In addition, unlike Toft et al.~who find that the emission lines are redshifted by 238 km~s${}^{-1}$ relative to the stars, we find no such velocity offset in the integrated spectrum ($v_{\rm em} - v_{\rm stars} = -12 \pm 30$ km~s${}^{-1}$), and we also do not detect \ion{He}{2} $\lambda$5413. Future integral field observations are desirable to investigate the possible origins of the differences.

\end{document}